\documentclass[12pt,preprint]{aastex}

\newcommand\iso[2]{$^{\rm #1}$#2}

\def\teff{\mbox{$T_{\rm {eff}}$}}

\def\vt{\mbox{v$_{\rm t}$}}
\def\BmV0{\mbox{(B-V)$^{\rm o}$}}
\def\VmK0{\mbox{(V-K)$^{\rm o}$}}
\def\MV0{\mbox{M$_{\rm V}^{\rm o}$}}

\def\Msun{\mbox{M$_{\odot}$}}

\def\deg{{$^{\circ}$}}
\def\bd17{\mbox{BD +17\deg 3248}}
\def\cs22{\mbox{BPS CS 22892-052}}

\begin{document}

\title{The Rise of the $s$-Process in the Galaxy}

\author{Jennifer Simmerer and Christopher Sneden}
\affil{Department of Astronomy and McDonald Observatory, University of Texas, Austin, TX 78712}
\email{jensim@astro.as.utexas.edu}
\email{chris@astro.as.utexas.edu}
\author{John J. Cowan and Jason Collier}
\affil{Department of Physics and Astronomy, University of Oklahoma, Room 131 Nielsen Hall, Norman, OK 73019}
\email{cowan@nhn.ou.edu}
\email{collier@nhn.ou.edu}
\author{Vincent M. Woolf} 
\affil{Astronomy Department,University of Washington, Box 351580, Seattle, WA 98195}
\email{vmw@astro.washington.edu}
\and
\author{James E. Lawler}
\affil{Department of Physics, University of Wisconsin, Madison, 1150 University Avenue, Madison, WI 53706}
\email{jelawler@wisc.edu}

\received{2004 March 25}
\accepted{2004 July 9}

\begin{abstract}
From newly-obtained high-resolution, high signal-to-noise ratio spectra the abundances of the elements La and Eu have been determined over the stellar metallicity range $-3<$[Fe/H]$<+0.3$ in 159 giant and dwarf stars. Lanthanum is predominantly made by the $s$-process in the solar system, while Eu owes most of its solar system abundance to the $r$-process. The changing ratio of these elements in stars over a wide metallicity range traces the changing contributions of these two processes to the Galactic abundance mix.  Large $s$-process abundances can be the result of mass transfer from very evolved stars, so to identify these cases, we also report carbon abundances in our metal-poor stars.  Results indicate that the $s$-process may be active as early as [Fe/H]$=-2.6$, alalthough we also find that some stars as metal-rich as [Fe/H]$=-1$ show no strong indication of $s$-process enrichment.  There is a significant spread in the level of $s$-process enrichment even at solar metallicity.
\end{abstract}


\section{Introduction}

Abundances of elements determined in stars over a very wide metallicity range contain vital clues to the chemical evolution of our Galaxy.  This is especially true for the neutron capture ($n$-capture) elements, those heavier than the Fe peak (Z$>$30). 
Syntheses of $n$-capture elements occur in a variety of fusion episodes in the late stages of stellar evolution, with neutron densities that range over factors of 10$^{15}$.

During quiescent He-burning, neutrons generated via the  \iso{13}{C}($\alpha$,n)\iso{16}{O} and \iso{22}{Ne}($\alpha$,n)\iso{25}{Mg} reactions are captured by heavy element seed nuclei.  The neutron densities are relatively low (N$_n\sim$10$^{8}$~cm$^{-3}$; \citealt{Pagel1997}), so nearly all possible $\beta$-decays will have time to occur between successive neutron captures.  Synthesis of successively heavier isotopes progresses along the ``valley of $\beta$-stability''.  This synthesis route is called the $s$-process, and it is responsible for about half of the isotopes of the $n$-capture elements.  Much larger neutron densities ($\geq$10$^{23}$~cm$^{-3}$; \citealt{Pagel1997}) 
are generated via p~+~e~$\rightarrow$ n~+~$\nu$ during the high-mass-star core collapse that results in Type~II supernovae (SNe).  Extremely neutron rich nuclei are formed in a matter of seconds; in this so-called $r$-process, it is not necessary to have pre-existing heavy-element seed nuclei.  The $n$-capture rates exceed $\beta$-decay rates, creating nuclei out to the neutron drip line.  These extremely neutron rich nuclei experience multiple $\beta$-decays back to the valley of $\beta$-stability after the very quick extinction of $n$-capture events as the Type~II SN envelopes are ejected.  The $r$-process is also responsible for about half of the solar system
$n$-capture isotopes but not always the same ones created in the $s$-process.
Thus, $n$-capture elements can be composed of some pure $r$-process, pure $s$-process, and some mixed-parentage isotopes.

Abundance studies of very low metallicity stars cannot measure isotopic abundances from typical absorption lines; the isotopic wavelength offsets are usually very small compared with other line-broadening effects. The only exceptions for the Z$>$30 elements to date are Eu \citep{Sneden2002, Aoki2003} and Ba \citep{Magain1995, Lambert2002}.  Therefore, to help understand the relative importance of the two $n$-capture synthesis mechanisms throughout the history of the Galaxy, stellar elemental abundance comparisons are made between elements whose solar system isotopic abundances are dominated by the $s$-process and those due mainly to the $r$-process.

Beginning with \citet{Spite1978}, such comparisons have nearly always focused on the relative abundances of Ba (Z=56, $\approx$85\% $s$-process in the solar system) and Eu (Z=63, $\approx$97\% $r$-process; see \citealt{Burris2000} and references therein).  The massive stars that are althought to provide the site of the $r$-process evolve on much shorter time scales than the lower mass stars that manufacture $s$-process elements.  On the average, then, older stars ought to show a pure or almost pure $r$-process signature \citep{Truran1981}.  While simple in principle, observational evidence for the onset of the $s$-process in a general way is still lacking.  A recent and extensive study of $n$-capture elements in metal-poor giant stars by \citet{Burris2000} offers a somewhat inconclusive answer to the question of when low-mass stars begin to contribute to the Galactic chemical mix.    They found that  the $s$-process may begin to contribute at lower metallicities than expected, an effect somewhat blunted by the considerable range in Ba/Eu abundance ratios at any given metallicity.   In contrast, \citet{Mashonkina2003} found that the onset of the $s$-process occurred at higher metallicities and that evolution of the $s$-/$r$-process ratio could be tied to Galactic stellar populations.  There is much less variation in the Ba/Eu ratio in the latter study.  

Unfortunately, the observational and analytical uncertainties in [Ba/Eu] abundance ratios
are substantial and may contribute considerably to the ambiguity of previous results.  Often, Ba abundances are derived from just a handful of \ion{Ba}{2} transitions (mainly the 4554~\AA\ resonance line, now sometimes accompanied by the 5853, 6141, and 6496~\AA\ lines), and Eu abundances from the \ion{Eu}{2} transitions
at 4129 and 4205~\AA ~\citep{Francois1996, McWilliam1998, Burris2000}. Detailed hyperfine and isotopic substructure analyses have been 
published for \ion{Eu}{2} features along with reliable transition probabilities, so accurate Eu abundances may be routinely determined \citep{LawlerEu}.  Barium abundance uncertainties now dominate the problem.  The near-ground-state \ion{Ba}{2} lines most commonly measured are usually very strong, while higher excitation lines (at 3891, 4130, and 4166~\AA) are almost all undetectable.
The low-excitation transitions are often saturated, yielding Ba abundances that are very sensitive to adopted values of stellar microturbulent velocities.
In In addition, Ba has five abundant isotopes (\iso{134}{Ba}-\iso{138}{Ba}) which are synthesized in different proportions by $r$- and $s$-processes.  Therefore, elemental abundances depend on the assumed $r$/$s$ fraction, which is what one is ultimately attempting to determine; these difficulties are especially acute for the 4554~\AA ~line.
From these effects, it is difficult to assess the Ba abundances in most stars to better than $\pm$0.2~dex. It is not at all obvious whether observed star-to-star scatter in [Ba/Eu]\footnote{We adopt the standard spectroscopic notation where   [A/B]=[log$_{10}\epsilon(A)-$log$_{10}\epsilon(B)]_*-[$log$_{10}\epsilon(A)-$log$_{10}\epsilon(B)]_{\odot}$, where log$_{10}\epsilon(X)=$log$_{10}(N_X/N_H)+12$ . However, for the purposes of this paper, it is usually easier to work in the log$~\epsilon$ notation, where  log$_{10}\epsilon(A/B)=$log$_{10}(N_A/N_H)-$log$_{10}(N_B/N_H)$. }
ratios at a given [Fe/H] metallicity is primarily indicative of astrophysical variations or Ba measurement uncertainties.

Fortunately, neighboring rare earth elements whose solar system abundances are mainly due to the $s$-process may be easily observed in stars over a wide metallicity range.
In particular, La (Z=57, $\approx$75\% $s$-process in the solar system;
\citealt{Burris2000}) has many detectable features of \ion{La}{2}. 
This species has enjoyed a recent laboratory study that yielded accurate transition probabilities and hyperfine structure constants \citep{LawlerLa}.
There are \ion{La}{2} lines of different strength throughout the near-UV to red spectral regions, and all of them yield consistent abundances from the solar spectrum and from $r$-process-rich, low-metallicity stars such as \bd17, \cs22, and BPS BPS CS 31082-001 \citep{LawlerLa, CS22892, Hill2002}.  Finally, \iso{139}{La} is the only abundant naturally-occurring isotope of La,
so $r$- and $s$-process contribution estimates are irrelevant to the determination of total elemental La abundances \citep{Lodders2003}.

In this paper we report new La and Eu abundances for 159 giant and dwarf stars.  Our aim is to refine measurements of the $s$-/$r$-process ratio to the point where measurement uncertainties no longer obscure the onset of nucleosynthetic contributions from asymptotic giant branch (AGB)  stars.  Our data also allow us to address $s$-process evolution in the Galactic stellar populations that we have sampled.  In \S \ref{obs} we present the spectroscopic data, in \S \ref{ana} we discuss the abundance analyses, in \S \ref{compare} we discuss our results in relation to previous work, and in \S \ref{disc} we discuss the Galactic trend in La/Eu ratios and the relationship between La/Eu and stellar kinematics.  Section \ref{con} contains some concluding remarks.

\section{Observations and Data Reduction \label{obs}}
Stars were initially selected to have $-2.5\leq$[Fe/H]$\leq-1.0$, alalthough our analysis scattered some stars to  higher or lower metallicities.  Program stars were taken from the objective prism survey of \citet{Bond1980} and the high proper-motion survey of \citet{Carney1994}.   Our sample therefore includes both main-sequence and evolved stars in roughly equal proportions.  There  is  a weak correlation between evolutionary state and [Fe/H] in our final sample, such that the more metal-poor stars ($-3.0\leq$[Fe/H]$\leq-1.7$) are more likely to be evolved stars and the more metal-rich stars ($-1.2\leq$[Fe/H]$\leq-0.5$) are more  likely to be unevolved stars.  In order to maintain a high signal-to-noise ratio (S/N) in all our spectra, our program was limited to stars with V$\leq$11.  Of the 98 stars observed for this program, 92 proved to have measurable lines of La or Eu or both.  A further four dwarf stars were rejected from the  sample after having been identified by \citet{Latham2002} as single-lined spectroscopic binaries.  The  final sample is composed of 88 metal-poor stars.

Spectra of these stars were obtained with the McDonald Observatory 2.7-m telescope, using the ``2-d coud\'{e}'' cross-dispersed echelle spectrograph configured to a 2 pixel resolving power of R$\sim$60,000 and a Tektronix 2048x2048 CCD.   This instrument allows full optical wavelength coverage in the blue, alalthough in some cases red spectral features (i.e., the 5797~\AA~\ion{La}{2} or the 6432~\AA~ \ion{Eu}{2} line)  were lost off the detector chip and could not be included in our analysis.  Most spectra have S/N$\sim$100 pixel$^{-1}$ at 4100~\AA,  with a few cool  stars having lower S/N in the blue part of the spectrum (where the majority of measurable La and Eu lines lie).  However, cooler stars have deeper lines, which aids the measurement process in these  cases.

In addition to the metal-poor stars observed, 67 Galactic disk stars observed by \citet{Woolf1995} were added to our sample.  These stars span a higher metallicity range, $-0.85\leq$[Fe/H]$\leq+0.20$, and are all unevolved stars.  Stellar parameters from \citet{Woolf1995} have been adopted without changes; the details of those derivations can be found in their paper.  The \citet{Woolf1995} study, although conducted with the same telescope and instrument, employed a different CCD chip and instrument configuration.  In order to investigate the effects of this and differing analysis methods, we re-observed five stars from the Woolf sample and analyzed them according to the same methods used for the stars in the lower metallicity sample.

Bias subtraction, flat-field corrections, and wavelength calibrations were performed using the standard routines available in IRAF \footnote{IRAF is distributed by the National Optical Astronomy Observatories, which are  operated by the Association of Universities for Research in Astronomy, Inc., under cooperative agreement with the
National Science Foundation.}.  We used the software package SPECTRE \citep{Fitzpatrick1987} for the remainder of the stellar data reductions, to eliminate anomalous radiation events and
co-add and normalize spectra with spline function fits to interactively chosen continuum points.  Spectra of lower quality or cool stars with lower flux in blue portions of the spectra were smoothed by convolving a Gaussian of 2 pixel FWHM with the observed spectra.    We also used SPECTRE to measure equivalent widths (EWs) of spectral lines, either by fitting a Gaussian profile to the line or by a simple integration.

\section{Analysis \label{ana}}
Our abundance analysis is a combination of spectrum synthesis and EW analysis, both of which require a model of the star's atmosphere.  The model atmospheres in turn are characterized by four input parameters: \teff, log$~g$, [Fe/H], and \vt.  It is common for these quantities to be derived from \ion{Fe}{1} and \ion{Fe}{2} EWs.  However,  we have used other methods to obtain independent estimates of these quantities, which were then checked against the standard spectroscopic constraints.  In the following sections we describe our initial estimates of stellar model parameters and the confirmation methods we used to determine final values and elemental abundances.   At this point we introduce our summary table, Table \ref{table1}, which contains initial stellar parameter estimates, final parameters, and the abundance ratios we have  ultimately derived.

\subsection{Stellar Models}
Abundances were derived using Kurucz stellar atmosphere models with no convective overshoot \citep{Castelli1997}.  Models were interpolated to a particular set of pre-determined stellar parameters (see next sections) using software provided by A. McWilliam (1990, private communication) and I. Ivans (2002, private communication).  For those stars for which model atmosphere parameters were chosen by EW analysis of \ion{Fe}{1} and \ion{Fe}{2} lines, models were tested in steps of 25 K in \teff~and  0.25 dex in log $g$.

\subsection{Line Lists}
\subsubsection{Iron \label{iron}}Stellar parameters were checked with EW measurements of a small list of \ion{Fe}{1} and \ion{Fe}{2} lines.  The 11 \ion{Fe}{2} and  21 \ion{Fe}{1} lines that were selected from \citet{Grevesse1999} are listed in Table \ref{table2}.  Since we have sampled a wide range of metallicity and temperature with the same line list, the number of \ion{Fe}{1} and \ion{Fe}{2} lines used to derive [Fe/H] for a particular star varies as indicated in Table \ref{table1}.  

\subsubsection{Lanthanum and Europium}
The \ion{La}{2} and \ion{Eu}{2} transitions measured in this study are also listed in Table \ref{table2}.  The spectral lines of La and Eu both have hyperfine components of varying separation and intensity, which make an EW analysis inappropriate.  In addition, Eu has two naturally occurring isotopes that are abundant in approximately equal proportions \citep{Lodders2003}, further contributing to the broadening of Eu lines \citep{LawlerEu,Sneden2002,Aoki2003}.  As mentioned earlier, La has only one abundant  naturally occurring isotope.  Abundances from strong Eu and La lines cannot be accurately measured by single-line equivalent widths and are therefore more properly obtained by synthesizing the various hyperfine and isotopic components of the lines and fitting the syntheses to the observed spectrum.   For the sake of uniformity, we have synthesized all La and Eu lines, even those weak lines usually deemed acceptable for EW analysis.  The transitions measured here are taken from those used to measure the solar photospheric abundances in \citet{LawlerLa, LawlerEu}.  Most of the stars in our temperature and metallicity range have at least two measurable \ion{Eu}{2} and four or five \ion{La}{2} lines.  Fewer \ion{Eu}{2} lines are typically measurable, as some of the strongest \ion{Eu}{2} transitions are often in the wings of very strong absorption features.  This is especially a problem with the  line at 3819~\AA,  and, although it is quite strong, an accurate abundance can rarely be easily derived from it.  In the Woolf et al. sample, only one line of La and one line of Eu were available for measurement.

We list in Table \ref{table3} and Table \ref{table4} the abundances derived from each La and Eu line in our metal-poor stars.  The final two columns of these tables list the average La or average Eu abundance and the $\sigma$ of each.  We use these average values in Table \ref{table2} and in our figures.  

\subsubsection{Carbon}
We derived carbon abundances from synthetic spectrum analyses of selected portions of the CH A$^2 \Delta - X^2$ G band, adopting the atomic and molecular line lists from the Kurucz web site.  We synthesized a 30~\AA~region centered on 4315~\AA, stepping the carbon abundance in increments of 0.50 dex.  This line list is the one employed by \citet{Westin2000}.  When used in conjunction with a Kurucz solar no-overshoot model atmosphere, it produces a solar carbon abundance of log$\epsilon $(C)=8.61.  We have used this value when calculating [C/Fe] (see Table \ref{table1}).   

\subsection{Stellar Parameters}
\subsubsection{Effective Temperature by Means of the Infrared Flux Method}
Effective temperature (\teff) was determined using the Infrared Flux Method (IRFM) calibrations of \citet{AlonsoDwarf} (dwarf stars) and \citet{AlonsoGiant} (giant stars).   Many program stars have published IRFM temperatures calculated by Alonso et al. with their own measured colors, which we adopt whenever they are consistent with our EW analysis (see next sections).  For program stars not observed by Alonso et al., \teff~was derived using the calibration $V-K$ (chosen because it is especially insensitive to the choice of metallicity).  We have used Two Micron All-Sky Survey (2MASS)  $K$ magnitudes and Tycho-2 $V$ magnitudes as these were available for all our program stars.  However, this raises the possibility of systematic offsets between those temperatures calculated  with \citet{AlonsoDwarf,AlonsoGiant} colors and those calculated with colors from other sources.  In order to determine and correct systematic offsets in our derived temperatures, we did the following:

1)  We  obtained $V$ and $K$ magnitudes for the entire \citet{Carney1994} and \citet{Bond1980} catalogues.  The Tycho-II $V$ magnitude was transformed according to $$V_J=V_T-0.09*(B_T-V_T)$$ \citep{tycho2}.  The 2MASS $K$ was not transformed onto the standard system, since the correction is smaller than the quoted measurement error for stars in this magnitude range.  Colors were corrected with reddenings from \citet{Twarog1994} when necessary.  For nearby (i.e., dwarf) stars, reddening is a negligible effect.  

2)  \teff~was derived for all stars with $V-K$ in the range covered by the calibration.  The \teff~calibrations also require an estimate of [Fe/H].  We adopted the metallicities provided by \citet{Carney1994} and \citet{Twarog1994}.  Although these were not always consistent with our final measurement of [Fe/H] (see \S \ref{metal}), gross disagreements were few and corrected iteratively when necessary.

3)  Our calculated temperatures  were compared with those published by \citet{AlonsoDwarf,AlonsoGiant} for overlapping stars.  \citet{AlonsoDwarf}  calibrations hold for unevolved stars; the mean offset for high-gravity stars is $-17$ K, in the sense \teff~(this study)$-$\teff~(Alonso).  For evolved stars, the \citet{AlonsoGiant} calibrations were used.  The mean offset for low-gravity stars is $+30$ K, again in the sense of \teff~(this study)$-$ \teff~(Alonso).

4) The mean offsets were applied to the calculated \teff values of all stars in our program.  We note that the mean offsets are quite small and well within the error of the calibrations themselves.  Also, for both giant and dwarf stars, $\sigma_{\rm{offset}} \approx 200$K.

\subsubsection{Spectroscopic Constraints on Effective Temperature}

Stellar effective temperatures can also be derived from spectroscopic constraints: specifically, the requirement that the abundances  from individual atomic lines be independent of the excitation potential of the line.  This is commonly done  (in this temperature and gravity regime) by measuring many EWs of  some iron peak element (usually \ion{Fe}{1}),  as a large number of these  lines at various excitation potentials are easily measured at optical wavelengths.  Stellar models are tested iteratively, until a set of parameters is found that eliminates any correlation between abundance value and excitation potential for the set of lines.  

Individual stars may have, for various reasons, IRFM temperatures that are not in agreement  with the spectroscopic constraints typically applied when \teff~is derived from EWs.  In order to identify errors in colors, IRFM \teff~values were checked with a small list of \ion{Fe}{1}  lines.  This same list, with the addition of \ion{Fe}{2} lines, served to verify log $g$ and determine [Fe/H] and stellar microturbulence (\vt) as described in the following sections.  In those cases in which a stellar model with an IRFM temperature showed a strong trend in \ion{Fe}{1} abundance with excitation potential, the temperature was modified until no trends appeared.  Our adopted temperatures, calculated IRFM temperatures, and the Alonso et al. published temperatures are given in Table \ref{table1}.

\subsubsection{Surface Gravity \label{logg}}

We have obtained two independent estimates of stellar surface gravity: the  so-called ``physical gravity",  calculated from the standard relation  
$$\rm{log}~g_*=0.4*(M_{V*}+BC-M_{Bol\odot})+log~g_{\odot} + 4*log (\frac{T_{eff *}}{T_{eff \odot}})+log(\frac{m_*}{m_{\odot}})$$
and a ``spectroscopic gravity",  set by requiring that [\ion{Fe}{1}/H]=[\ion{Fe}{2}/H].  Each method has its drawbacks.  The physical gravity is limited by the accuracy of the distance measurement, and tends to be most useful for nearby stars with accurate parallaxes.  The spectroscopic method is plagued by the possibility of departures from LTE.  Several recent studies have called into question the validity of forcing ionization equilibrium for nearby stars where both methods can be used.  \citet{Thevenin1999} have suggested that the non-LTE (NLTE) effects on Fe lines in dwarf stars can be severe, resulting in large discrepancies in standard analyses between the derived abundance of [\ion{Fe}{1}/H] and [\ion{Fe}{2}/H] for very metal-poor stars.  The extent of NLTE effects on derived Fe abundances in evolved stars has not yet been determined.  Therefore, for those stars with accurate parallaxes, we set log~$g$ from the standard relation shown above.  The stellar mass, $m_*$, is either taken from \citet{Carney1994} (for high-gravity stars) or assumed to be 0.8 $m_{\odot}$ (low-gravity stars), the bolometric correction BC is from \citet{BCdwarf, AlonsoGiant}, and we have adopted 4.75 as the bolometric magnitude (M$_{Bol}$) of the Sun.  The value of M$_{V*}$ can be derived from
$$\rm{M}_{V*}=\rm{m}_V + 5 -5*\rm{log}(\frac{1}{\pi}) - 3.2*E(B-V)$$  where $\pi$ was taken from Hipparchos  \citep{hipp}.  

For the more luminous (and usually more distant) stars, Hipparchos parallaxes are insufficient, since the quoted error is often larger than the measurement itself.  In these cases, we have used the distances and reddenings derived from Str\"{o}mgren colors by \citet{Twarog1994}.  There are some cases in which gravities derived from the distances given by \citet{Twarog1994} do not appear to give sensible results, in that the abundance of Fe derived from \ion{Fe}{1} is significantly different from the abundance derived from \ion{Fe}{2} (and there is no reason to suppose that this is limited to Fe).  In those cases where the abundance difference between [\ion{Fe}{1}/H] and [\ion{Fe}{2}/H] was large (larger than has yet been attributed to NLTE effects; this is $\sim$ 0.40 dex or more), a new gravity was adopted, chosen by forcing [\ion{Fe}{1}/H] and [\ion{Fe}{2}/H] into agreement.   These cases are likely due to very high reddening values, which would make \citet{Twarog1994} distances more uncertain.  

Some stars are in neither the Hipparchos catalogue nor the \citet{Twarog1994} work; these are primarily mildly metal-poor dwarfs and subgiants.  Whenever it was necessary to force agreement, we accepted the value of log$~g$ that put the difference in [\ion{Fe}{1}/H] and [\ion{Fe}{2}/H] within 2$\sigma$ of the  [\ion{Fe}{2}/H] abundance.  Except for the coolest, most metal-poor stars (for which Fe lines were weak or blended), this amounted to less than 0.3 dex (with one exception: see \S \ref{metal}), and no star had a difference exceeding 0.35 dex.  Those stars for which log~$g$ was not derived directly from a parallax or distance measurement have been indicated in Table \ref{table1}.  

\subsubsection{Metallicity \label{metal}}
We have derived stellar metallicities from a small list of \ion{Fe}{1} and \ion{Fe}{2} lines (see section \ref{iron}).  Because we have not forced the \ion{Fe}{1} and \ion{Fe}{2} abundances to agree, there may be some ambiguity as to the actual value of the Fe abundance.  In choosing stellar models, the adopted Fe abundance is the mean Fe value, weighted by the number of lines measured.  For stars more metal-poor than [Fe/H]$\sim -1$, model metallicities were increased by 0.1-0.25 dex to simulate $\alpha$-enhancements \citep{Fulbright1999}.  

On the basis of the work of \citet{Thevenin1999}, some analyses have adopted [\ion{Fe}{2}/H] as the ``true'' Fe abundance (for stars in the F, G, and K temperature ranges) owing to the smaller influence of NLTE effects on this species and the fact that \ion{Fe}{2} is the dominant Fe species. In particular, \citet{Kraft2003} adopt this convention and note that although \citet{Thevenin1999} attribute some of the discrepancy in \ion{Fe}{1} and [\ion{Fe}{2}/H] to overionization by UV photons in dwarf stars, the sign of $\Delta$ [Fe/H](=[\ion{Fe}{1}/H]$-$[\ion{Fe}{2}/H]) in giant stars is not always consistent with this.  According to \citet{Thevenin1999}, the derived abundance of \ion{Fe}{2} should be consistently higher than that of \ion{Fe}{1}, but \citet{Kraft2003} have found that this is not necessarily the case.  \citet{Kraft2003} attribute this to the failure of one-dimensional LTE model atmospheres to accurately represent the abundance of \ion{Fe}{1}. 

By setting log$~g$ from distance measurements, we have largely avoided this issue.  However, we do find that there is a systematic offset in $\Delta$ [Fe/H] that is most strongly dependent on \teff.  In Fig. \ref{params} we plot the difference in Fe as measured from [\ion{Fe}{1}/H] and [\ion{Fe}{2}/H].  At low temperatures, the derived [\ion{Fe}{1}/H] is as much as 0.30 dex lower than the derived [\ion{Fe}{2}/H].  The difference (on average) effectively disappears for stars with \teff$>5200$ K, although even at higher temperatures there are individual stars with large $\Delta$ [Fe/H].  \citet{Thevenin1999} found that the more metal-poor stars tended to have larger disagreements in $\Delta$ [Fe/H], although there were temperature dependencies as well.  While it is tempting to attribute these trends solely to temperature effects, the coolest stars are almost all giants stars and it is at these temperatures that we see trends in $\Delta$[Fe/H].  We can not therefore conclusively distinguish either \teff~or log~$g$ as the controlling variable.  It is also not clear that the same mechanism is responsible for the discrepancies in the dwarf stars in \citet{Thevenin1999} and the giant stars studied here.   We note that \citet{Yong2003} found that metal-poor globular cluster stars showed a similar behavior in \teff, with $\Delta$ [Fe/H] becoming more pronounced in the coolest giant stars. 

With our Fe line list and for \teff=5780 K, log~$g$=4.45, and \vt=0.90 km/s, we derive $\Delta$[Fe/H]=0 in the Sun.  However, Fig. \ref{params} reveals that we ought not to expect anything different for those parameters.  One star in particular stands out in this figure, the dwarf star HD 188510.  It has almost the largest $\Delta$ [Fe/H] (+0.3 dex) in our sample, yet it also has a well-determined Hipparchos parallax ($\pi=25.17$ mas).  While this star is at the high \teff, high log$~g$ end of our sample, other stars with similar parameters have a much smaller $\Delta$ [Fe/H].  This star does not appear to be unusual in any other respect, and \citet{Latham2002} does not identify this star as a binary. There is, however, a significant difference in the $K$ magnitudes reported by 2MASS and and that of \citet{AlonsoDwarf}.  2MASS finds $K$=7.854 (V-K=1.62) and Alonso et al. use $K$=7.13 (V-K=1.69).  The value of Alonso et al. makes the star 200 K hotter than the 2MASS colors would imply, and the lower  temperature results in a higher metallicity overall (and a smaller $\Delta$ [Fe/H]).  However, our methods find that both temperatures are acceptable for this star, and we have adopted the published Alonso value whenever possible for consistency.  We do not find any significant offset in K between Alonso et al. and 2MASS, but values for individual stars may vary.  However, La/Eu is not affected by Fe uncertainties of this magnitude (see \S \ref{errors}).  

We have chosen to use [\ion{Fe}{2}/H] in most of our figures and in our analysis discussions, mainly because we have measured \ion{La}{2} and \ion{Eu}{2}.  However, our conclusions are not altered by this choice; trends in La/Eu and [\ion{Fe}{1}/H] are very similar.  We report both [\ion{Fe}{1}/H] and [\ion{Fe}{2}/H] in Table \ref{table1}.  

\subsubsection{Microturbulence}

The microturbulence was set by requiring that the derived Fe abundance be independent of the quantity log $(EW/\lambda)$.  Derived values of \vt~less than $\sim 0.40~$km/s were not accepted, as this is equal to a typical error on derived velocity shifts in our spectra.  Generally, more \ion{Fe}{1} lines are measurable, and they cover a wider range in $(EW/\lambda)$ than the measured \ion{Fe}{2} lines.  For this reason, the microturbulence derived from \ion{Fe}{1} was given more weight whenever mild disagreements arose.  

\subsubsection{M$_V$}

The majority of the dwarf stars have good Hipparchos parallaxes, and for these stars M$_V$ can be found from the standard relation given in \S \ref{logg} (for stars within $\approx$1 kpc, reddening was ignored).  \citet{Twarog1994} report M$_V$ for the stars for which they derived reddenings.  As these giant stars, while luminous, are generally very distant and have poorly determined parallaxes, the Twarog \& Twarog value of M$_V$ was adopted wherever it did not conflict with the log$~g$ implied by our Fe line list.  The remaining stars have no reliable distance measurement, so an M$_V$ was inferred from the adopted stellar parameters--we simply invert the relation in \S \ref{logg} to recover M$_V$.  While \teff~for these stars may be set either from Fe lines or from the IRFM, log$~g$ must be set from Fe lines.  This method of model selection is less precise and the error in the implied M$_V$ is much greater than for those stars with distance measurements.  

Our sample covers a wide range of evolutionary states, from unevolved stars on the main sequence to red giants, as shown in Fig. \ref{mv}.  \citet{Behr2003} identified seven of our sample stars as horizontal branch (HB) stars (HD 025532, HD 082590, HD 105546, HD 166161, HD 119516, HD 184266, and HD 214362).  These stars are boxed in Fig. \ref{mv}.

\subsubsection{Kinematics Calculations}

The majority of our stars (even the evolved stars) have proper motions available from the Tycho-2 catalogue.  Radial velocity measurements were obtained through the SIMBAD database.  Most of the bright, nearby dwarf stars have Hipparchos parallaxes, but many of the evolved stars in the sample are too distant to have good measurements.  For these stars, we use the ``implied" distance from \teff~and log$~g$, as described in \S \ref{dist}.  Stars with ``implied" distances will necessarily have larger errors in their kinematics (see \S \ref{kinerr}).  

With this information, we have computed the stellar velocities $U$, $V$, and $W$ according to \citet{Johnson1987}.  \citet{Johnson1987} adopt a right-handed coordinate system, where $U$ (motion radial to the Galactic Center) is defined as positive away from the GC, $V$ (orbital motion around the GC) is defined as positive counter-clockwise, and $W$ (motion in and out of the plane of the disk) is defined as positive northward.  Our values and errors are given for all the stars in our sample in Table \ref{table5}.  The velocities listed have been corrected for the solar motion (U$-9$,V$+12$,W$+7$;\citealt{Mihalas1968}).

\subsection{Error Analysis \label{errors}}

\subsubsection{Stellar Parameters}

Uncertainties in the stellar model parameters may directly affect other derived stellar parameters,  as in the case of \teff~and log$~g$.  Other effects may be more subtle, and errors in each stellar input parameter can propagate through to affect the final elemental abundance.  We find that the particular element ratio La/Eu is insensitive to these uncertainties, and we illustrate this with the following examples.

For the typical giant star HD 105546, a change of $\pm$100 K in \teff~(a $\sim 2 \sigma$ error in the \citet{AlonsoDwarf, AlonsoGiant} calibrations) produces a change in the calculated log$~g$ of $\pm$0.04 dex, resulting in a cumulative change of $\pm 0.10$ dex in [\ion{Fe}{1}/H],  an $\mp 0.02$ dex in [\ion{Fe}{2}/H], and an increase of 0.20 km/s in the  chosen value of \vt.  In this particular star, the change in log$~g$ increased the difference between [\ion{Fe}{1}/H] and [\ion{Fe}{2}/H].  Cooler stars show a larger change in log$~g$, reaching $\pm 0.1$ dex for \teff$\approx 4000$K. For HD 105546, a change of +100 K in \teff~and 0.04 dex in log$~g$, along with the corresponding changes in [Fe/H] (+0.06 dex) and \vt (+0.20 km/s), produce +0.05 dex change in log$~\epsilon$({\ion{La}{2}) and +0.06 dex change in log$~\epsilon$(\ion{Eu}{2}), which is a $-0.01$ dex change in log$~\epsilon$(La/Eu).  For the dwarf star G102-020 (at similar \teff~and [Fe/H]), an increase  of +100K in \teff~and +0.03 dex in log$~g$ gave almost identical results, making [Fe/H] +0.05 dex larger (a change of +0.08 and $-0.02$ dex in [\ion{Fe}{1}/H] and [\ion{Fe}{2}/H], respectively).  The result was an increase in both log$~\epsilon$(\ion{La}{2}) and log$~\epsilon$(\ion{Eu}{2}) of +0.05 dex, making no change in log$~\epsilon$(La/Eu).  

For those stars for which distance estimates were unreliable, unavailable, or that produced a log$~g$ inconsistent with the $\Delta$ [Fe/H] we require, log$~g$ was tested in 0.25 dex increments.  An increase in log$~g$ of this magnitude (again for HD 105546, and keeping \teff~constant) drives [\ion{Fe}{2}/H] +0.10 dex higher while leaving [\ion{Fe}{1}/H] unchanged.  This alters the average [Fe/H] very little, only +0.03 dex--well within the line-to-line spread of each ion.  Our choice of \vt~is unchanged.  The cumulative changes increase  log$~\epsilon$(\ion{La}{2}) by 0.09 dex and  log$~\epsilon$(\ion{Eu}{2}) by 0.11 dex, decreasing log$~\epsilon$(La/Eu) overall by 0.02 dex.  For those stars with an IRFM \teff~inconsistent with our EW Fe  analysis, \teff~was tested in increments of 25 K, which produces negligible changes in log$~g$ and [Fe/H].  The result was small changes in log$~\epsilon$(\ion{La}{2}) and log$~\epsilon$(\ion{Eu}{2}).   

 Our Fe line list is not extensive and NLTE effects have not been accounted for, so it is unlikely that the errors in stellar parameters are as small as those investigated in this section.  Our parameters are internally consistent to $\Delta \teff=\pm$100 K (the difference in \teff~at which trends in excitation potential cannot be reasonably corrected by a different choice of \vt) and $\Delta$ log$~g= \pm$0.25 dex (the difference in log$~g$ at which [\ion{Fe}{2}/H] changes by 1 $\sigma_{[Fe~II/H]}$), and  $\Delta$ \vt=$\pm$0.1 km/s (the difference in \vt for which a correlation in [\ion{Fe}{1}/H] as a function of log (EW/$\lambda$) appears).  Note that the cumulative change in the calculated log$~g$ arising from a 100 K change in \teff~and 10\% changes in the assumed mass, M$_{V}$, and BC$_{V}$ is smaller, $\approx 0.1$ dex (for stars with \teff~$\geq$ 5200 K; in cooler stars the calculated log$~g$ is more sensitive to changes in \teff).   Errors in [Fe/H] vary according to the number of measurable [\ion{Fe}{1}/H] and [\ion{Fe}{2}/H] lines (see Table \ref{table1}).  Typically, $\sigma_{[Fe~I/H]}=0.07$ dex and $\sigma_{[Fe~II/H]}=0.13$ dex.  

\subsubsection{Absolute Magnitude and Distance \label{dist}}

Some stars (mainly giant and sub-giant stars) have poorly determined distances.  Distance and $M_V$ can be inferred from stellar parameters chosen with an EW analysis, but then small errors in the input values affect the final answer.  $M_V$ depends on \teff, $m_*$, log$~g$, and the BC.  In order to discern the effect of typical errors in these parameters on $M_V$, we make the following changes: $\teff-100$ K, $m_*-0.1 m_{\odot}$, BC$-0.01$ mag,  and log$~g+0.25$ dex.  This produces a cumulative change in $M_V$ of $\sim$+0.9 mag.  Reversing the sense of the input errors produces a change in $M_V$  of similar magnitude but opposite sign.  

The effect of input errors on the derived distance is much greater, and depends also on the adopted reddening, $A_V$ (from \citealt{Schlegel1998}), and the apparent magnitude m$_V$,  which we change by +0.05 mag  and --0.1 mag to maximize the change in the distance.  The value of the cumulative change in the distance  varies widely, but is typically $\sim 40-50\%$ of the original value.  Ultimately, we make use of the  implied distance in \S \ref{corr}, when we deal with space velocities, and there only the absolute error is relevant to our purposes (we reject stars with velocity errors greater than 100 km/s, an error that may arise  from sources other than the distance).   Nevertheless, this ensures that the adopted stellar parameters and adopted stellar velocities are self-consistent. 

\subsubsection{Kinematics \label{kinerr}}
Each of the input quantities carries an error that propagates through the velocity calculations.  For stars with implied distances, errors are calculated by propagating errors in individual quantities as described in \S \ref{dist}, with the addition of quoted errors in proper motion (from Tycho-2) and radial velocity (typically 1-2 km/s, taken from SIMBAD).  As with $M_V$, errors are not entirely symmetric, and we have chosen to adopt the larger error.  Although we ultimately eliminate stars with greather than 100 km/s errors  in any of the three velocity components, the inclusion of stars with implied distances greatly enlarges our sample. 

\subsubsection{Carbon}
Difficulties inherent in synthesizing large stretches of a spectrum (continuum placement, incomplete or inconsistent line lists, saturated features, etc.) limit the precision of our C abundances (although there is some gain in fitting so many features at once).  Fitting errors amount to $\pm 0.1$ dex.  In addition, we are inferring the C abundance from molecular features that may be quite sensitive to \teff~and log$~g$ changes.  However, our focus in this work is the relative abundances of La and Eu.  The C abundances we measure serve only to identify specific cases where unusual pollution (i.e., from a more evolved binary companion) may have occurred.  For this we do not require high precision or accuracy, since more detailed and comprehensive C studies (e.g., \citealt{Kraft1982,Gratton2000}) find that enhancements of this kind are significant: [C/Fe]$\sim$1.0 dex for BD$-01^{\circ} 2582$, a well-known CH giant.  Nevertheless, we have re-identified this and other known CH stars (see \S \ref{carbon}). Our abundances are generally in good agreement with both \citet{Kraft1982} and \citet{Gratton2000}, with average [C/Fe] offsets of +0.18 dex ($\sigma=0.24$) and +0.06 dex ($\sigma=0.16$), respectively.

\subsubsection{Lanthanum and Europium}
As evidenced in the preceding sections, errors in stellar parameters have virtually no effect on the measurement of log$~\epsilon$(La/Eu).  Values of log$~\epsilon$(\ion{La}{2}) and log$~\epsilon$(\ion{Eu}{2}) are sensitive to stellar parameters, log$~g$ most especially, but changes in the model atmosphere choice affect only the  absolute measurement of these abundances.  The relative abundance of La and Eu is constant so long as the stellar parameters are obtained in a self-consistent manner.     

Quoted error bars in La/Eu are $$\sigma_{La/Eu}=\sqrt{\sigma_{La}^2+\sigma_{Eu}^2}$$. This represents the line to line spread in derived abundances--the line measurement error.  Deviations between the observed and synthesized spectrum are very typically $\pm$0.05 dex--this is the accumulated error in continuum fitting, wavelength centering, line profile fitting, etc.  We therefore take $\pm$ 0.1 dex as the (2$\sigma$) fitting error of a single line.  Ultimately, however, if only one line is measurable in a star, that line is very likely to be either  weak or blended, and the fitting error may be larger.  Since \ion{La}{2} and \ion{Eu}{2} have similar ionization potentials, the transitions used here have similar excitation potentials, and the structure of the La and Eu atoms are similar, changes in stellar parameters tend to alter log$~\epsilon$(La) and log$~\epsilon$(\ion{Eu}{2}) in concert; these abundances move in the same direction and by almost the same amount in an LTE analysis.  

The uncertainties on the log(gf) values of the strong blue lines of \ion{La}{2} and \ion{Eu}{2} in Table 2 are generally $\pm$ 0.02 dex or 5\% with high
confidence, because these uncertainties are dominated by the uncertainties of the laser induced fluorescence (LIF) radiative lifetime measurements.  The accuracy of LIF radiative lifetime measurements is well documented by multiple comparisons with independent LIF measurements (Table 1 in Lawler et al. 2001a \& b) and with regular re-measurement of benchmark lifetimes using the exact same apparatus and procedure (Section 2 in Lawler et al. 2001a \& b).  The log(gf) of the weak red lines in Table 2 do have larger uncertainties, typically $\pm$ 0.04 dex or 10\%, because these uncertainties are dominated by the uncertainties of branching fraction measurements for widely separated lines.  In metal-poor stars where spectral features are weak,  La and Eu abundances were measured predominantly from the stronger blue lines.  In the coolest metal-rich stars, our measurements came mainly from the less saturated red lines.  Despite the intrinsic errors in log $gf$, Lawler et al. (2001a \& b) found that red and blue lines alike gave the same solar abundance.  We have found that that holds true for stars in which all our lines were measurable.

\section{Comparison With previous studies \label{compare}}

The most recent and comprehensive measurements of $n$-capture elements are those of \citet{Burris2000}, which encompassed 70 metal-poor giant stars, many of which are included in our sample.  In addition, \citet{Johnson2002} conducted an extensive abundance analysis of several of our most metal-poor stars.  Three elements are common to all three studies and of interest here: La, Eu, and Fe.

There is only a small offset between [\ion{Fe}{2}/H] values derived in this study and those found in \citet{Burris2000} and \citet{Johnson2002}, as shown in Fig. \ref{comp4}.  On average, our [\ion{Fe}{2}/H] values are larger than those of both \citet{Johnson2002} and \citet{Burris2000}: [\ion{Fe}{2}/H] (this study)$-$[\ion{Fe}{2}/H](Johnson)=0.16 dex and [\ion{Fe}{2}/H] (this study)$-$[\ion{Fe}{2}/H](Burris)=0.04 dex.  The [\ion{Fe}{1}/H] values we derive are, on average, smaller than those of \citet{Burris2000} but slightly larger than those found in \citet{Johnson2002}: [\ion{Fe}{1}/H] (this study)$-$[\ion{Fe}{1}/H](Johnson)=0.03 dex and [\ion{Fe}{1}/H] (this study)$-$[\ion{Fe}{1}/H](Burris)=$-0.08$ dex.  These offsets appear to be constant over the range of metallicities sampled and are well within our line-to-line abundance spreads for our stars ($\sigma_{[Fe~I/H]} \approx 0.07$ dex and $\sigma_{[Fe~II/H]} \approx 0.13$ dex, typically).

In 27 of their stars \citet{Burris2000} were able to measure only Ba, and at that time the improved atomic parameters of \citet{LawlerLa} and \citet{LawlerEu} were not available.  Despite this (and given allowances for errors on the \citealt{Burris2000} measurements), there is a fairly good agreement between the log$~\epsilon$(\ion{La}{2}) and log$~\epsilon$(\ion{Eu}{2})  derived in that study and this one, as shown for La (Fig. \ref{comp1}), Eu (Fig. \ref{comp2}), and La/Eu (Fig. \ref{comp3}).  This agreement can be attributed to two factors: (1), that \citet{Burris2000} find very similar stellar parameters to ours (one notable exception is HD 171496, the outlier in Figs. \ref{comp1} and \ref{comp2}, for which \citet{Burris2000} finds a \teff~300K cooler) and log$~\epsilon$ (X) is sensitive to this where [X/Fe] is not and (2), that \citet{Burris2000} had lower quality spectra and necessarily report La and Eu in stars where these elements are easily measured.

\citet{Johnson2002} also measured $n$-capture elements in a sample of metal-poor stars, and we find here an offset in both La and Eu (see Fig. \ref{comp1} and Fig. \ref{comp2}).  However, log$~\epsilon $(La/Eu) is, on the average, quite similar in our study and theirs (see Fig. \ref{comp3}).  It should be noted that in some cases--most particularly HD 122563--our log$~\epsilon$(La/Eu) is significantly different.  

For this star, our values of La and Eu differ from not only \citet{Burris2000} and \citet{Johnson2002}, but also \citet{Westin2000}.  Variances in log$~\epsilon$(\ion{La}{2}) and log$~\epsilon$(\ion{Eu}{2}) are to be expected with variances in \teff~and log$~g$, however, log$~\epsilon$(La/Eu) might well be expected to be consistent (see \S \ref{errors}).  This is not the case for HD 122563--values of log$~\epsilon$(La/Eu) range from 0.41 \citep{Johnson2002} to 0.37 \citep{Westin2000} to 0.30 \citep{Burris2000}.  Our adopted log$~\epsilon$(La/Eu)=0.24 is the lowest reported among these studies.  Interestingly, the log$~\epsilon$(\ion{La}{2}) we find is quite similar to that measured by \citet{Westin2000} and \citet{Burris2000}, with log$~\epsilon$(\ion{La}{2})=$-2.20,-2.22,$ and$ -2.20$, respectively.  The value given by \citet{Johnson2002} is different, log$~\epsilon$(\ion{La}{2})=$-2.44$, a change likely due to the very much lower surface gravity adopted there (compare log$~g=0.50$, \citealt{Johnson2002}, with log$~g \approx 1.4$ \citealt{Westin2000, Burris2000}, and this study).  The discrepancies in log$~\epsilon$(La/Eu)  are therefore due to log$~\epsilon$(\ion{Eu}{2}), and the source of these discrepancies is not easily explained.  We  find a higher metallicity for HD 122563 than any of the other three studies (which all use [Fe/H]$=-2.70$) although our \teff~and log$~g$ are comparable (with the exception of \citealt{Johnson2002}, whose difference in adopted log$~g$ is much larger than the difference in adopted \teff~would indicate).  \citet{Burris2000} measure two  \ion{Eu}{2} lines also used here, the 4129 and 4205~\AA~lines.  \citet{Westin2000} use five lines, three of which (3724, 3930, and 3971~\AA)  are located in the wings of very strong absorption features (\ion{Ca}{2}  H and K and a \ion{Fe}{1} line), and this very probably affects the abundances derived from them.  

The particular case of HD 122563 can  serve  as an illustration of the general situation: one of the main differences between this study and previous work is not simply the quality of atomic parameters for individual lines, but also the number of lines we have been able to measure in a particular star.  \ion{Eu}{2} lines are often contaminated by the presence of strong nearby lines, but in many stars we have been able to obtain multiple independent measurements of both \ion{Eu}{2} and \ion{La}{2}.  

Much of our data on La/Eu at near-solar metallicities is from the \citet{Woolf1995} spectra.  We find that log$~g$ as derived from Hipparchos parallaxes tend to be significantly smaller than those used by \citet{Woolf1995}, originally from \citet{Edvardsson1993} (on average our log$~g$ is 0.27 dex smaller; only in HR 235 have we reproduced their answer), so much so that we find that $\Delta$ [Fe/H] is positive in each star but one, HR 8354 (see Table \ref{table6}).   On the average, we find that [\ion{Fe}{1}/H] is 0.06 dex smaller ($\sigma_{[Fe I/H]}=0.11$ dex) and [\ion{Fe}{2}/H] is 0.16 dex smaller ($\sigma_{[Fe II/H]}= 0.11$ dex) when our spectra and analysis methods are used.  While there are significant offsets in \ion{La}{2} and \ion{Eu}{2} ($-0.13$ and $-0.16$ dex, respectively; see Tables \ref{table7} and \ref{table8} for the line-by-line abundances),  the net effect is only a small positive offset in log$\epsilon$(La/Eu) ( such that the La/Eu measured from our spectra  is 0.03 dex larger on average).  This appears to be independent of metallicity (although we do find a rather lower metallicity for HR 458 than did \citealt{Woolf1995}).  We  have added these corrections to the Fe, La, and Eu abundances listed in Table \ref{table9} (derived from Woolf et al. spectra), and in all the figures we use the altered values.

\section{Results \label{disc}}

Our main purpose was to determine at what metallicity the $s$-process begins to contribute significantly to the Galactic chemical mix.  That is, to discover at what point low- and intermediate-mass stars start to contribute their nucleosynthetic products and how soon those products dominate.   The abundance distributions of the lowest metallicity stars ought to reflect the earliest Galactic nucleosynthesis processes, as these stars were formed from gas that had undergone very little enrichment.  In the oldest stars, the abundance pattern might be dominated by the $r$-process, since the high mass stars with which the $r$-process is associated evolve on shorter time scales than the lower mass stars that house the $s$-process (and therefore contribute to the Galactic mix sooner, e.g., \citealt{Truran1981}).  In a general way, we can use the iron abundance of a star as an indication of its formation age (that is, low [Fe/H] stars are older than solar-metallicity stars), and so we plot in Fig. \ref{feh} the La/Eu ratio as a function of ``time".  Interestingly, this figure suggests that there is no unambiguous value of [Fe/H] at which the $s$-process begins.  Not only is there considerable scatter even near solar metallicities (with stars at [Fe/H]$\simeq -0.6$ and log$~\epsilon$(La/Eu)$\simeq 0.25$ (a decidedly non-solar value), but at low metallicities ([Fe/H]$<-2$) we find stars with log$~\epsilon$(La/Eu)$\simeq 0.4$, nearly 0.20 dex higher than other stars at that metallicity, and few of our metal-poor stars have abundances  consistent with a ``pure $r$-process" distribution.  Although there is an overall upward trend with [Fe/H], at any particular [Fe/H] larger than $\simeq -2$ we find stars with very different La/Eu ratios.  These differences are larger than the abundance uncertainties, and we explore interpretations in the next sections. 

 \subsection{Setting the Baseline: The $s$-and $r$-Process at Low Metallicity}

The question of when the $s$-process becomes a significant nucleosynthetic contributor can only be reasonably answered if we can probe the time before, when the $r$-process might have dominated heavy element production.  For this purpose we again use [Fe/H] as a measure of time and return to Fig. \ref{feh}, where we have superimposed two horizontal lines - the first is a dotted line indicating the pure $r$-process log(La/Eu) ratio of 0.09 given by Burris et al. (2000). Also shown as a dashed line is a very recent prediction for this $r$-process ratio of 0.12 based upon the \citet{Arlandini1999} stellar model calculations and new experimental cross-section measurements on $^{139}$La from \citet{OBrien2003}.  Some stars at the very lowest metallicities, [Fe/H] $\simeq$ --3, have La/Eu ratios that seem to be consistent with a pure $r$-process origin.  Thus, for example, log $\epsilon$(La/Eu) = 0.11 in \cs22 \citep{CS22892}, consistent with the new values determined by O'Brien et al.  Nevertheless, there is significant scatter in the data at all [Fe/H].  A number of the La/Eu ratios, even for some low-metallicity stars  below --2.5, fall above  both the O'Brien et al. and Arlandini et al. predictions for pure  $r$-process synthesis.  Abundance errors are highly unlikely to produce systematically higher La/Eu ratios in such a fashion, and we note that we have employed the same line list as that used for \cs22 (which reproduces the $r$-process value; see \citealt{CS22892}).   Interestingly, the older (larger) value for the La cross section as reported in \citealt{Arlandini1999} leads to the larger predicted pure $r$-process ratio of 0.26, which would appear to be more consistent with much of the lower metallicity stellar data.  

An alternate explanation of these data is that there is some $s$-process synthesis contributing to the La production, even at very low metallicities.  Some very low-metallicity stars, such as \cs22 \citep{CS22892}, have heavy element  abundance patterns that are consistent with the scaled solar system $r$-process curve.  It is possible, however, that other low-metallicity stars have been ``dusted'' with a small $s$-process contribution that might have slightly increased the La value above the pure $r$-process value. (Recall that Eu is almost totally an $r$-process element, while La is mostly produced in the $s$-process in solar system material; see the Appendix.) 

Burris et al. (2000) found some indications of $s$-process production in stars  even at metallicities as low as [Fe/H] = --2.75, with the more significant processing  appearing closer to a metallicity of [Fe/H] = $\simeq$ --2.3.  These new abundance data seem to be consistent with those conclusions,  and may offer some clues about when,  and in what types of stars, the $s$-process occurs.  They might, for example, suggest that there is a somewhat wide stellar mass range for the sites of the $s$-process, encompassing  more massive intermediate-mass stars with shorter evolutionary times than the main low-mass, slower evolving  sites \citep{Travaglio1999}. 

\subsection{The Abundance Spread\label{trend}}
Despite the overall slow rise, the  spread in log$~\epsilon$(La/Eu) is significant across almost the  entire range in metallicity, and  perhaps becomes even larger near solar metallicities.   Near [Fe/H]$\simeq-1$, the differences in log$~\epsilon$(La/Eu) are real.  In at least one case, this difference is due almost exclusively to the varying influence of the $s$-process.  In Figure \ref{trio}, we show three stars with essentially identical stellar parameters, BD $+19^{\circ} $1185A, G 113-022, and G 126-036.  From Fig. \ref{feh} it is apparent that G126-036  has the highest La/Eu ratio at that metallicity. BD$ +19^{\circ}$ 1185A has one of the  lowest La/Eu ratios, and G 113-022 falls between them (however, this star still has a higher ratio than the majority of stars in the sample).  Since these stars are all at the same \teff, log$~g$, and metallicity, a simple inspection of the spectra may reveal abundance differences.  In Fig. \ref{trio},  it is evident that the depths of the \ion{Eu}{2} lines match very well in all three stars, as do other singly-ionized atomic features (\ion{Ti}{2}, \ion{Fe}{2}, and \ion{Sc}{2} are marked).   In contrast, the depths of the \ion{La}{2} lines are quite different, in the  sense that the lines in G126-36 are the strongest, and the lines in BD $+19^{\circ} $1185A are the weakest.  Also, the abundance variations are not restricted to La; neighboring features show that several $s$-process elements  (marked with asterisks)  change in concert with La, and in the same sense.  There  are other examples of this in our sample, although at very low metallicities the  paucity of stars makes it difficult to find parameter pairs.  Nevertheless, it is clear that the  abundance of $s$-process products at a given metallicity is not single-valued, and can cover a significant range. 

While the case for those three stars is quite clear, we find no evidence that the $s$-process alone contributes to the scatter in La/Eu.  In Fig. \ref{logeps}, we plot log$\epsilon$ (\ion{La}{2}) and log$\epsilon$ (\ion{Eu}{2}) separately as functions of [\ion{Fe}{2}/H].  The lowest metallicity stars show widely varying amounts of \ion{La}{2} and \ion{Eu}{2} (although their ratio is roughly constant);  however,  the dispersion in each is constant with [\ion{Fe}{2}/H].  There is no indication that there is intrinsically more spread in either the $s$-process or $r$-process products at a given iron abundance.  As an example, we take the two stars G 102-027 and HR 0033.  While these two stars are not similar enough in temperature for their spectra to be visually compared, they have the same overall metallicity ([\ion{Fe}{2}/H]$\simeq -0.55$).  They also have the same La abundance, log$\epsilon$~(\ion{La}{2})$\simeq$ 0.66.  However, G 102-027 has a log $\epsilon$ Eu abundance more than 0.3 dex higher than that of HR 0033 (0.40 dex and 0.06 dex, respectively).  The star HR 4039 also has a low Eu abundance at that metallicity (0.08 dex).  In fact, at high metallicities we find a spread in Eu that is entirely consistent with that reported by \citet{Reddy2003} for the thin disk--even as other  elements (most notably, the iron-peak elements) are virtually single-valued.  In addition, our Eu scatter is consistent with the combined thick disk and thin disk population  Eu abundances derived by \citet{Mashonkina2001}, as well as the abundances found by \citet{Woolf1995} (for their total sample).  The implication is that that $s$- and $r$-process products are incompletely mixed to a very similar degree even in younger stars.

The overall abundance scatter might point to an early chemically unmixed Galaxy at low metallicity or stellar population contamination  at high metallicity, discussed further in the following sections.  However, we emphasize the important overall abundance trend  --while there is considerable scatter,  the La/Eu ratios seem to show a generally  rising abundance trend,  with increasing Galactic $s$-processing, as a function of metallicity.   Despite low ratios in some mildly metal-poor stars, the average log$\epsilon$(La/Eu) does increase with increasing metallicity.   These data are consistent with a  gradual and continually increasing  synthesis in the Galaxy,  rather than an abrupt turn-on of the $s$-process.

\subsection{ The $s$-enhanced Stars}
Three metal-poor stars in our sample have very large (super-solar) La/Eu abundance ratios.  These stars, BD $-01^{\circ} 2582$, G 140-046, and G 126-036, are not representative of the general Galactic trend.  All show substantially increased log$~\epsilon$(\ion{La}{2}) relative to other stars at the same [Fe/H] (see Fig. \ref{logeps}) as well.  That these stars differ from the other stars in our sample is also apparent in Fig. \ref{flin}, where they share a quadrant with the lead stars--stars in which $s$-process enhancements are confirmed by a measurable overabundance of lead \citep{Aoki2002}.  These three stars are overabundant in La for their [Eu/Fe] (we also note in Fig. \ref{flin} HD 122563, which is curiously Eu-poor).

\subsubsection{Carbon Abundances \label{carbon}}
The $s$-process operates in AGB stars and a particular star's abundances  may simply reflect local pollution.  Unusually high $s$-process abundances may also be the result of mass accretion from a more evolved (and now unseen) binary companion.  Carbon abundances were measured for our metal-poor stars, and we have compared stars at similar evolutionary stages (as measured by the inferred M$_V$). We find that two of the three  stars with very large $s$-process enhancements  also display large carbon abundances, although the case of G 126-036 is less clear than the others.   The reverse need not be true, as Fig. \ref{c} illustrates--several stars with high carbon abundances show no evidence of $s$-process enhancement (this phenomenon is known, if not necessarily understood; \citealt{Dominy1985, Preston2001} cite other examples).   The three stars with high $s$-process abundances show varying degrees of carbon enhancement.   The star G 126-036, in particular, has a very modest carbon enhancement compared to other unevolved stars, although it and the other dwarf C-enhanced star, G 140-046, both have much smaller carbon abundances than the giant star BD $-01^{\circ} 2582$.  In contrast, the dwarf star HD 25329 shows no $s$-process enhancement yet has a large C abundance.

 BD $-01^{\circ} 2582$ and HD 25329 are well-known  C-enhanced stars.  Three  other stars in this study have been tentatively identified as CH stars--G 095-57A , G 095-57B \citep{Tomkin1999} and HD 135148 \citep{Shetrone1999, Carney2003}.  While we find no evidence  of [C/Fe] overabundances in either  G 095-57A  or G 095-57B (and have rejected the latter as a single-lined spectroscopic binary), G  095-57A  may show a higher $s$-process abundance than is typical for its metallicity.  Unfortunately, in this star,  as in HD 25329, Eu was particularly difficult to measure (due mainly to the presence of unidentified blends in the lines, not necessarily overall line weakness).  The presence of $s$-process enrichment must therefore be judged almost entirely on log$\epsilon~$(La), which is sensitive to the choice of stellar parameters, and is therefore not particularly reliable by itself.  HD 135148 is a more recent discovery, although it and BD  $-01^{\circ} 2582$, both C-rich for their M$_V$, have been identified by \citet{Carney2003} as binary stars.  HD 135148 also shows no evidence for $s$-process enhancement.

\subsection{La/Eu and Kinematics}
We have found that even at high (near solar)  metallicities, a substantial spread in the $s-/r-$process ratio exists.  In the solar metallicity regime, our sample ought to be dominated by thin disk stars, since \citet{Woolf1995} selected their sample from the solar-neighborhood study of \citet{Edvardsson1993}.  Although the local fraction of thick disk stars may not be as low as originally believed (see \citealt{Beers2002}), nearby stars are still overwhelmingly thin disk members.  However, \citet{Edvardsson1993} did find rather large star-to-star abundance spreads, which have since been attributed to thick disk contamination.  \citet{Reddy2003} have re-examined the issue of thin disk abundance variations, and found virtually no scatter.  Given this, it is possible that the  0.4 dex difference in La/Eu in stars with [\ion{Fe}{2}/H]$>-0.5$ is due to intrusion of thick disk stars into our sample.

\subsubsection{Correlations \label {corr}}

 Although the full picture of thick disk stellar abundances is still evolving, elemental abundance work so far indicates that high-mass stellar nucleosynthesis products are enhanced in thick disk stars relative to thin disk stars \citep{Fuhrmann1998, Prochaska2000, Reddy2003, Mashonkina2003}.  Although none of these studies include La as an $s$-process marker, \citet{Mashonkina2003} measured the Eu/Ba ratio in halo, thick disk, and thin disk stars identified by \citet{Fuhrmann1998}.  They found that thick disk stars were almost indistinguishable from halo stars in [Eu/Ba] but distinctly overabundant in Eu with respect to thin disk stars ($s$-process products are suppressed relative to $r$-process products, and the $s-/r-$ ratio is significantly sub-solar).  The thick disk is generally althought to be older and kinematically hotter than the thin disk, and its relationship to the thin disk and the halo is uncertain.  Although thick disk stars are typically found in the metallicity range  $-1.2<$[Fe/H]$<-0.5$, \citet{Bensby2003} finds solar and super-solar metallicity stars that fit the kinematic criterion of the thick disk.  Thus, it is possible that the low La/Eu stars may well be indicative of this population.

In Fig. \ref{mot} we plot La/Eu as functions of U, V, and W.   We note a few features revealed in this figure.  First, we find much stronger relationships with velocity than with metallicity, and the large scatter so evident in Fig. \ref{feh} has largely (but not totally) disappeared. With very few exceptions, low-velocity stars have a higher La/Eu ratio than high-velocity stars.  This is especially evident in the U and W velocity distributions, where the stars with the highest $s$-process abundances group near 0 km/s, regardless of their overall iron content.  In the V velocity distribution, although there is an overall increase in La/Eu with decreasing V (also largely independent of [Fe/H]), there still exists a low La/Eu group of stars with solar V motion and near-solar metallicities.
 
Secondly, in U and W, the velocity dispersion increases with decreasing $s$-process contributions, regardless of overall iron content.  There are stars with only mild iron deficiencies ($-1.20<$[Fe/H]$<-0.51$) that have low U or W velocities and low La/Eu, but the range of U and W velocity values increases dramatically as La/Eu becomes smaller.

Thirdly, there is no metallicity bin that will only encompass the high La/Eu, low-velocity stars.  In Fig. \ref{mot}, lowering the boundary on the highest [\ion{Fe}{2}/H] bin (say, to include all the stars in the low V velocity clump) introduces a significant tail of high-V, low-La/Eu stars in that metallicity bin.  Many of the lower-metallicity stars in the low-V clump are not artifacts of the choice of metallicity bin.

Finally, the spread in La/Eu in the low V velocity ``clump" stars is larger than the spread in La/Eu in the high V velocity ``tail" stars ($\sim 0.4$ and $\sim 0.2$ dex, respectively).  This may be coincidental, although the spread at low V is certainly real.  From \citet{Edvardsson1993} we can get ages for most of the stars in the ``clump", and, as shown in Fig. \ref{bdpvin1} there is a distinct relationship between stellar age and La/Eu.  This correlation is also present, although more muted, in [\ion{Fe}{2}/H] and \ion{Eu}{2} and \ion{La}{2}, as shown in Fig. \ref{bdpvin2}.  Since the thick disk is characterized by high velocities and a $\sim$50 km/s rotation lag, it is more likely that these  low-La, low-V stars are part of the old thin disk.  Indeed, we can find no distinct contrast between thick disk and halo stars in log$\epsilon$(La/Eu), although several of our stars have been identified as such by other studies (see \citealt{Fuhrmann1998}).

{\section{Conclusion} \label{con}}

In this study we observed 159 giant and dwarf stars across a wide range of metallicity in order to measure the evolution of the abundance ratio La/Eu, a proxy for the $s$-/$r$-process ratio.  We   have found that the $s$-/$r$-process ratio does not increase monotonically with [Fe/H].  However, there is evidence for evolution in this ratio.  At low metallicity, the abundances of La and Eu are approximately equal; near solar metallicity, La is consistently more abundant.   The $s$-process contribution to individual stars varies widely even at near-solar metallicities.  However, we find that on the average the dispersion in La and the dispersion in Eu individually are equal, indicating that the $r$-process contribution, while smaller overall, also varies.    

Some of the variation in La/Eu can be attributed to the intrusion of different Galactic stellar populations.  We  find that when stars are separated by velocity, very little scatter in La/Eu remains.  Rather, stars separate into  high-velocity and low-velocity groups.  The former group has essentially a single value of La/Eu (although there is some spread, and a few stars stand out in this respect; the overall spread in abundances is about 0.2 dex here).  The latter group shows considerable dispersion still (about 0.4 dex overall), although at a higher overall La/Eu value. This variation is further reduced when stellar age is considered.  The age correlation and the velocity data argue  most strongly for evolution in the $s$-/$r$-ratio in the  disk, with very little change throughout the halo.  This sample also includes thick disk stars identified by other studies, although these are not readily distinguishable from the halo based solely on La/Eu.

Neither stellar population considerations nor measurement  uncertainties can account for the persistently high La/Eu ratio at low metallicities, where the $r$-process is believed to be dominant.  None  of the stars studied here has a ``pure $r$-process" value of La/Eu,  in contrast to the $r$-process rich star \cs22, studied with the same La and Eu atomic data.  It is still unclear whether this indicates  that the $s$-process was indeed active in the halo at very low metallicities and established a ``floor" in La/Eu for old stars or simply that the ``pure $r$-process" value of La/Eu, despite extensive recent work, is in error.

\acknowledgments
We wish to thank the anonymous referee for helpful suggestions and comments.  J.S. wishes to thank D. Yong, D. Paulson, and D. Lambert for many instructive discussions during the course of this work.
This work has been supported in part by NSF grants AST 99-86974 (J. J. C.), AST03-07279 (J. J. C.),  AST 02-05124 (J.E.L.), and AST 03-07495 (C.S., J.S.).  This research has made use of the SIMBAD database, operated at CDS, Strasbourg, France.  This publication also makes use of data products from the Two Micron All Sky Survey, which is a joint project of the University of Massachusetts and the Infrared Processing and Analysis Center/California Institute of Technology, funded by the National Aeronautics and Space Administration and the National Science Foundation.  

\newpage

\appendix

\section{$s$- AND $r$-PROCESS SOLAR  SYSTEM ABUNDANCES}

The $s$- and $r$-neutron capture processes are responsible for the synthesis of almost all of the isotopes above iron - the exceptions being the relatively rare  $p$-process nuclei. While a few  of those isotopes are formed solely in one of those processes, most isotopes are a combination of the products of the $s$- and the $r$-process.   The deconvolution of the solar system material into the individual isotopic contributions from the $s$-process and $r$-process has traditionally relied upon reproducing the (smooth behavior) of the  ``$\sigma$ N$_s$'' curve (i.e., the product of the $n$-capture cross-section and $s$-process abundance).  This so-called ``classical'' fit  to the $s$-process is empirical and by definition  model-independent.  Extensive neutron capture cross section measurements (see \citealt{Kappeler1989})
thus allow the determination of the  $s$-process abundance, N$_s$, contributions to each isotope.   (Experimental determinations of individual $r$-process contributions are, in general, not experimentally possible at this time.)  Subtracting these N$_s$ values from the total solar abundances  determines  the residual isotopic $r$-process contributions, N$_r$, which are then summed to obtain solar elemental $r$- (and $s$-) process abundance distributions (see also the reviews by \citealt{Cowan1991, Truran2002, Sneden2003} for further discussion). 

Earlier tabulations of this solar deconvolution,  based on this classical approach, were included in \citet{Sneden1996}  and more recently in \citet{Burris2000}.  We have slightly revised and updated the Burris et al. values and list these elemental abundance distributions in Table \ref{tablea}.  In particular 
the Nd values have been  revised to incorporate more recent measurements from \citet{Wisshak1998}.  We note that these cross section experiments assumed total solar abundances from earlier compilations including \citet{Anders1982} and \citet{Anders1989}.   Thus the total abundances,  based on a scale of Si = 10$^6$  and listed in column (3), are approximately equal to, but slightly different than, the most recent solar abundance determinations from  \citet{Lodders2003}.  The elemental N$_r$ and N$_s$ values, listed in columns (4) and (6) respectively, are the summation of all of the isotopic contributions from these two processes. (In some cases due to small contributions from the $p$-process and uncertainties in the experimental cross sections and hence the $s$-process contributions, there are a few cases in which the sums of  N$_S$ and N$_r$ are slightly different than the total abundances.) We also note that we have not included Zn  (Z = 30) in  this tabulation, in contrast to previous versions. This is due to  Zn having a significant non-$n$-capture component from explosive charged-particle nucleosynthesis. 
 
For each element we have also listed the fractional contribution of the $s$- and $r$-process. Thus it is seen that Eu is overwhelmingly (97\%) synthesized by the $r$-process, while Ba is predominantly (85\%) produced by the $s$-process in solar system material.  In addition we have tabulated  the abundances in spectroscopic units where 
log~$\epsilon$(A)~$\equiv$ log$_{\rm 10}$(N$_{\rm A}$/N$_{\rm H}$)~+~12.0, for elements A and B.  The spectroscopic units are then related to the abundance units by
$$ Log \ \epsilon(El)  = Log \ N(El) + 1.54 $$ (see \citealt{Lodders2003}). 

In addition to the classical approach for understanding the $s$-process, more sophisticated abundance predictions, based on $s$-process nucleosynthesis models in low-mass AGB stars, have also been developed \citep{Arlandini1999}.  (The primary site for $s$-process nucleosynthesis is identified with  low- or intermediate-mass stars, i.e., $M \simeq 0.8-8$ \Msun; see \citealt{Busso1999}.)  For comparison purposes we have tabulated the $s$- and $r$-process solar abundances determined for one particular set of predictions (i.e., the ``stellar model'') from \citet{Arlandini1999}.  We have made one modification to those predictions by  updating the La value on the basis of new neutron capture cross sections on $^{139}$La \citep{OBrien2003}. 

A comparison of these latter abundance predictions  with those obtained in the classical approach is shown in Figure~\ref{newfig}.   It is seen that in general there is a good overall agreement between the literature  values for the $r$-process and both of the predictions. Nevertheless, there are some important differences: for example Te, Nd, and Bi. We also note  a significant difference in the abundances determined from the classical approach and the stellar model calculations for the element Sn.  Also, while  not plotted there are also significant differences in the Y predictions. Interestingly, the abundance determined for this element from \citet{Arlandini1999} seems to give a much better fit to the  observed $n$-capture abundances in some metal-poor halo stars (see e.g., \citealt{Cowan2002}).

\clearpage
\begin{deluxetable}{cccccccccccccccc}
\rotate
\tabletypesize{\scriptsize}
\tablecaption{Derived Stellar Parameters and Abundances \label{table1}}
\tablewidth{0pt}
\tablehead{
\colhead{Star}&\colhead{note}&\colhead{\teff (K)}&\colhead{\teff (K)}&\colhead{\teff (K)}&\colhead{log~$g$}&\colhead{log~$g$}&\colhead{\vt (km sec$^-1$)}&\colhead{[Fe I/H]}&\colhead{\# lines}&\colhead{[Fe II/H]}&\colhead{\# lines}&\colhead{M$_V$}&\colhead{d (pc)}&\colhead{log$\epsilon$(La/Eu)}&\colhead{[C/Fe]}\\
&&\colhead{(Alonso)}&\colhead{(IRFM)}&\colhead{(Final)}&\colhead{(calc)}&&&&&&&&&&
}
\startdata
B+191185&	1&	\nodata&	5328&	5500&	4.19&	4.19&	1.10&	-1.09&	14&	-1.17&	7&	5.04&	67&	0.29&	-0.25\\
B+521601&	&	4911&	4816&	4911&	2.10&	2.10&	2.05&	-1.40&	15&	-1.34&	7&	0.13&	541&	0.38&	-0.3\\
B-010306&	1&	\nodata&	5680&	5550&	4.19&	4.19&	1.50&	-1.13&	10&	-1.13&	8&	5.04&	62&	0.39&	-0.15\\
B-012582&	&	5148&	5100&	5148&	2.86&	2.86&	1.20&	-2.21&	7&	-2.09&	7&	1.78&	364&	0.98&	0.8\\
G005-001&	1&	\nodata&	5612&	5500&	4.32&	4.32&	0.80&	-1.24&	15&	-1.28&	7&	5.26&	91&	0.29&	-0.05\\
G009-036&	&	5625&	5265&	5625&	4.57&	4.57&	0.65&	-1.17&	10&	-1.01&	8&	5.81&	167&	0.42&	-0.25\\
G017-025&	3&	4966&	4856&	4966&	4.26&	4.26&	0.80&	-1.54&	16&	-1.44&	9&	5.88&	48&	0.54&	-0.05\\
G023-014&	1,2&	\nodata&	4529&	5025&	4.02&	3.00&	1.30&	-1.64&	15&	-1.57&	7&	2.68*&	312&	0.25&	-0.2\\
G028-043&	2&	5061&	\nodata&	5061&	\nodata&	4.50&	0.80&	-1.64&	11&	-1.55&	3&	6.39&	48&	0.31&	-0.15\\
G029-025&	1&	\nodata&	5115&	5225&	4.28&	4.28&	0.80&	-1.09&	14&	-0.88&	8&	5.51&	112&	0.42&	-0.05\\
G040-008&	1,3&	\nodata&	5133&	5200&	4.08&	4.08&	0.50&	-0.97&	18&	-0.88&	9&	5.03&	83&	0.32&	-0.1\\
G058-025&	&	6001&	5996&	6001&	4.21&	4.21&	1.00&	-1.40&	8&	-1.53&	7&	4.56&	52&	0.63&	0.05\\
G059-001&	&	\nodata&	5299&	5922&	3.98&	3.98&	0.40&	-0.95&	16&	-0.99&	8&	4.23&	113&	0.42&	-0.15\\
G063-046&	2&	5705&	5701&	5705&	3.69&	4.25&	1.30&	-0.90&	14&	-0.89&	8&	4.94*&	74&	0.32&	0\\
G068-003&	1,2&	\nodata&	4787&	4975&	4.59&	3.50&	0.95&	-0.76&	19&	-0.76&	10&	3.88*&	104&	0.33&	0\\
G074-005&	&	\nodata&	5668&	5668&	4.24&	4.24&	1.50&	-1.05&	13&	-1.23&	8&	4.96&	57&	0.34&	0.05\\
G090-025&	&	5441&	5303&	5303&	4.46&	4.46&	1.20&	-1.78&	9&	-1.79&	4&	5.98&	28&	0.45&	-0.05\\
G095-57A&	&	\nodata&	4965&	4965&	4.40&	4.40&	0.90&	-1.22&	17&	-1.15&	6&	6.15&	24&	0.66&	-0.05\\
G095-57B&	1,3&	\nodata&	4540&	4800&	4.57&	4.57&	0.60&	-1.06&	18&	-0.95&	5&	6.78&	24&	0.46&	-0.1\\
G102-020&	&	5254&	5223&	5254&	4.44&	4.44&	0.90&	-1.25&	15&	-1.21&	7&	5.90&	70&	0.30&	-0.2\\
G102-027&	1,2&	\nodata&	5286&	5600&	2.88&	3.75&	1.05&	-0.59&	19&	-0.53&	12&	3.80*&	58&	0.29&	-0.05\\
G113-022&	1,2&	\nodata&	5616&	5525&	\nodata&	4.25&	1.10&	-1.18&	14&	-1.00&	8&	5.15*&	75&	0.55&	-0.15\\
G122-051&	&	\nodata&	4864&	4864&	4.51&	4.51&	1.40&	-1.43&	15&	-1.29&	6&	6.59&	9&	0.17&	-0.05\\
G123-009&	2&	\nodata&	5487&	5487&	\nodata&	4.75&	1.50&	-1.25&	14&	-1.22&	7&	6.44*&	63&	0.42&	-0.25\\
G126-036&	2&	\nodata&	5487&	5487&	\nodata&	4.50&	0.60&	-1.06&	15&	-0.92&	9&	5.75*&	57&	0.83&	0.15\\
G126-062&	3&	5941&	5998&	5941&	3.98&	3.98&	2.00&	-1.59&	5&	-1.61&	7&	4.07&	119&	0.40&	-0.05\\
G140-046&	&	4980&	4959&	4980&	4.42&	4.42&	0.70&	-1.30&	16&	-1.34&	4&	6.25&	59&	0.95&	-0.2\\
G153-021&	1&	\nodata&	5190&	5700&	4.36&	4.36&	1.40&	-0.70&	14&	-0.65&	10&	5.33&	92&	0.25&	-0.2\\
G176-053&	&	5593&	5710&	5593&	4.50&	4.50&	1.20&	-1.34&	9&	-1.39&	7&	5.72&	66&	0.24&	-0.05\\
G179-022&	&	\nodata&	5082&	5082&	3.20&	3.20&	1.20&	-1.35&	15&	-1.27&	8&	3.05&	332&	0.27&	-0.25\\
G180-024&	&	6059&	5993&	6059&	4.09&	4.09&	0.50&	-1.34&	6&	-1.30&	8&	4.22&	125&	0.39&	-0.5\\
G188-022&	&	\nodata&	5827&	5827&	4.27&	4.27&	1.20&	-1.52&	7&	-1.35&	8&	4.83&	109&	0.45&	-0.35\\
G191-055&	2&	\nodata&	5770&	5770&	\nodata&	4.50&	1.00&	-1.63&	7&	-1.68&	5&	5.55*&	78&	0.48&	-0.1\\
G192-043&	2&	6085&	6101&	6085&	3.73&	4.50&	1.50&	-1.50&	7&	-1.39&	7&	5.20*&	97&	0.23&	0.07\\
G221-007&	&	\nodata&	5016&	5016&	3.37&	3.37&	0.90&	-0.98&	20&	-0.90&	10&	3.56&	115&	0.39&	0.05\\
HD002665&	&	4990&	5015&	4990&	2.34&	2.34&	2.00&	-1.99&	16&	-2.04&	9&	0.66&	236&	0.23&	-0.05\\
HD003008&	1,2&	4047&	4141&	4250&	1.03&	0.25&	2.60&	-2.08&	18&	-2.02&	11&	-3.49*&	3868&	0.30&	-0.25\\
HD006755&	&	\nodata&	5105&	5105&	2.93&	2.93&	2.50&	-1.68&	14&	-1.57&	10&	1.98&	129&	0.21&	-0.2\\
HD006833&	2&	4402&	4392&	4402&	\nodata&	1.50&	1.20&	-0.85&	12&	-0.83&	10&	-0.65*&	163&	0.24&	-0.2\\
HD008724&	&	4535&	4467&	4535&	1.40&	1.40&	1.40&	-1.91&	15&	-1.69&	10&	-1.11&	732&	0.37&	-0.25\\
HD021581&	&	4870&	4866&	4870&	2.27&	2.27&	1.40&	-1.71&	10&	-1.68&	7&	0.61&	390&	0.39&	-0.15\\
HD023798&	1&	\nodata&	4294&	4450&	1.06&	1.06&	2.50&	-2.26&	7&	-2.17&	7&	-1.81&	1057&	0.35&	-0.55\\
HD025329&	&	4842&	4571&	4842&	4.66&	4.66&	0.60&	-1.67&	15&	-1.56&	2&	7.18&	18&	\nodata&	0.35\\
HD025532&	2&	5396&	\nodata&	5396&	2.57&	2.00&	1.20&	-1.34&	15&	-1.31&	9&	-0.64*&	454&	0.51&	-0.25\\
HD026297&	&	4322&	4271&	4322&	1.11&	1.11&	1.80&	-1.98&	16&	-1.76&	9&	-1.48&	620&	0.37&	-0.45\\
HD029574&	1&	4020&	3952&	4250&	0.80&	0.80&	2.20&	-2.00&	14&	-1.80&	8&	-2.11&	1165&	0.26&	-0.65\\
HD037828&	1,2&	\nodata&	4299&	4350&	\nodata&	1.50&	1.85&	-1.62&	17&	-1.42&	9&	-0.56*&	282&	0.41&	-0.2\\
HD044007&	2&	4851&	5007&	4851&	2.75&	2.00&	2.00&	-1.72&	16&	-1.71&	9&	-0.04*&	298&	0.43&	0\\
HD063791&	1,2&	\nodata&	4556&	4675&	\nodata&	2.00&	2.00&	-1.90&	15&	-1.67&	10&	0.19*&	325&	0.37&	-0.25\\
HD074462&	1,2&	\nodata&	4427&	4700&	\nodata&	2.00&	1.90&	-1.52&	7&	-1.51&	9&	0.15*&	471&	0.22&	-0.3\\
HD082590&	&	6005&	5945&	6005&	2.75&	2.75&	3.00&	\nodata&	0&	-1.32&	6&	0.69&	528&	0.35&	-1.05\\
HD085773&	2&	\nodata&	4268&	4268&	0.87&	0.50&	2.00&	-2.62&	13&	-2.39&	7&	-2.90*&	2729&	0.28&	-0.45\\
HD101063&	1&	\nodata&	4984&	5150&	3.25&	3.25&	1.70&	-1.33&	16&	-1.27&	7&	2.74&	210&	0.21&	-0.15\\
HD103036&	1,2&	\nodata&	4103&	4200&	1.14&	0.25&	3.00&	-2.04&	12&	-1.83&	7&	-3.39*&	1990&	0.47&	-0.35\\
HD103545&	&	4666&	4528&	4666&	1.64&	1.64&	2.00&	-2.45&	10&	-2.16&	5&	-0.68&	1047&	0.38&	-0.4\\
HD105546&	&	5190&	5147&	5190&	2.49&	2.49&	1.60&	-1.48&	11&	-1.41&	6&	0.79&	365&	0.42&	-0.45\\
HD105755&	&	\nodata&	5701&	5701&	3.82&	3.82&	1.20&	-0.83&	17&	-0.84&	10&	4.01&	78&	0.31&	0\\
HD106516&	&	\nodata&	6166&	6166&	4.21&	4.21&	1.10&	-0.81&	13&	-0.78&	10&	4.31&	23&	0.35&	0\\
HD107752&	&	\nodata&	4649&	4649&	1.63&	1.63&	2&	-2.78&	8&	-2.59&	5&	-0.68&	1364&	0.40&	-0.55\\
HD108317&	&	5234&	5230&	5234&	2.68&	2.68&	2.00&	-2.18&	6&	-2.28&	7&	1.26&	221&	0.23&	-0.05\\
HD110184&	2&	4250&	4185&	4250&	0.79&	0.50&	2.50&	-2.72&	11&	-2.50&	6&	-2.87*&	1662&	0.24&	-0.3\\
HD115444&	&	4721&	4661&	4721&	1.74&	1.74&	2.00&	-2.90&	8&	-2.71&	5&	-0.49&	784&	0.26&	-0.55\\
HD119516&	&	\nodata&	5382&	5382&	2.47&	2.47&	2.50&	-2.11&	7&	-1.85&	8&	0.56&	507&	0.31&	-1.15\\
HD121135&	&	4934&	4885&	4934&	1.91&	1.91&	1.60&	-1.54&	15&	-1.37&	8&	-0.36&	869&	0.37&	-0.45\\
HD122563&	&	4572&	4537&	4572&	1.36&	1.36&	2.90&	-2.72&	8&	-2.61&	7&	-1.24&	308&	0.24&	-0.6\\
HD122956&	&	\nodata&	4508&	4508&	1.55&	1.55&	1.60&	-1.95&	14&	-1.69&	10&	-0.69&	356&	0.31&	-0.3\\
HD124358&	&	4688&	4645&	4688&	1.57&	1.57&	2.10&	-1.91&	12&	-1.64&	7&	-0.89&	1128&	0.26&	-0.75\\
HD128279&	&	5290&	5316&	5290&	2.95&	2.95&	1.50&	-2.01&	8&	-2.13&	7&	1.86&	158&	0.42&	-0.1\\
HD132475&	&	5788&	5425&	5425&	3.56&	3.56&	2.30&	-1.86&	10&	-1.68&	6&	3.61&	92&	0.54&	-0.15\\
HD135148&	2&	\nodata&	4183&	4183&	1.24&	0.25&	2.90&	-2.17&	12&	-2.07&	7&	-3.36*&	3333&	0.19&	0.8\\
HD141531&	&	4461&	4356&	4356&	1.14&	1.14&	2.20&	-1.79&	16&	-1.62&	9&	-1.46&	1292&	0.27&	-0.4\\
HD165195&	&	4237&	4316&	4237&	0.78&	0.78&	2.30&	-2.60&	13&	-2.28&	5&	-2.14&	646&	0.18&	-0.5\\
HD166161&	1&	4974&	5179&	5350&	2.56&	2.56&	2.25&	-1.23&	16&	-1.22&	9&	0.79&	197&	0.71&	-0.1\\
HD171496&	&	4485&	4952&	4952&	2.37&	2.37&	1.40&	-0.67&	13&	-0.64&	6&	0.75&	246&	0.40&	-0.15\\
HD184266&	1&	5587&	5565&	6000&	2.74&	2.74&	3.00&	-1.43&	7&	-1.40&	9&	0.68&	223&	0.32&	-0.55\\
HD186478&	&	4598&	4565&	4598&	1.43&	1.43&	2.00&	-2.56&	10&	-2.44&	7&	-1.12&	1025&	0.21&	-0.4\\
HD187111&	&	4271&	4276&	4271&	1.05&	1.05&	1.90&	-1.97&	11&	-1.69&	9&	-1.54&	615&	0.31&	-0.4\\
HD188510&	&	5564&	5373&	5564&	4.51&	4.51&	1.00&	-1.32&	14&	-1.62&	7&	5.82&	39&	0.38&	-0.1\\
HD193901&	&	5750&	5768&	5750&	4.46&	4.46&	1.50&	-1.08&	12&	-1.16&	8&	5.43&	44&	0.29&	-0.2\\
HD194598&	&	\nodata&	6044&	6044&	4.19&	4.19&	1.00&	-1.08&	15&	-1.16&	8&	4.46&	56&	0.36&	-0.05\\
HD201891&	&	5909&	5883&	5909&	4.19&	4.19&	1.00&	-1.09&	9&	-1.10&	7&	4.59&	35&	0.34&	0.05\\
HD204543&	&	4672&	4600&	4672&	1.49&	1.49&	2.00&	-1.87&	16&	-1.72&	9&	-1.09&	725&	0.44&	-0.55\\
HD206739&	&	4647&	4548&	4647&	1.78&	1.78&	1.90&	-1.72&	16&	-1.61&	9&	-0.33&	574&	0.30&	-0.2\\
HD210295&	1,2&	\nodata&	4574&	4750&	\nodata&	2.50&	1.55&	-1.46&	19&	-1.25&	9&	1.33*&	441&	0.24&	-0.2\\
HD214362&	&	5727&	5780&	5727&	2.62&	2.62&	2.00&	-1.87&	4&	-1.69&	9&	0.62&	493&	0.34&	-1.05\\
HD218857&	&	\nodata&	5103&	5103&	2.44&	2.44&	1.90&	-1.90&	10&	-2.01&	8&	0.81&	410&	0.26&	-0.05\\
HD221170&	&	4410&	4402&	4410&	1.09&	1.09&	1.70&	-2.35&	12&	-2.03&	8&	-1.67&	689&	0.13&	-0.5\\
HD232078&	1,2&	3654&	3628&	3875&	0.93&	0.50&	2.10&	-1.69&	11&	-1.40&	8&	-1.89*&	120&	0.34&	-0.15\\
HD233666&	2&	\nodata&	5157&	5157&	\nodata&	2.00&	1.70&	-1.79&	9&	-1.81&	8&	-0.39*&	867&	0.35&	-0.35\\
\enddata
\tablecomments{In Column 2, a (1)  indicates a star for which \teff was adopted based solely on Fe Ews, a (2) indicates a star for which log$~g$ was set by forcing agreement between [Fe I/H] and [Fe II/H], and a (3) indicates a star identified as a spectrscopic binary by Latham et al. (2002).  A * marks M$_V$ derived from an adopted \teff and log$~g$.}
\end{deluxetable}

\clearpage

\clearpage
\begin{deluxetable}{ccc}
\tabletypesize{\scriptsize}
\tablecaption{Line Parameters \label{table2}}
\tablewidth{0pt}
\tablehead{
\colhead{wavelength(\AA)}&\colhead{Excitation Potential (EV)}&\colhead{log$gf$}
}
\startdata
\sidehead{Fe I}4445.48&0.087&-5.44\\
5225.53&0.110&-4.79\\
5247.06&0.087&-4.95\\
5250.22&0.121&-4.94\\
5326.15&3.570&-2.07\\
5412.79&4.440&-1.72\\
5491.84&4.190&-2.29\\
5600.23&4.260&-1.42\\
5855.08&4.610&-1.48\\
6120.26&0.910&-5.97\\
6151.62&2.176&-3.30\\
6481.88&2.279&-2.98\\
6498.95&0.958&-4.70\\
6518.37&2.830&-2.45\\
6609.12&2.559&-2.69\\
6625.03&1.010&-5.34\\
6739.52&1.560&-4.79\\
6750.16&2.424&-2.62\\
6752.71&4.640&-1.20\\
7189.16&3.070&-2.77\\
7723.21&2.280&-3.62\\
\sidehead{Fe II}4620.52&2.828&-3.19\\
4656.98&2.891&-3.57\\
5234.63&3.221&-2.22\\
5264.79&3.250&-3.23\\
5414.08&3.221&-3.48\\
5525.13&3.267&-3.94\\
6432.68&2.891&-3.51\\
6516.08&2.891&-3.38\\
7224.46&3.890&-3.28\\
7515.84&3.903&-3.37\\
7711.73&3.903&-2.45\\
\sidehead{La II}3988.52&0.400&0.21\\
3995.75&0.170&-0.06\\
4086.71&0.000&-0.07\\
4123.22&0.320&0.13\\
4333.75&0.170&-0.06\\
4662.50&0.000&-1.24\\
5122.99&0.320&-0.85\\
5303.53&0.320&-1.35\\
5797.57&0.240&-1.36\\
6390.48&0.320&-1.41\\
\sidehead{Eu II}3819.67&0.000&0.51\\
3907.11&0.207&0.17\\
4129.72&0.000&0.22\\
4205.04&0.000&0.21\\
6437.64&1.319&-0.32\\
6645.06&1.379&0.12\\
7217.56&1.229&-0.35\\
\enddata
\end{deluxetable}

\begin{deluxetable}{ccccccccccccc}
\tabletypesize{\scriptsize}
\tablecaption{La Abundances \label{table3}}
\tablewidth{0pt}
\tablehead{
\colhead{Star}&\colhead{3988 \AA}&\colhead{3995 \AA}&\colhead{4086 \AA}&\colhead{4123 \AA}&\colhead{4333 \AA}&\colhead{4662 \AA}&\colhead{5122 \AA}&\colhead{5303 \AA}&\colhead{5797 \AA}&\colhead{6930 \AA}&\colhead{log$\epsilon$(La$_{avg}$)}&\colhead{$\sigma$}
}
\startdata
BD +191185&	0&	0.13&	0.08&	0.08&	0.03&	\nodata&	\nodata&	\nodata&	\nodata&	\nodata&	0.06&	0.05\\
BD +511696&	-0.08&	-0.03&	-0.1&	-0.01&	\nodata&	\nodata&	\nodata&	\nodata&	\nodata&	\nodata&	-0.06&	0.04\\
BD +521601&	-0.19&	-0.14&	-0.24&	-0.19&	-0.14&	-0.12&	-0.09&	-0.09&	-0.04&	-0.06&	-0.13&	0.06\\
BD -010306&	0.13&	0.11&	0.09&	0.16&	0.16&	0.06&	\nodata&	\nodata&	\nodata&	\nodata&	0.12&	0.04\\
BD -012582&	-0.09&	-0.12&	-0.04&	-0.19&	\nodata&	0.04&	0.01&	\nodata&	\nodata&	\nodata&	-0.05&	0.08\\
G005-001&	-0.13&	-0.15&	-0.01&	-0.03&	-0.13&	\nodata&	\nodata&	\nodata&	\nodata&	\nodata&	-0.09&	0.06\\
G009-036&	\nodata&	0.29&	0.39&	0.12&	0.22&	\nodata&	\nodata&	\nodata&	\nodata&	\nodata&	0.26&	0.11\\
G023-014&	-0.36&	-0.32&	-0.37&	-0.3&	-0.33&	-0.35&	-0.25&	\nodata&	\nodata&	\nodata&	-0.33&	0.04\\
G028-043&	\nodata&	-0.17&	-0.27&	\nodata&	-0.22&	\nodata&	\nodata&	\nodata&	\nodata&	\nodata&	-0.22&	0.05\\
G029-025&	\nodata&	0.18&	0.21&	0.26&	0.11&	\nodata&	\nodata&	\nodata&	\nodata&	\nodata&	0.19&	0.06\\
G058-025&	0.01&	-0.04&	-0.04&	0.03&	\nodata&	\nodata&	\nodata&	\nodata&	\nodata&	\nodata&	-0.03&	0.03\\
G059-001&	\nodata&	0.09&	0.11&	0.21&	0.06&	0.16&	\nodata&	\nodata&	\nodata&	\nodata&	0.13&	0.06\\
G063-046&	0.22&	0.27&	0.22&	0.32&	0.32&	\nodata&	\nodata&	\nodata&	\nodata&	\nodata&	0.27&	0.05\\
G068-003&	0.44&	0.43&	0.61&	0.61&	0.36&	0.51&	\nodata&	\nodata&	\nodata&	0.46&	0.49&	0.09\\
G074-005&	0.1&	0.07&	0.1&	0.12&	0.17&	\nodata&	\nodata&	\nodata&	\nodata&	\nodata&	0.11&	0.04\\
G090-025&	-0.56&	-0.46&	\nodata&	-0.51&	\nodata&	\nodata&	\nodata&	\nodata&	\nodata&	\nodata&	-0.51&	0.05\\
G095-57A&	0.37&	0.38&	0.39&	0.39&	0.37&	0.47&	\nodata&	0.62&	\nodata&	\nodata&	0.43&	0.09\\
G102-020&	-0.02&	0.08&	0.03&	0.03&	-0.22&	\nodata&	\nodata&	\nodata&	\nodata&	\nodata&	-0.02&	0.12\\
G102-027&	0.7&	0.67&	0.77&	0.65&	0.62&	0.7&	0.7&	\nodata&	\nodata&	\nodata&	0.69&	0.05\\
G113-022&	0.41&	0.34&	0.47&	0.36&	0.44&	0.48&	0.46&	\nodata&	\nodata&	\nodata&	0.42&	0.05\\
G122-051&	-0.17&	-0.07&	-0.07&	-0.07&	-0.1&	\nodata&	\nodata&	\nodata&	\nodata&	\nodata&	-0.1&	0.04\\
G123-009&	0.23&	0.18&	0.28&	0.16&	0.18&	\nodata&	\nodata&	\nodata&	\nodata&	\nodata&	0.21&	0.05\\
G126-036&	0.86&	0.91&	0.81&	0.81&	0.84&	0.96&	0.94&	\nodata&	\nodata&	0.81&	0.87&	0.06\\
G140-046&	0.53&	0.47&	0.45&	0.50&	0.7&	0.55&	0.4&	0.7&	\nodata&	\nodata&	0.54&	0.11\\
G153-021&	0.61&	0.57&	0.56&	0.64&	0.54&	0.64&	\nodata&	\nodata&	\nodata&	\nodata&	0.59&	0.04\\
G176-053&	-0.02&	-0.1&	-0.08&	-0.1&	\nodata&	\nodata&	\nodata&	\nodata&	\nodata&	\nodata&	-0.08&	0.04\\
G179-022&	0&	0.02&	-0.03&	-0.05&	-0.02&	0.1&	0.1&	\nodata&	\nodata&	\nodata&	0.02&	0.06\\
G180-024&	0&	-0.3&	-0.25&	-0.2&	-0.2&	\nodata&	\nodata&	\nodata&	\nodata&	\nodata&	-0.19&	0.11\\
G188-022&	\nodata&	\nodata&	-0.18&	-0.08&	-0.11&	\nodata&	\nodata&	\nodata&	\nodata&	\nodata&	-0.12&	0.05\\
G191-055&	-0.28&	\nodata&	-0.48&	-0.48&	\nodata&	\nodata&	\nodata&	\nodata&	\nodata&	\nodata&	-0.41&	0.12\\
G192-043&	0.07&	-0.03&	0.02&	-0.13&	\nodata&	\nodata&	\nodata&	\nodata&	\nodata&	\nodata&	-0.02&	0.09\\
G221-007&	0.24&	0.27&	0.42&	0.14&	0.22&	0.24&	0.32&	\nodata&	\nodata&	0.37&	0.28&	0.09\\
HD002665&	-0.99&	-0.94&	-1.02&	-0.89&	-0.84&	-0.89&	-0.84&	\nodata&	\nodata&	\nodata&	-0.92&	0.07\\
HD003008&	-0.84&	-0.64&	-0.74&	-0.89&	\nodata&	-0.77&	-0.79&	-0.84&	\nodata&	-0.84&	-0.79&	0.08\\
HD006755&	-0.28&	-0.34&	-0.38&	-0.46&	-0.26&	-0.14&	-0.16&	\nodata&	\nodata&	\nodata&	-0.29&	0.12\\
HD006833&	0.11&	0.23&	0.48&	0.28&	0.23&	0.38&	0.45&	0.41&	\nodata&	0.45&	0.34&	0.13\\
HD008724&	-0.5&	-0.47&	-0.45&	-0.47&	-0.47&	-0.51&	-0.55&	-0.5&	\nodata&	-0.5&	-0.49&	0.03\\
HD021581&	-0.48&	-0.43&	-0.37&	-0.38&	-0.51&	-0.28&	-0.38&	-0.43&	-0.41&	-0.53&	-0.42&	0.07\\
HD023798&	-1&	-1&	-1.15&	-1.15&	-1.1&	-1&	-1&	-0.77&	\nodata&	-0.95&	-1.01&	0.12\\
HD025329&	\nodata&	-0.09&	-0.01&	-0.04&	&	&	&	&	&	&	-0.05&	0.04\\
HD025532&	-0.11&	-0.16&	-0.06&	-0.16&	-0.13&	-0.11&	-0.08&	-0.11&	\nodata&	-0.21&	-0.13&	0.05\\
HD026297&	-0.93&	-0.81&	-0.93&	-0.88&	-0.88&	-0.83&	-0.83&	-0.83&	-0.76&	-0.81&	-0.85&	0.05\\
HD029574&	-0.51&	-0.31&	-0.26&	-0.46&	-0.34&	-0.36&	-0.39&	-0.38&	\nodata&	-0.31&	-0.37&	0.08\\
HD037828&	-0.13&	-0.06&	-0.18&	-0.16&	-0.18&	-0.13&	\nodata&	-0.06&	\nodata&	-0.02&	-0.12&	0.06\\
HD044007&	-0.51&	-0.53&	-0.57&	-0.55&	-0.53&	-0.51&	-0.5&	-0.45&	\nodata&	-0.4&	-0.51&	0.05\\
HD063791&	-0.54&	-0.58&	-0.69&	-0.62&	-0.56&	-0.47&	\nodata&	-0.39&	\nodata&	\nodata&	-0.55&	0.1\\
HD074462&	-0.26&	-0.16&	-0.24&	-0.24&	-0.19&	-0.1&	-0.12&	-0.09&	\nodata&	-0.09&	-0.17&	0.07\\
HD082590&	-0.1&	-0.1&	-0.1&	-0.15&	-0.12&	\nodata&	\nodata&	\nodata&	\nodata&	\nodata&	-0.11&	0.02\\
HD085773&	-1.47&	-1.59&	-1.65&	-1.52&	-1.67&	-1.43&	\nodata&	\nodata&	\nodata&	\nodata&	-1.56&	0.1\\
HD101063&	0.26&	0.26&	0.11&	0.11&	0.23&	0.28&	\nodata&	\nodata&	\nodata&	\nodata&	0.21&	0.08\\
HD103036&	-0.59&	-0.49&	-0.54&	-0.57&	-0.72&	-0.61&	\nodata&	-0.71&	\nodata&	-0.74&	-0.62&	0.09\\
HD103545&	-1.2&	-1.16&	-1.2&	-1.2&	-1.16&	\nodata&	\nodata&	\nodata&	\nodata&	\nodata&	-1.18&	0.02\\
HD105546&	-0.17&	-0.17&	-0.11&	-0.29&	-0.16&	-0.1&	-0.14&	-0.01&	0.01&	\nodata&	-0.13&	0.09\\
HD105755&	0.32&	0.32&	0.34&	0.39&	0.29&	\nodata&	\nodata&	\nodata&	\nodata&	\nodata&	0.33&	0.04\\
HD106516&	0.32&	0.27&	0.42&	0.32&	0.22&	\nodata&	\nodata&	\nodata&	\nodata&	\nodata&	0.31&	0.07\\
HD107752&	\nodata&	\nodata&	-1.54&	-1.64&	\nodata&	\nodata&	\nodata&	\nodata&	\nodata&	\nodata&	-1.59&	0.07\\
HD108317&	-1.01&	-1.01&	-1.06&	-1.01&	\nodata&	\nodata&	\nodata&	\nodata&	\nodata&	\nodata&	-1.02&	0.03\\
HD110184&	-1.49&	-1.42&	-1.49&	-1.47&	-1.47&	\nodata&	-1.47&	\nodata&	\nodata&	\nodata&	-1.47&	0.03\\
HD115444&	-1.31&	-1.31&	-1.29&	-1.46&	-1.36&	\nodata&	\nodata&	\nodata&	\nodata&	\nodata&	-1.35&	0.07\\
HD119516&	-0.55&	-0.55&	-0.62&	-0.65&	-0.6&	\nodata&	\nodata&	\nodata&	\nodata&	\nodata&	-0.59&	0.04\\
HD121135&	-0.39&	-0.41&	-0.31&	-0.36&	-0.36&	-0.31&	\nodata&	-0.31&	\nodata&	-0.21&	-0.33&	0.06\\
HD122563&	\nodata&	-2.25&	-2.45&	-2.35&	\nodata&	\nodata&	\nodata&	\nodata&	\nodata&	\nodata&	-2.35&	0.1\\
HD122956&	-0.52&	-0.49&	-0.49&	-0.62&	-0.52&	-0.39&	-0.42&	-0.4&	\nodata&	\nodata&	-0.48&	0.08\\
HD124358&	-0.65&	-0.7&	-0.8&	-0.72&	-0.75&	-0.6&	-0.65&	-0.57&	\nodata&	\nodata&	-0.68&	0.08\\
HD128279&	\nodata&	-1.35&	-1.05&	-1.05&	-1.25&	\nodata&	\nodata&	\nodata&	\nodata&	\nodata&	-1.18&	0.15\\
HD132475&	-0.37&	-0.37&	-0.35&	-0.4&	-0.4&	\nodata&	\nodata&	\nodata&	\nodata&	\nodata&	-0.38&	0.02\\
HD135148&	-0.66&	-0.66&	-0.91&	-0.91&	-0.68&	-0.76&	\nodata&	-0.81&	\nodata&	-0.71&	-0.76&	0.1\\
HD141531&	\nodata&	-0.42&	-0.53&	-0.51&	-0.44&	\nodata&	-0.46&	-0.42&	-0.41&	-0.41&	-0.45&	0.05\\
HD165195&	-1.18&	-1.14&	-1.24&	-1.22&	-1.14&	-1.09&	-1.19&	-1.14&	\nodata&	-1.14&	-1.16&	0.05\\
HD166161&	0.23&	0.24&	0.14&	0.14&	0.19&	\nodata&	0.32&	0.24&	0.29&	0.29&	0.23&	0.06\\
HD171496&	0.4&	0.5&	0.6&	0.33&	0.52&	0.54&	0.6&	0.6&	\nodata&	0.53&	0.51&	0.09\\
HD184266&	-0.14&	-0.14&	-0.04&	-0.14&	-0.09&	\nodata&	\nodata&	\nodata&	\nodata&	\nodata&	-0.11&	0.04\\
HD186478&	-1.29&	-1.24&	-1.26&	-1.46&	-1.29&	\nodata&	-1.19&	\nodata&	\nodata&	\nodata&	-1.29&	0.09\\
HD187111&	-0.57&	-0.52&	-0.72&	-0.56&	-0.62&	-0.5&	-0.57&	-0.47&	-0.57&	-0.57&	-0.57&	0.07\\
HD188510&	\nodata&	-0.1&	-0.2&	-0.13&	\nodata&	\nodata&	\nodata&	\nodata&	\nodata&	\nodata&	-0.14&	0.05\\
HD193901&	0.16&	0.06&	0.26&	0.24&	0.21&	\nodata&	\nodata&	\nodata&	\nodata&	\nodata&	0.19&	0.08\\
HD194598&	0.04&	0.06&	0.11&	0.11&	0.08&	\nodata&	\nodata&	\nodata&	\nodata&	\nodata&	0.08&	0.03\\
HD201891&	0.02&	0.09&	0.17&	0.22&	0.09&	\nodata&	\nodata&	\nodata&	\nodata&	\nodata&	0.12&	0.08\\
HD204543&	-0.51&	-0.55&	-0.6&	-0.63&	-0.6&	-0.5&	-0.51&	-0.46&	-0.62&	-0.45&	-0.54&	0.07\\
HD206739&	-0.41&	-0.3&	-0.46&	-0.41&	-0.36&	-0.26&	-0.23&	-0.23&	\nodata&	-0.26&	-0.32&	0.09\\
HD210295&	-0.17&	-0.15&	-0.2&	-0.2&	-0.04&	0.03&	\nodata&	0.05&	\nodata&	\nodata&	-0.1&	0.11\\
HD214362&	-0.43&	-0.48&	-0.53&	-0.53&	-0.45&	\nodata&	\nodata&	\nodata&	\nodata&	\nodata&	-0.48&	0.05\\
HD218857&	-1.13&	-1.18&	-1.06&	-1.18&	-1.23&	\nodata&	\nodata&	\nodata&	\nodata&	\nodata&	-1.16&	0.06\\
HD221170&	-0.74&	-0.7&	-0.7&	-0.8&	-0.7&	-0.64&	-0.72&	-0.72&	\nodata&	-0.73&	-0.72&	0.04\\
HD232078&	\nodata&	\nodata&	\nodata&	-0.32&	-0.45&	-0.4&	\nodata&	-0.38&	\nodata&	-0.33&	-0.38&	0.05\\
HD233666&	-0.74&	-0.7&	-0.67&	-0.67&	-0.68&	-0.62&	\nodata&	\nodata&	\nodata&	\nodata&	-0.68&	0.04\\
\enddata
\end{deluxetable}

\begin{deluxetable}{cccccccccc}
\tabletypesize{\scriptsize}
\tablecaption{Eu Abundances \label{table4}}
\tablewidth{0pt}
\tablehead{
\colhead{Star}&\colhead{3918 \AA}&\colhead{3907 \AA}&\colhead{4129 \AA}&\colhead{4205 \AA}&\colhead{6437 \AA}&\colhead{6645 \AA}&\colhead{7217 \AA}&\colhead{log$\epsilon$(Eu$_{avg}$)}&\colhead{$\sigma$}
}
\startdata
BD +191185&	\nodata&	\nodata&	-0.23&	-0.23&	\nodata&	\nodata&	\nodata&	-0.23&	0\\
BD +511696&	-0.36&	\nodata&	-0.39&	-0.37&	\nodata&	\nodata&	\nodata&	-0.37&	0.02\\
BD +521601&	\nodata&	-0.6&	-0.55&	-0.5&	\nodata&	-0.4&	\nodata&	-0.51&	0.09\\
BD -010306&	\nodata&	-0.35&	-0.25&	-0.2&	\nodata&	\nodata&	\nodata&	-0.27&	0.08\\
BD -012582&	-1.05&	-1.07&	-1.03&	-0.98&	\nodata&	\nodata&	\nodata&	-1.03&	0.04\\
G005-001&	\nodata&	-0.34&	-0.41&	-0.39&	\nodata&	\nodata&	\nodata&	-0.38&	0.04\\
G009-036&	\nodata&	-0.09&	-0.22&	-0.17&	\nodata&	\nodata&	\nodata&	-0.16&	0.07\\
G023-014&	\nodata&	-0.64&	-0.68&	-0.59&	\nodata&	-0.41&	\nodata&	-0.58&	0.12\\
G028-043&	\nodata&	\nodata&	-0.58&	-0.48&	\nodata&	\nodata&	\nodata&	-0.53&	0.07\\
G029-025&	\nodata&	\nodata&	-0.25&	-0.2&	\nodata&	\nodata&	\nodata&	-0.23&	0.04\\
G058-025&	\nodata&	-0.65&	-0.75&	-0.58&	\nodata&	\nodata&	\nodata&	-0.66&	0.09\\
G059-001&	\nodata&	\nodata&	-0.35&	-0.22&	\nodata&	\nodata&	\nodata&	-0.29&	0.09\\
G063-046&	\nodata&	-0.02&	-0.09&	-0.04&	\nodata&	\nodata&	\nodata&	-0.05&	0.04\\
G068-003&	\nodata&	\nodata&	0.15&	0.16&	\nodata&	0.18&	\nodata&	0.16&	0.02\\
G074-005&	\nodata&	-0.26&	-0.26&	-0.19&	\nodata&	\nodata&	\nodata&	-0.23&	0.06\\
G090-025&	\nodata&	\nodata&	\nodata&	-0.97&	\nodata&	\nodata&	\nodata&	-0.97&	\nodata\\
G095-57A&	\nodata&	-0.34&	\nodata&	-0.16&	\nodata&	-0.19&	\nodata&	-0.23&	0.1\\
G102-020&	\nodata&	\nodata&	-0.36&	-0.28&	\nodata&	\nodata&	\nodata&	-0.32&	0.06\\
G102-027&	\nodata&	0.37&	0.31&	0.41&	\nodata&	0.49&	\nodata&	0.4&	0.08\\
G113-022&	\nodata&	\nodata&	-0.12&	-0.13&	\nodata&	\nodata&	\nodata&	-0.13&	0.01\\
G122-051&	\nodata&	\nodata&	-0.25&	-0.28&	\nodata&	\nodata&	\nodata&	-0.27&	0.02\\
G123-009&	\nodata&	\nodata&	-0.18&	-0.23&	\nodata&	\nodata&	\nodata&	-0.21&	0.04\\
G126-036&	\nodata&	0.03&	0.03&	0.05&	\nodata&	\nodata&	\nodata&	0.04&	0.01\\
G140-046&	\nodata&	\nodata&	-0.44&	-0.38&	\nodata&	\nodata&	\nodata&	-0.41&	0.04\\
G153-021&	\nodata&	0.33&	0.33&	0.33&	\nodata&	0.36&	\nodata&	0.34&	0.01\\
G176-053&	\nodata&	-0.31&	-0.36&	-0.29&	\nodata&	\nodata&	\nodata&	-0.32&	0.04\\
G179-022&	\nodata&	-0.26&	-0.28&	-0.22&	\nodata&	-0.11&	\nodata&	-0.22&	0.08\\
G180-024&	\nodata&	-0.61&	-0.51&	-0.61&	\nodata&	\nodata&	\nodata&	-0.58&	0.06\\
G188-022&	-0.72&	-0.52&	-0.52&	-0.62&	\nodata&	\nodata&	\nodata&	-0.6&	0.1\\
G191-055&	\nodata&	\nodata&	\nodata&	-0.89&	\nodata&	\nodata&	\nodata&	-0.89&	\nodata\\
G192-043&	-0.27&	\nodata&	-0.24&	-0.24&	\nodata&	\nodata&	\nodata&	-0.25&	0.02\\
G221-007&	\nodata&	-0.11&	-0.14&	-0.07&	\nodata&	\nodata&	\nodata&	-0.11&	0.04\\
HD002665&	\nodata&	-1.16&	-1.15&	-1.13&	\nodata&	\nodata&	\nodata&	-1.15&	0.02\\
HD003008&	\nodata&	-1.35&	-0.92&	-1&	\nodata&	-1.08&	\nodata&	-1.09&	0.19\\
HD006755&	\nodata&	-0.53&	-0.54&	-0.52&	\nodata&	-0.42&	\nodata&	-0.5&	0.06\\
HD006833&	\nodata&	\nodata&	\nodata&	\nodata&	0.13&	0.14&	0.04&	0.1&	0.06\\
HD008724&	\nodata&	-0.9&	-0.86&	-0.86&	-0.88&	-0.81&	\nodata&	-0.86&	0.03\\
HD021581&	\nodata&	-0.86&	-0.79&	-0.79&	\nodata&	-0.81&	\nodata&	-0.81&	0.03\\
HD023798&	\nodata&	\nodata&	-1.41&	-1.36&	\nodata&	-1.31&	\nodata&	-1.36&	0.05\\
HD025329&	&	&	&	&	&	&	&	&	\\
HD025532&	-0.71&	-0.64&	-0.65&	-0.59&	\nodata&	-0.62&	\nodata&	-0.64&	0.05\\
HD026297&	\nodata&	-1.29&	-1.19&	-1.21&	\nodata&	-1.19&	\nodata&	-1.22&	0.05\\
HD029574&	\nodata&	\nodata&	-0.64&	-0.67&	\nodata&	-0.59&	\nodata&	-0.63&	0.04\\
HD037828&	\nodata&	\nodata&	-0.54&	-0.59&	\nodata&	-0.51&	-0.49&	-0.53&	0.04\\
HD044007&	-1.08&	-0.96&	-0.94&	-0.9&	\nodata&	-0.81&	\nodata&	-0.94&	0.1\\
HD063791&	\nodata&	-0.99&	-0.93&	-0.9&	\nodata&	-0.85&	\nodata&	-0.92&	0.06\\
HD074462&	\nodata&	-0.5&	-0.4&	-0.4&	-0.3&	-0.35&	\nodata&	-0.39&	0.07\\
HD082590&	-0.54&	-0.41&	-0.44&	-0.44&	\nodata&	\nodata-&	\nodata&	-0.46&	0.06\\
HD085773&	\nodata&	-1.86&	-1.83&	-1.83&	\nodata&	\nodata&	\nodata&	-1.84&	0.02\\
HD101063&	\nodata&	-0.05&	-0.05&	0.05&	\nodata&	0.1&	-0.03&	0&	0.07\\
HD103036&	\nodata&	-0.95&	-1.15&	-1.15&	\nodata&	-1.1&	\nodata&	-1.09&	0.09\\
HD103545&	\nodata&	-1.64&	-1.55&	-1.54&	\nodata&	-1.5&	\nodata&	-1.56&	0.06\\
HD105546&	\nodata&	-0.53&	-0.6&	-0.54&	\nodata&	\nodata&	\nodata&	-0.56&	0.04\\
HD105755&	\nodata&	\nodata&	-0.03&	0.01&	\nodata&	0.08&	\nodata&	0.02&	0.06\\
HD106516&	\nodata&	-0.11&	-0.04&	0.04&	\nodata&	\nodata&	\nodata&	-0.04&	0.08\\
HD107752&	\nodata&	\nodata&	-1.9&	-2.07&	\nodata&	\nodata&	\nodata&	-1.99&	0.12\\
HD108317&	-1.32&	-1.22&	-1.24&	-1.22&	\nodata&	\nodata&	\nodata&	-1.25&	0.05\\
HD110184&	-1.83&	-1.61&	-1.68&	-1.65&	\nodata&	-1.78&	\nodata&	-1.71&	0.09\\
HD115444&	-1.65&	-1.57&	-1.62&	-1.61&	\nodata&	\nodata&	\nodata&	-1.61&	0.03\\
HD119516&	-0.96&	-0.86&	\nodata&	-0.89&	\nodata&	\nodata&	\nodata&	-0.9&	0.05\\
HD121135&	\nodata&	-0.77&	-0.7&	-0.7&	\nodata&	-0.62&	\nodata&	-0.7&	0.06\\
HD122563&	\nodata&	\nodata&	-2.56&	-2.61&	\nodata&	\nodata&	\nodata&	-2.59&	0.04\\
HD122956&	\nodata&	-0.86&	-0.83&	-0.8&	\nodata&	-0.68&	\nodata&	-0.79&	0.08\\
HD124358&	\nodata&	-0.96&	-0.96&	-0.96&	\nodata&	-0.87&	\nodata&	-0.94&	0.04\\
HD128279&	-1.63&	\nodata&	-1.51&	-1.56&	\nodata&	\nodata&	\nodata&	-1.57&	0.06\\
HD132475&	-0.93&	\nodata&	-0.98&	-0.85&	\nodata&	\nodata&	\nodata&	-0.92&	0.07\\
HD135148&	\nodata&	-0.95&	-0.92&	-0.87&	\nodata&	-1.05&	\nodata&	-0.95&	0.08\\
HD141531&	\nodata&	\nodata&	-0.65&	-0.74&	\nodata&	-0.77&	-0.72&	-0.72&	0.05\\
HD165195&	-1.37&	-1.27&	-1.35&	-1.33&	\nodata&	-1.4&	\nodata&	-1.34&	0.05\\
HD166161&	\nodata&	-0.52&	-0.52&	-0.47&	\nodata&	-0.42&	\nodata&	-0.48&	0.05\\
HD171496&	\nodata&	\nodata&	0.1&	0.04&	0.14&	0.14&	\nodata&	0.11&	0.05\\
HD184266&	-0.47&	-0.5&	-0.4&	-0.35&	\nodata&	\nodata&	\nodata&	-0.43&	0.07\\
HD186478&	-1.55&	-1.53&	-1.45&	-1.45&	\nodata&	\nodata&	\nodata&	-1.5&	0.05\\
HD187111&	\nodata&	-0.93&	-0.88&	-0.9&	-0.88&	-0.8&	\nodata&	-0.88&	0.05\\
HD188510&	\nodata&	-0.41&	-0.54&	-0.61&	\nodata&	\nodata&	\nodata&	-0.52&	0.1\\
HD193901&	\nodata&	-0.05&	-0.15&	-0.1&	\nodata&	\nodata&	\nodata&	-0.1&	0.05\\
HD194598&	\nodata&	-0.33&	-0.27&	-0.25&	\nodata&	\nodata&	\nodata&	-0.28&	0.04\\
HD201891&	\nodata&	-0.29&	-0.19&	-0.19&	\nodata&	\nodata&	\nodata&	-0.22&	0.06\\
HD204543&	\nodata&	-1.01&	-0.98&	-0.96&	\nodata&	\nodata&	\nodata&	-0.98&	0.03\\
HD206739&	\nodata&	-0.72&	-0.64&	-0.58&	\nodata&	-0.55&	\nodata&	-0.62&	0.07\\
HD210295&	\nodata&	\nodata&	-0.38&	-0.45&	-0.28&	-0.25&	\nodata&	-0.34&	0.09\\
HD214362&	\nodata&	-0.89&	-0.79&	-0.77&	\nodata&	\nodata&	\nodata&	-0.82&	0.06\\
HD218857&	-1.44&	-1.29&	-1.44&	-1.49&	\nodata&	\nodata&	\nodata&	-1.42&	0.09\\
HD221170&	\nodata&	-0.88&	-0.89&	-0.86&	-0.66&	-0.89&	-0.91&	-0.85&	0.09\\
HD232078&	\nodata&	\nodata&	\nodata&	-0.86&	-0.56&	-0.66&	-0.81&	-0.72&	0.14\\
\enddata
\end{deluxetable}

\begin{deluxetable}{ccccccc}
\tabletypesize{\scriptsize}
\tablecaption{Kinematics  \label{table5}}
\tablewidth{0pt}
\tablehead{
\colhead{Star}&\colhead{U$_{LSR}$}&\colhead{error}&\colhead{V$_{LSR}$}&\colhead{error}&\colhead{W$_{LSR}$}&\colhead{error}\\
\colhead{}&\colhead{km/s}&\colhead{km/s}&\colhead{km/s}&\colhead{km/s}&\colhead{km/s}&\colhead{km/s}}
\startdata
BD +191185&	234.3&	9.3&	-240.3&	45.8&	90.5&	14.3\\
BD +521601&	-42.1&	9.9&	-21.3&	2.5&	-60&	3.9\\
BD -010306&	-201.1&	31.9&	-203.2&	34.4&	65.2&	13.7\\
BD -012582&	66&	11.2&	-163.7&	27.4&	-106.7&	18\\
G 005-001&	34.5&	2.9&	-125.8&	19.7&	-88.7&	16.8\\
G 009-036&	179.8&	24.1&	-173.3&	31.4&	52&	13.9\\
G 017-025&	-88.3&	10.1&	-172.1&	23.6&	-137.1&	9.6\\
G 023-014&	\nodata&	\nodata&	\nodata&	\nodata&	\nodata&	\nodata\\
G 028-043&	180.7&	78.1&	-263.1&	84.4&	-22.4&	49.9\\
G 029-025&	-112.4&	19&	-108.6&	7.8&	17.5&	7.5\\
G 040-008&	57.8&	3.6&	-131.1&	23.8&	-64.3&	7.5\\
G 058-025&	-34.5&	2.2&	-139.3&	20.1&	7.4&	8.1\\
G 059-001&	-191.3&	32.1&	-78.1&	13.3&	-41.8&	7.3\\
G 063-046&	78.8&	30.8&	-52.3&	19.5&	-56.3&	11.7\\
G 068-003&	13.9&	3.7&	-143.8&	24.9&	-37.1&	34.1\\
G 074-005&	-59.8&	7.3&	-77.1&	15&	-47.1&	5.9\\
G 090-025&	265.2&	9.8&	-220.5&	43.1&	-91.6&	3.1\\
G 095-57A&	-94.2&	8.5&	-114.7&	22.6&	-77.4&	11.4\\
G 095-57B&	-95.7&	8.5&	-116.2&	22.9&	-80&	11.8\\
G 102-020&	-17.6&	0.9&	-72.7&	11.2&	60.2&	10.7\\
G 102-027&	-24.4&	4.6&	-59.9&	18.6&	-16.8&	4.3\\
G 113-022&	32&	25&	-89.6&	19.8&	56.2&	14\\
G 122-051&	279.6&	42.1&	-158.6&	25.5&	-12.6&	13.6\\
G 123-009&	-66.8&	33&	-107.2&	40&	-22.4&	4.9\\
G 126-036&	85.4&	32.2&	-89&	1.7&	-29&	19.7\\
G 126-062&	-303.4&	41.9&	-276.9&	5.3&	5.2&	23.8\\
G 140-046&	114.4&	19.2&	-192.6&	31.9&	41.7&	7\\
G 153-021&	-99.1&	7.6&	-56.2&	8.8&	40.8&	12.7\\
G 176-053&	-210.1&	31.1&	-245.6&	43.7&	53.6&	0.7\\
G 179-022&	290.9&	43.3&	-147.3&	31.1&	75.9&	1.6\\
G 180-024&	108.7&	25&	-268.5&	29&	-28.4&	15.4\\
G 188-022&	128.7&	24.2&	-106.8&	3.1&	54.7&	4\\
G 191-055&	252.4&	1.4&	-116.7&	30.6&	48&	35\\
G 192-043&	-265.2&	37.2&	-116.3&	65.3&	12.3&	21.3\\
G 221-007&	-140.8&	22.7&	-105.9&	19&	-51.5&	9\\
HD 002665&	158.6&	5.1&	-352.5&	4.7&	-37.3&	12.6\\
HD 003008&	-153.6&	61&	-286.1&	93.5&	14.6&	22.5\\
HD 006755&	-129.4&	52.1&	-478.4&	36.2&	79.9&	12.1\\
HD 006833&	127&	5.7&	-202.5&	4.4&	62.3&	10\\
HD 008724&	2.2&	9.2&	-330.4&	46.2&	-81.6&	26.8\\
HD 018768&	\nodata&	\nodata&	\nodata&	\nodata&	\nodata&	\nodata\\
HD 021581&	-103.4&	2.3&	-194.3&	31.7&	-104.2&	1.9\\
HD 022879&	-105.2&	1.5&	-84.9&	1.5&	-41.4&	1.6\\
HD 023798&	-71.1&	6.8&	-108.3&	11.6&	-13.6&	9.9\\
HD 025329&	-37.2&	11&	-190.9&	30.2&	19.8&	2.2\\
HD 025532&	63.3&	14.2&	-331.6&	109.8&	4.1&	13.3\\
HD 026297&	-45.8&	7.1&	-98.8&	16&	72.6&	14.5\\
HD 029574&	204.1&	36.2&	-164.1&	25.5&	-172.8&	26.2\\
HD 030649&	-57.2&	2.2&	-81&	3.1&	-10&	0.4\\
HD 037828&	-100.7&	16.4&	-168.4&	25.2&	-49&	7\\
HD 038007&	-72.3&	5.5&	-19.1&	2.6&	9.8&	1.2\\
HD 044007&	-64.8&	19.4&	-197.1&	31&	37.3&	27.6\\
HD 062301&	-7.7&	1.8&	-108.5&	3.6&	-22.7&	1.1\\
HD 063791&	-7.4&	32.5&	-144.8&	37.6&	-117.4&	22\\
HD 074462&	108.7&	2.7&	-270.6&	69.8&	41.1&	49.4\\
HD 078558&	-66.1&	1.6&	-67.4&	1.8&	-66.5&	3\\
HD 082590&	186&	43.1&	-341.6&	26.6&	-43.4&	22.7\\
HD 085773&	-38.3&	16.8&	-281.6&	52.6&	-269&	116.8\\
HD 091347&	50.6&	1.5&	27.7&	1&	-2.5&	1.7\\
HD 101063&	-228.3&	45&	-284&	37.7&	-2.8&	25.5\\
HD 103036&	\nodata&	\nodata&	\nodata&	\nodata&	\nodata&	\nodata\\
HD 103545&	-130.7&	21.5&	-325.8&	43.2&	54.8&	18.7\\
HD 105546&	-16.4&	1.9&	-113.2&	20.3&	68.5&	8.9\\
HD 105755&	\nodata&	\nodata&	\nodata&	\nodata&	\nodata&	\nodata\\
HD 106516&	54&	8.7&	-73.8&	11.5&	-58.7&	10.9\\
HD 107752&	-143.6&	27.4&	-415.4&	58.6&	110.7&	16.7\\
HD 108317&	-137.3&	23.2&	-110.6&	18.1&	-20.2&	4.6\\
HD 110184&	-43.6&	23.8&	-159.5&	43.7&	104.2&	10.5\\
HD 114762&	-83&	5.3&	-69.6&	3.8&	57.6&	2\\
HD 115444&	147.8&	24.4&	-171&	27.4&	5.8&	7.4\\
HD 119516&	-154.9&	12.6&	-89.3&	17.2&	-257.5&	3.2\\
HD 121135&	-8.2&	11.1&	-169.5&	24.4&	111.3&	1.9\\
HD 122563&	-151.1&	23.6&	-252.3&	42.2&	21.9&	7.4\\
HD 122956&	12.1&	14.9&	-213.9&	25.3&	111.3&	1.4\\
HD 124358&	-97.7&	50.7&	-532&	72.4&	295.6&	11.8\\
HD 126512&	85.1&	4.1&	-84.2&	3&	-78&	2.2\\
HD 128279&	4.8&	11.5&	-85.8&	20.5&	-261.5&	37\\
HD 132475&	34.1&	16.7&	-363.3&	51.8&	48.5&	7.2\\
HD 135148&	-319.6&	99&	-273.8&	96.3&	165&	88.3\\
HD 141531&	171.3&	30.5&	-307.6&	53.9&	-59.4&	14.6\\
HD 159307&	\nodata&	\nodata&	\nodata&	\nodata&	\nodata&	\nodata\\
HD 165195&	143.2&	24.1&	-223.5&	37.5&	-37.5&	6.7\\
HD 166161&	133.3&	11.7&	-164.5&	31.4&	0.7&	1.4\\
HD 171496&	-41.3&	9.8&	-8.4&	2.1&	17.2&	2.9\\
HD 184266&	-311.6&	1.2&	-275.4&	24.5&	-103.4&	33.9\\
HD 184499&	-63.6&	0.8&	-159&	1.8&	58.6&	1.6\\
HD 186478&	168.4&	23.3&	-381.9&	64.3&	-71.4&	11.4\\
HD 187111&	-146.8&	3&	-200.8&	19.9&	-105.5&	26.9\\
HD 188510&	-152.7&	5&	-113.7&	5.3&	62.3&	5.5\\
HD 193901&	-156.7&	5&	-244.9&	31.3&	-73.7&	26.4\\
HD 194598&	-76&	11.3&	-275.6&	14.6&	-31.9&	16.6\\
HD 201891&	91.4&	18&	-115.6&	12.9&	-58.8&	12.4\\
HD 204543&	23.1&	13.1&	-187.3&	21.3&	-9.1&	11.9\\
HD 206739&	-83.6&	9.3&	-113.7&	14.4&	-61.9&	17.2\\
HD 208906&	73.1&	1.5&	-2.9&	1.9&	-10.9&	0.7\\
HD 210295&	125.8&	48.3&	-150.1&	50.7&	12.1&	8\\
HD 214362&	-332.9&	49.4&	-244.7&	36.7&	-136.9&	36.3\\
HD 218857&	118.7&	8.4&	-189.1&	6.4&	150.8&	2\\
HD 221170&	148.1&	20.9&	-147.2&	8.8&	-71.3&	21.6\\
HD 221830&	-67.4&	2.4&	-105.3&	4.3&	57.8&	2.4\\
HD 232078&	-224.9&	3.5&	-319.9&	2.9&	-0.9&	5.6\\
HD 233666&	76.7&	12.4&	-124.7&	39.7&	14.3&	22.4\\
HR 0033&	19&	0.3&	-13.2&	0.3&	-17.7&	0.9\\
HR 0219&	-30.3&	0.1&	-9.2&	0.1&	-17&	0.1\\
HR 0235&	21.4&	0.4&	-2.7&	0.2&	-12&	0.9\\
HR 0244&	-6.5&	0.5&	21&	0.8&	14.3&	0.2\\
HR 0366&	-34.1&	0.8&	21.8&	0.6&	-8.9&	1.9\\
HR 0368&	-26.8&	0.9&	41.9&	1.4&	-4.2&	1.9\\
HR 0448&	-8.4&	0.7&	-26.6&	0.6&	13.5&	1.8\\
HR 0458&	28.6&	0.6&	-22.3&	0.6&	-14.2&	0.4\\
HR 0483&	-38.1&	0.7&	-30.4&	0.7&	-2.4&	0.3\\
HR 0646&	-20.4&	0.7&	-12&	0.5&	3.8&	0.6\\
HR 0672&	-65.3&	1.3&	9.2&	0.2&	13.5&	1.1\\
HR 0720&	-24.7&	1.5&	31.9&	1&	-14.2&	1\\
HR 0740&	30.3&	0.6&	-4.5&	0.2&	18.6&	0.8\\
HR 0784&	15.7&	2.6&	3.4&	0.2&	-4.9&	4.3\\
HR 0962&	-19.8&	0.6&	-19.5&	0.3&	-6.2&	0.7\\
HR 1101&	1.6&	0.7&	-15.1&	0.2&	-41.7&	0.7\\
HR 1489&	-55.4&	1.9&	-20.5&	1.1&	12.8&	0.5\\
HR 1536&	-53.2&	1.6&	-73.4&	1.3&	-22.1&	1.1\\
HR 1545&	25.8&	1.6&	-5.4&	0.9&	-23.5&	1.3\\
HR 1673&	-9.2&	0.7&	-5.8&	0.4&	2.4&	0.4\\
HR 1729&	-74.9&	0.9&	-35.3&	0.6&	3.9&	0.1\\
HR 1983&	18.3&	0.6&	4.7&	0.6&	-11.7&	0.4\\
HR 2047&	13.7&	0.9&	1.9&	0.1&	-7.2&	0.1\\
HR 2220&	-32.9&	0.9&	-18.3&	0.3&	-16.7&	0.3\\
HR 2233&	46.4&	1.8&	2.1&	1&	-35.5&	1.1\\
HR 2493&	-24.1&	1.2&	25.6&	1.7&	16.8&	0.5\\
HR 2530&	25.8&	1.7&	-13.2&	1.2&	-9.4&	0.3\\
HR 2601&	-17.2&	2&	31.8&	1.4&	-22.4&	1.1\\
HR 2721&	-80.3&	1.8&	-1.5&	0.4&	32.4&	0.8\\
HR 2835&	-59.4&	4.6&	-2.9&	1.4&	-28&	1.9\\
HR 2883&	\nodata&	\nodata&	\nodata&	\nodata&	\nodata&	\nodata\\
HR 2906&	-39.1&	0.5&	-47.4&	0.8&	-3&	0.1\\
HR 2943&	4.7&	0.7&	-8.8&	0.5&	-18.6&	0.2\\
HR 3018&	-145.9&	1&	-58&	0.9&	39.2&	0.4\\
HR 3262&	-24.1&	0.7&	-38.4&	0.6&	6.9&	0.5\\
HR 3271&	\nodata&	\nodata&	\nodata&	\nodata&	\nodata&	\nodata\\
HR 3538&	-38.2&	2.8&	-16.9&	3.7&	-13&	2.1\\
HR 3578&	-48.8&	0.4&	-91.6&	0.8&	70&	0.7\\
HR 3648&	8.3&	0.6&	-7.5&	0.3&	-9&	0.6\\
HR 3775&	-57.5&	0.8&	-34.3&	0.4&	-24.5&	0.7\\
HR 3881&	11.6&	0.6&	-5.6&	0.1&	17.4&	0.7\\
HR 3951&	-55.9&	1.2&	-43.5&	0.5&	20.8&	1.6\\
HR 4039&	-51.7&	1.1&	-29.2&	0.7&	5&	1.7\\
HR 4067&	-10.6&	1.2&	-29.6&	0.8&	-14.3&	1.7\\
HR 4158&	68.3&	1.2&	-35.1&	1.7&	-35.5&	1.4\\
HR 4277&	-24.7&	0.4&	-2.3&	0.1&	1.8&	0.8\\
HR 4533&	-29.4&	1&	-15.7&	2.6&	-4.4&	4.3\\
HR 4540&	40.3&	0.3&	3.3&	0.4&	6.9&	0.8\\
HR 4657&	\nodata&	\nodata&	\nodata&	\nodata&	\nodata&	\nodata\\
HR 4785&	-30.8&	0.3&	-3.4&	0.2&	1.9&	0.9\\
HR 4845&	-41.9&	0.4&	7&	0.3&	75.8&	2\\
HR 4983&	-50.1&	0.3&	11.5&	0.1&	8.4&	0.9\\
HR 5011&	-38.2&	0.5&	1.2&	0.2&	-16.8&	0.9\\
HR 5019&	-23.7&	0.4&	-46.9&	0.6&	-32.1&	0.6\\
HR 5235&	9.3&	0.3&	-17&	0.2&	-2.4&	0.9\\
HR 5447&	2&	0.2&	16&	0.3&	-5.1&	0.8\\
HR 5723&	-4.8&	0.7&	-23.3&	0.8&	-13.6&	0.6\\
HR 5914&	-41.3&	0.3&	11.3&	0.7&	-67.5&	0.7\\
HR 5933&	56.5&	0.7&	-33.2&	0.4&	-24.3&	0.7\\
HR 6458&	25.7&	7.5&	-80.8&	11&	-64.3&	8.5\\
HR 7061&	37.6&	0.6&	2.1&	0.7&	-7.8&	0.2\\
HR 8354&	13.6&	0.3&	15.8&	1.9&	-7.4&	0.7\\
HR 8969&	-7.8&	0.1&	-26.6&	0.7&	-26.1&	0.8\\
\enddata
\end{deluxetable}

\begin{deluxetable}{ccccccccccccccccc}
\tabletypesize{\tiny}
\tablecaption{Comparison with Woolf et al. 1995 \label{table6}}
\tablewidth{0pt}
\center
\tablehead{
\colhead{Star}&\colhead{\teff (K)}&\colhead{\teff (K)}&\colhead{log g}&\colhead{log g}&\colhead{\vt (km sec$^-1$)}&\colhead{\vt (km sec$^-1$)}&\colhead{[Fe I/H]}&\colhead{[Fe II/H]}
&\colhead{[Fe I/H]}&\colhead{[Fe II/H]}&\colhead{log$\epsilon$(La/Eu)}&\colhead{log$\epsilon$(La/Eu)}\\
\colhead{}&\colhead{IRFM}&\colhead{Woolf}&\colhead{}&\colhead{Woolf}&\colhead{}&\colhead{Woolf}&\colhead{}&\colhead{}
&\colhead{Woolf}&\colhead{Woolf}&\colhead{}&\colhead{Woolf}\\
}
\startdata
HR 0235&	6301&	6254&	4.24&	4.32&	1.50&	1.69&	-0.12&	-0.29&	-0.22&	-0.28&	0.72&	0.67\\
HR 0458&	6100\tablenotemark {a}&	6212&	3.86&	4.17&	2.20&	1.85&	-0.12&	-0.23&	0.09&	0.08&	0.61&	0.49\\
HR 0646&	6407&	6358&	3.69&	4.07&	1.90&	2.10&	-0.35&	-0.45&	-0.29&	-0.25&	0.69&	0.70\\
HR 8354&	6259&	6285&	3.82&	4.09&	1.00&	2.02&	-0.73&	-0.70&	-0.62&	-0.57&	0.62&	0.51\\
HR 8969&	6200\tablenotemark{b}&	6255&	3.87&	4.16&	1.60&	1.90&	-0.22&	-0.38&	-0.20&	-0.23&	0.51&	0.44\\
\enddata
\tablenotetext{a}{\teff~IRFM=6202, 6100 is the adopted temperature}
\tablenotetext{b}{\teff~IRFM=6291, 6200 is the adopted temperature}
\end{deluxetable}

\begin{deluxetable}{ccccccccccccc}
\tabletypesize{\scriptsize}
\tablecaption{La Line Abundances in \citet{Woolf1995} Stars \label{table7}}
\tablewidth{0pt}
\tablehead{
\colhead{Star}&\colhead{3988 \AA}&\colhead{3995 \AA}&\colhead{4086 \AA}&\colhead{4123 \AA}&\colhead{4333 \AA}&\colhead{4662 \AA}&\colhead{5122 \AA}&\colhead{5303 \AA}&\colhead{5797 \AA}&\colhead{6930 \AA}&\colhead{log$\epsilon$(La$_{avg}$)}&\colhead{$\sigma$}
}
\startdata
HR 0235&	1.09&	1.07&	1.07&	1.04&	1.04&	1.17&	\nodata&	\nodata&	\nodata&	\nodata&	1.08&	0.05\\
HR 0458&	0.99&	0.99&	0.84&	0.94&	0.94&	1.04&	\nodata&	\nodata&	\nodata&	1.04&	0.98&	0.08\\
HR 0646&	0.96&	0.96&	0.85&	0.85&	0.86&	0.93&	\nodata&	\nodata&	\nodata&	\nodata&	0.90&	0.05\\
HR 8354&	0.21&	0.16&	0.31&	0.26&	0.16&	\nodata&	\nodata&	\nodata&	\nodata&	\nodata&	0.22&	0.07\\
HR 8969&	0.99&	0.94&	0.94&	0.91&	0.89&	0.99&	\nodata& 0.99&	\nodata&	0.89&	0.94&	0.04\\
\enddata
\end {deluxetable}

\begin{deluxetable}{cccccccccc}
\tabletypesize{\scriptsize}
\tablecaption{Eu  Line Abundances For  \citet{Woolf1995} Stars \label{table8}}
\tablewidth{0pt}
\tablehead{
\colhead{Star}&\colhead{3918 \AA}&\colhead{3907 \AA}&\colhead{4129 \AA} &\colhead{4205 \AA}&\colhead{6437 \AA}&\colhead{6645 \AA}&\colhead{7217 \AA}&\colhead{log$\epsilon$(Eu$_{avg}$)}&\colhead{$\sigma$}
}
\startdata
HR 0235&	\nodata&	\nodata&	0.36&		0.38&	\nodata& 0.33&	\nodata&	0.36&	0.03\\
HR 0458&	\nodata&	0.33&	0.33&	0.38&	\nodata&	0.43&	\nodata&	0.37&	0.05\\
HR 0646&	\nodata&	0.22&	0.17&	0.25&	\nodata&	0.12&	\nodata&	0.19&	0.06\\
HR 8354&	-0.21&	-0.21&	-0.11&	-0.08&	\nodata&	\nodata&	\nodata&	-0.15&	0.07\\
HR 8969&	\nodata&	0.46&	0.40&	0.43&	\nodata&	0.43&	\nodata&	0.43&	0.02\\
\enddata
\end{deluxetable}

\begin{deluxetable}{cccccc}
\tabletypesize{\scriptsize}
\tablecaption{La and Eu Abundances in \citet{Woolf1995} Stars \label{table9}}
\tablewidth{0pt}
\tablehead{
\colhead{Star}&\colhead{[Fe I/H]}&\colhead{[Fe II/H]}&\colhead{log$\epsilon$(La/Eu)}&\colhead{log$\epsilon$(La II)}&\colhead{log$\epsilon$(Eu II)}
}
\startdata
HD 018768&	-0.65&	-0.71&	0.52&	0.45&	-0.07\\
HD 022879&	-0.91&	-1.03&	0.51&	0.21&	-0.3\\
HD 030649&	-0.59&	-0.714&	0.38&	0.38&	0\\
HD 038007&	-0.4&	-0.49&	0.52&	0.55&	0.03\\
HD 062301&	-0.71&	-0.77&	0.51&	0.44&	-0.07\\
HD 078558&	-0.49&	-0.61&	0.34&	0.49&	0.15\\
HD 091347&	-0.55&	-0.66&	0.56&	0.45&	-0.11\\
HD 114762&	-0.75&	-0.8&	0.44&	0.2&	-0.24\\
HD 126512&	-0.63&	-0.66&	0.36&	0.42&	0.06\\
HD 159307&	-0.76&	-0.85&	0.62&	0.44&	-0.18\\

HD 184499&	-0.72&	-0.67&	0.39&	0.38&	-0.01\\
HD 208906&	-0.75&	-0.82&	0.54&	0.3&	-0.24\\
HD 221830&	-0.56&	-0.63&	0.38&	0.64&	0.26\\
HR 0033&	-0.45&	-0.56&	0.59&	0.65&	0.06\\
HR 0219&	-0.36&	-0.44&	0.67&	0.82&	0.15\\
HR 0235&	-0.28&	-0.44&	0.7&	0.9&	0.2\\
HR 0244&	-0.13&	-0.29&	0.75&	0.93&	0.18\\
HR 0366&	-0.4&	-0.51&	0.68&	0.82&	0.14\\
HR 0368&	-0.31&	-0.41&	0.63&	0.84&	0.21\\
HR 0448&	0&	-0.14&	0.46&	0.67&	0.21\\
HR 0458&	0.03&	-0.08&	0.59&	0.93&	0.34\\
HR 0483&	-0.11&	-0.23&	0.61&	0.91&	0.3\\
HR 0646&	-0.35&	-0.41&	0.73&	0.92&	0.19\\
HR 0672&	-0.06&	-0.23&	0.59&	0.89&	0.3\\
HR 0720&	-0.25&	-0.32&	0.6&	0.71&	0.11\\
HR 0740&	-0.32&	-0.43&	0.65&	0.84&	0.19\\
HR 0784&	-0.02&	-0.11&	0.67&	1.05&	0.38\\
HR 0962&	0.04&	-0.05&	0.59&	0.99&	0.4\\
HR 1101&	-0.15&	-0.24&	0.58&	0.82&	0.24\\
HR 1489&	-0.03&	-0.16&	0.58&	0.94&	0.36\\
HR 1536&	0.04&	-0.1&	0.44&	0.84&	0.4\\
HR 1545&	-0.48&	-0.67&	0.64&	0.72&	0.08\\
HR 1673&	-0.43&	-0.41&	0.69&	0.82&	0.13\\
HR 1729&	-0.14&	-0.29&	0.57&	0.89&	0.32\\
HR 1983&	-0.11&	-0.19&	0.53&	0.96&	0.43\\
HR 2047&	-0.05&	-0.11&	0.72&	1.13&	0.41\\
HR 2220&	-0.05&	-0.19&	0.66&	1.07&	0.41\\
HR 2233&	-0.25&	-0.37&	0.49&	0.85&	0.36\\
HR 2493&	-0.51&	-0.67&	0.45&	0.47&	0.02\\
HR 2530&	-0.49&	-0.6&	0.44&	0.56&	0.12\\
HR 2601&	-0.61&	-0.7&	0.71&	0.71&	0\\
HR 2721&	-0.33&	-0.42&	0.51&	0.82&	0.31\\
HR 2835&	-0.61&	-0.7&	0.51&	0.44&	-0.07\\
HR 2883&	-0.77&	-0.83&	0.41&	0.38&	-0.03\\
HR 2906&	-0.18&	-0.21&	\nodata&	0.92&	\nodata\\
HR 2943&	-0.11&	-0.24&	0.51&	0.84&	0.33\\
HR 3018&	-0.82&	-0.91&	0.39&	0.36&	-0.03\\
HR 3262&	-0.27&	-0.32&	0.51&	0.83&	0.32\\
HR 3271&	0.02&	-0.07&	0.42&	0.92&	0.5\\
HR 3538&	0.04&	-0.36&	0.59&	0.89&	0.3\\
HR 3578&	-0.89&	-1&	0.39&	0.36&	-0.03\\
HR 3648&	-0.11&	-0.17&	0.58&	0.81&	0.23\\
HR 3775&	-0.22&	-0.28&	0.62&	0.98&	0.36\\
HR 3881&	0&	-0.08&	0.51&	0.92&	0.41\\
HR 3951&	-0.05&	-0.24&	0.59&	0.85&	0.26\\
HR 4039&	-0.49&	-0.64&	0.58&	0.66&	0.08\\
HR 4067&	0.11&	0&	0.49&	1.06&	0.57\\
HR 4158&	-0.3&	-0.4&	0.44&	0.68&	0.24\\
HR 4277&	-0.02&	-0.09&	0.49&	0.88&	0.39\\
HR 4533&	0.12&	-0.04&	0.47&	0.87&	0.4\\
HR 4540&	0.05&	-0.07&	0.49&	0.9&	0.41\\
HR 4657&	-0.76&	-0.87&	0.29&	0.24&	-0.05\\
HR 4785&	-0.21&	-0.27&	0.54&	0.79&	0.25\\
HR 4845&	-0.73&	-0.91&	0.42&	0.17&	-0.25\\
HR 4983&	-0.05&	-0.16&	0.54&	0.9&	0.36\\
HR 5011&	0.03&	-0.08&	0.52&	1.01&	0.49\\
HR 5019&	-0.13&	-0.26&	0.39&	0.67&	0.28\\
HR 5235&	0.15&	-0.41&	0.54&	1&	0.46\\
HR 5447&	-0.47&	-0.56&	0.68&	0.79&	0.11\\
HR 5723&	-0.19&	-0.29&	0.81&	0.96&	0.15\\
HR 5914&	-0.48&	-0.48&	0.44&	0.57&	0.13\\
HR 5933&	-0.2&	-0.27&	0.74&	0.9&	0.16\\
HR 6458&	-0.43&	-0.48&	0.34&	0.57&	0.23\\
HR 7061&	-0.16&	-0.24&	\nodata&	0.99&	\nodata\\
HR 8354&	-0.68&	-0.73&	0.54&	0.42&	-0.12\\
HR 8969&	-0.26&	-0.39&	0.47&	0.89&	0.42\\
\enddata
\end{deluxetable}

\newpage
\begin{deluxetable}{ccccccccccc} 
\tabletypesize{\scriptsize}
\tablecolumns{11} 
\tablewidth{0pt} 
\tablecaption{Solar System s- and r-process abundances\label{tablea}} 
\tablehead{\colhead{Element} & \colhead{Z}   & \colhead{N$_{tot}$}    & \colhead{N$_r$} &	\colhead{Log $\epsilon$$_r$\tablenotemark{a}}  
  & \colhead{N$_s$}   & \colhead{Log $\epsilon$$_s$\tablenotemark{a}} & \colhead{r-fraction}
	& \colhead{s-fraction}	& \colhead{Stellar}	& \colhead{Stellar}\\
\colhead{} & \colhead{}   & \colhead{}    & \colhead{} &	\colhead{}    & \colhead{} 
  & \colhead{}    & \colhead{}	& \colhead{}	& \colhead{Log $\epsilon$$_r$\tablenotemark{b}}
	& \colhead{Log $\epsilon$$_s$\tablenotemark{b}}\\ 
\colhead{(1)} & \colhead{(2)}   & \colhead{(3)}    & \colhead{(4)} &	\colhead{(5)}
    & \colhead{(6)}   & \colhead{(7)}    & \colhead{(8)}	& \colhead{(9)}	&
 \colhead{(10)}	& \colhead{(11)}
}

\startdata
Ga	&	31	&	37.850	&	16.300	&	2.752	&	21.550	&	2.873	&	0.431	&	0.569	&	-	&	1.772	\\
Ge	&	32	&	108.757	&	56.170	&	3.290	&	52.587	&	3.261	&	0.516	&	0.484	&	-	&	2.395	\\
As	&	33	&	6.786	&	5.330	&	2.267	&	1.456	&	1.703	&	0.785	&	0.215	&	-	&	1.020	\\
Se	&	34	&	61.443	&	40.260	&	3.145	&	21.183	&	2.866	&	0.655	&	0.345	&	-	&	2.279	\\
Br	&	35	&	5.569	&	4.640	&	2.207	&	0.929	&	1.508	&	0.833	&	0.167	&	-	&	1.567	\\
Kr	&	36	&	51.952	&	22.680	&	2.896	&	29.272	&	3.006	&	0.437	&	0.563	&	-	&	2.461	\\
Rb	&	37	&	5.794	&	2.890	&	2.001	&	2.904	&
	2.003	&	0.499	&	0.501	&	-	&	1.739	\\
Sr	&	38	&	23.090	&	2.550	&	1.947	&	20.540	&	2.853	&	0.11	&	0.89	&	-	&	2.836	\\
Y	&	39	&	4.654	&	1.310	&	1.657	&	3.344	&	2.064	&	0.281	&	0.719	&	1.111	&	2.170	\\
Zr	&	40	&	10.703	&	2.040	&	1.850	&	8.663	&	2.478	&	0.191	&	0.809	&	1.798	&	2.414	\\
Nb	&	41	&	0.339	&	0.110	&	0.581	&	0.229	&	0.900	&	0.324	&	0.676	&	0.549	&	1.315	\\
Mo	&	42	&	1.968	&	0.635	&	1.343	&	1.333	&	1.665	&	0.323	&	0.677	&	1.365	&	1.467	\\
Tc	&	43	&	0.178	&	0.172	&	0.776	&	0.006	&	-0.668	&	0.965	&	0.035	&	-	&	-	\\
Ru	&	44	&	1.543	&	0.941	&	1.514	&	0.602	&	1.319	&	0.61	&	0.39	&	1.585	&	1.319	\\
Rh	&	45	&	0.344	&	0.289	&	1.001	&	0.055	&	0.284	&	0.839	&	0.161	&	1.013	&	0.209	\\
Pd	&	46	&	1.387	&	0.770	&	1.426	&	0.617	&	1.330	&	0.555	&	0.445	&	1.407	&	1.225	\\
Ag	&	47	&	0.552	&	0.435	&	1.178	&	0.117	&	0.608	&	0.788	&	0.212	&	1.130	&	0.524	\\
Cd	&	48	&	1.526	&	0.761	&	1.421	&	0.765	&	1.424	&	0.499	&	0.501	&	1.407	&	1.461	\\
In	&	49	&	0.178	&	0.121	&	0.623	&	0.057	&	0.299	&	0.678	&	0.322	&	0.589	&	0.348	\\
Sn	&	50	&	3.378	&	0.761	&	1.421	&	2.617	&	1.958	&	0.225	&	0.775	&	1.530	&	1.932	\\
Sb	&	51	&	0.292	&	0.245	&	0.929	&	0.047	&	0.213	&	0.839	&	0.161	&	0.907	&	0.420	\\
Te	&	52	&	4.920	&	3.952	&	2.137	&	0.968	&	1.526	&	0.803	&	0.197	&	1.905	&	1.452	\\
I	&	53	&	0.901	&	0.851	&	1.470	&	0.050	&	0.241	&	0.944	&	0.056	&	1.471	&	0.217	\\
Xe	&	54	&	4.793	&	3.816	&	2.122	&	0.977	&
	1.530	&	0.796	&	0.204	&	2.079	&	1.442	\\
Cs	&	55	&	0.371	&	0.315	&	1.038	&	0.056	&	0.285	&	0.85	&	0.15	&	1.042	&	0.272	\\
Ba	&	56	&	5.470	&	0.806	&	1.446	&	4.664	&	2.209	&	0.147	&	0.853	&	1.470	&	2.055	\\
La	&	57	&	0.447	&	0.110	&	0.581	&	0.337	&	1.067	&	0.246	&	0.754	&	0.619\tablenotemark{c}	&	1.053\tablenotemark{c}	\\
Ce	&	58	&	1.098	&	0.204	&	0.850	&	0.894	&	1.491	&	0.186	&	0.814	&	0.967	&	1.476	\\
Pr	&	59	&	0.161	&	0.082	&	0.454	&	0.079	&	0.440	&	0.508	&	0.492	&	0.473	&	0.450	\\
Nd	&	60	&	0.836	&	0.352	&	1.086	&	0.484	&	1.225	&	0.421	&	0.579	&	1.024	&	1.201	\\
Sm	&	62	&	0.260	&	0.174	&	0.781	&	0.086	&	0.474	&	0.669	&	0.331	&	0.781	&	0.422	\\
Eu	&	63	&	0.093	&	0.090	&	0.494	&	0.003	&	-1.062	&	0.973	&	0.027	&	0.502	&	-0.710	\\
Gd	&	64	&	0.337	&	0.276	&	0.981	&	0.061	&	0.326	&	0.819	&	0.181	&	0.985	&	0.245	\\
Tb	&	65	&	0.064	&	0.060	&	0.318	&	0.004	&	-0.827	&	0.933	&	0.067	&	0.287	&	-0.821	\\
Dy	&	66	&	0.410	&	0.360	&	1.096	&	0.050	&	0.235	&	0.879	&	0.121	&	1.065	&	0.305	\\
Ho	&	67	&	0.089	&	0.083	&	0.459	&	0.006	&	-0.704	&	0.936	&	0.064	&	0.454	&	-0.618	\\
Er	&	68	&	0.251	&	0.209	&	0.860	&	0.042	&	0.165	&	0.832	&	0.168	&	0.857	&	0.168	\\
Tm	&	69	&	0.037	&	0.031	&	0.031	&	0.006	&	-0.654	&	0.829	&	0.171	&	0.056	&	-0.758	\\
Yb	&	70	&	0.239	&	0.163	&	0.751	&	0.076	&	0.420	&	0.682	&	0.318	&	0.762	&	0.406	\\
Lu	&	71	&	0.039	&	0.031	&	0.030	&	0.008	&	-0.562	&	0.796	&	0.204	&	0.008	&	-0.659	\\
Hf	&	72	&	0.157	&	0.080	&	0.443	&	0.077	&	0.425	&	0.51	&	0.49	&	0.373	&	0.472	\\
Ta	&	73	&	0.023	&	0.013	&	-0.336	&	0.009	&	-0.492	&	0.588	&	0.412	&	-0.374	&	-0.528	\\
W	&	74	&	0.135	&	0.063	&	0.336	&	0.073	&	0.402	&	0.462	&	0.538	&	0.308	&	0.410	\\
Re	&	75	&	0.052	&	0.047	&	0.215	&	0.005	&	-0.797	&	0.911	&	0.089	&	0.235	&	-0.775	\\
Os	&	76	&	0.711	&	0.651	&	1.353	&	0.060	&	0.318	&	0.916	&	0.084	&	1.324	&	0.340	\\
Ir	&	77	&	0.658	&	0.650	&	1.353	&	0.008	&	-0.568	&	0.988	&	0.012	&	1.354	&	-0.502	\\
Pt	&	78	&	1.369	&	1.299	&	1.654	&	0.070	&	0.384	&	0.949	&	0.051	&	1.644	&	0.376	\\

Au	&	79	&	0.186	&	0.176	&	0.785	&	0.010	&	-0.443	&	0.944	&	0.056	&	0.786	&	-0.423	\\
Hg	&	80	&	0.347	&	0.146	&	0.703	&	0.201	&	0.843	&	0.42	&	0.58	&	0.661	&	0.779	\\
Tl	&	81	&	0.154	&	0.053	&	0.262	&	0.102	&	0.547	&	0.341	&	0.659	&	0.188	&	0.685	\\
Pb	&	82	&	2.905	&	0.622	&	1.334	&	2.283	&	1.899	&	0.214	&	0.786	&	-	&	1.699	\\
Bi	&	83	&	0.144	&	0.093	&	0.508	&	0.051	&
	0.246	&	0.647	&	0.353	&	0.677	&	-0.611	\\
Th	&	90	&	0.042	&	0.042	&	0.163	&	0.000	&	-	&	1.000	&	0.000	&	-	&	-	\\
U	&	92	&	0.027	&	0.027	&	-0.033	&	0.000	&	-	&	1.000	&	0.000	&	-	&	-	\\

\enddata

\tablenotetext{a}{\hspace*{0.10in}Log $\epsilon$(El)  = Log N(El) + 1.54}
\tablenotetext{b}{\hspace*{0.10in}Stellar model predictions from \citet{Arlandini1999}}
\tablenotetext{c}{\hspace*{0.10in}The Stellar model neutron capture values for La have been updated with the La value from \citet{OBrien2003}.
}

\end{deluxetable}

\begin{figure}
\plotone{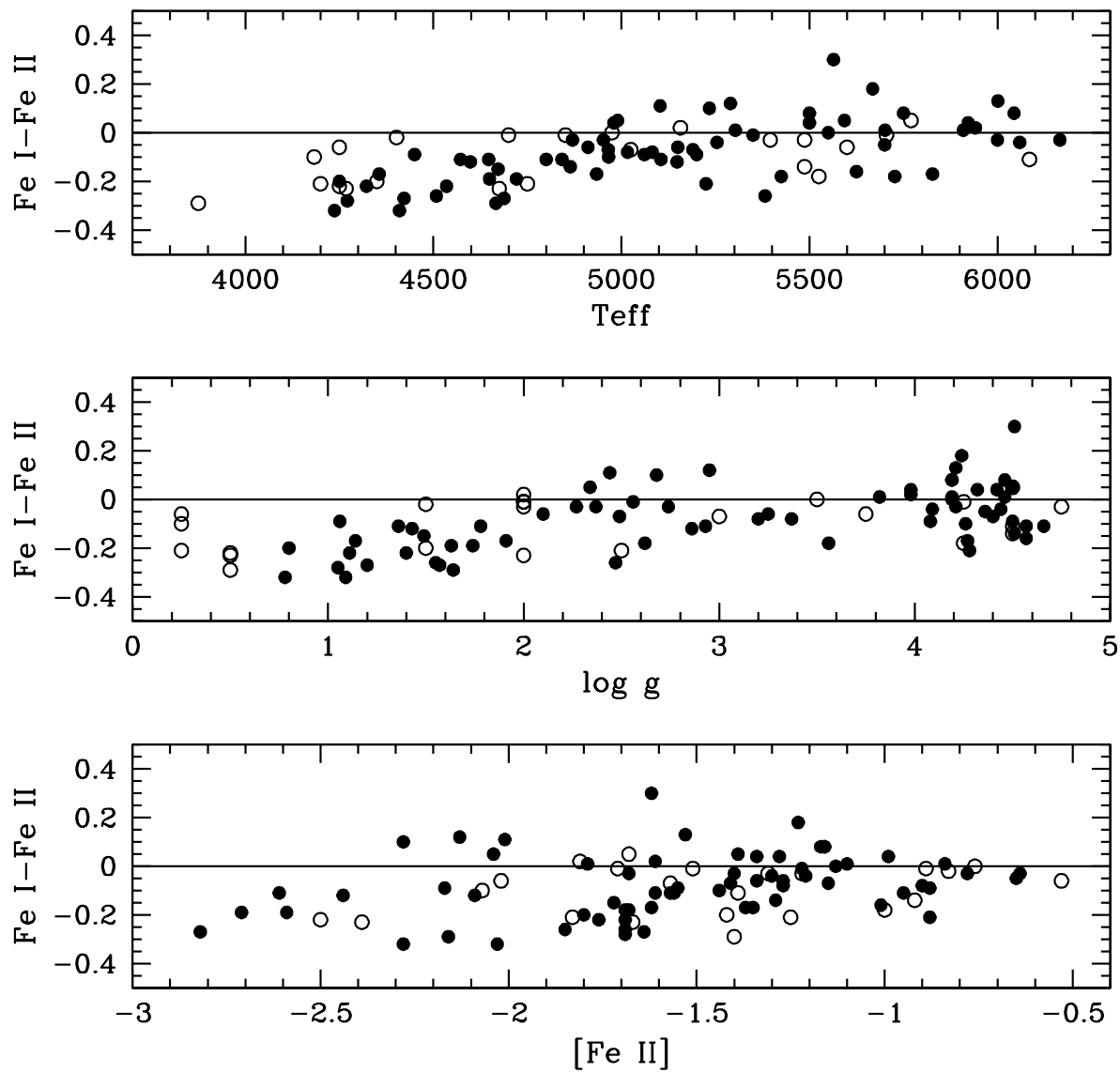}
\caption{$\Delta$ [Fe/H] as a function of \teff (top), log$~g$ (middle), or [\ion{Fe}{2}/H](bottom).  These quantities are interrelated in our sample, but the strongest correlation is with \teff, such that cooler stars show an exclusively negative difference.  Open symbols are for those stars where M$_V$ was determined from stellar parameters chosen in an EW analysis. 
\label{params}}
\end{figure}

\begin{figure}
\plotone{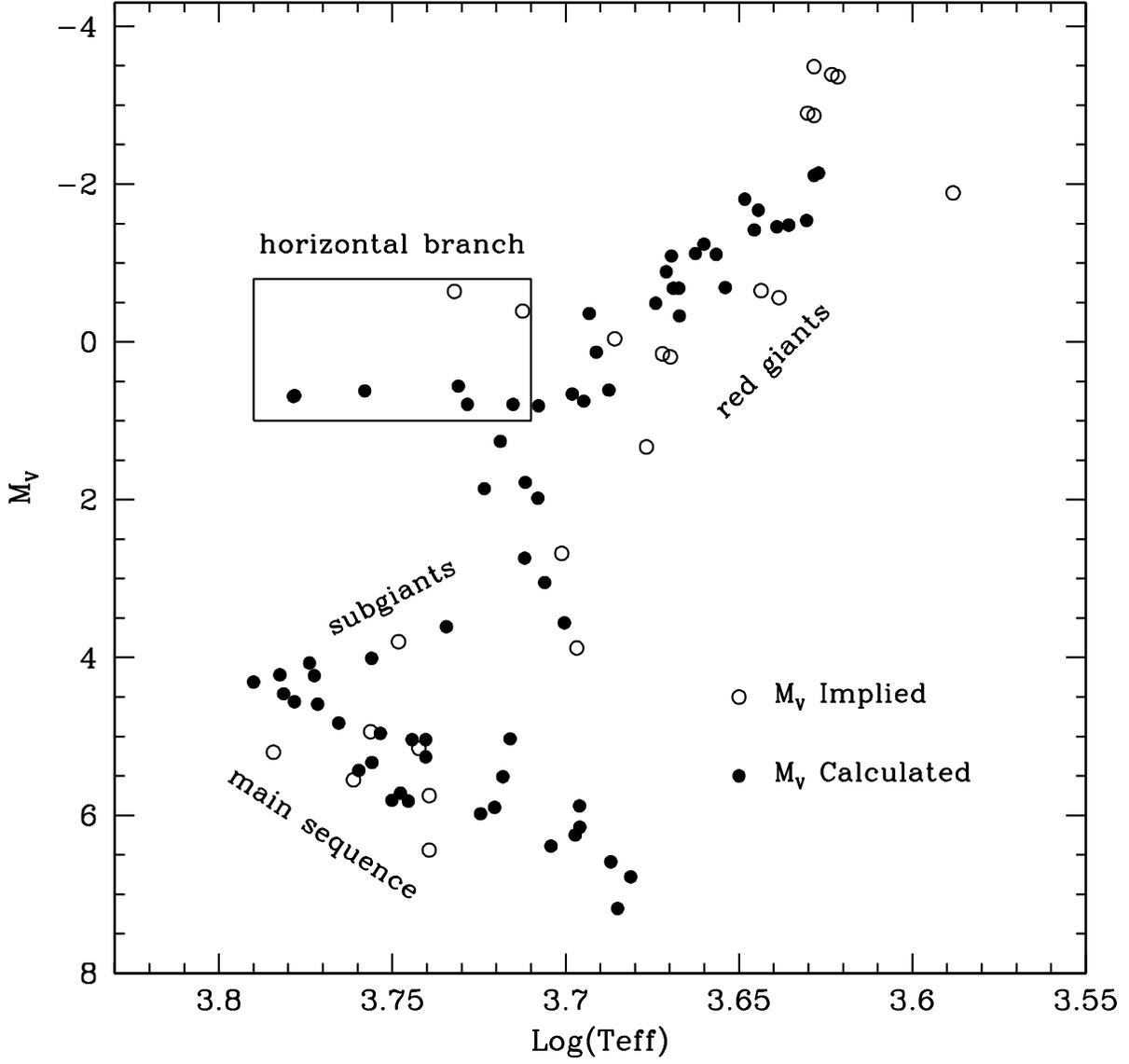}
\caption{Derived absolute magnitude (M$_V$) as a function of log(\teff). Symbols are as in Fig. \ref{params}. \label{mv}}
\end{figure}

\begin{figure}
\plotone{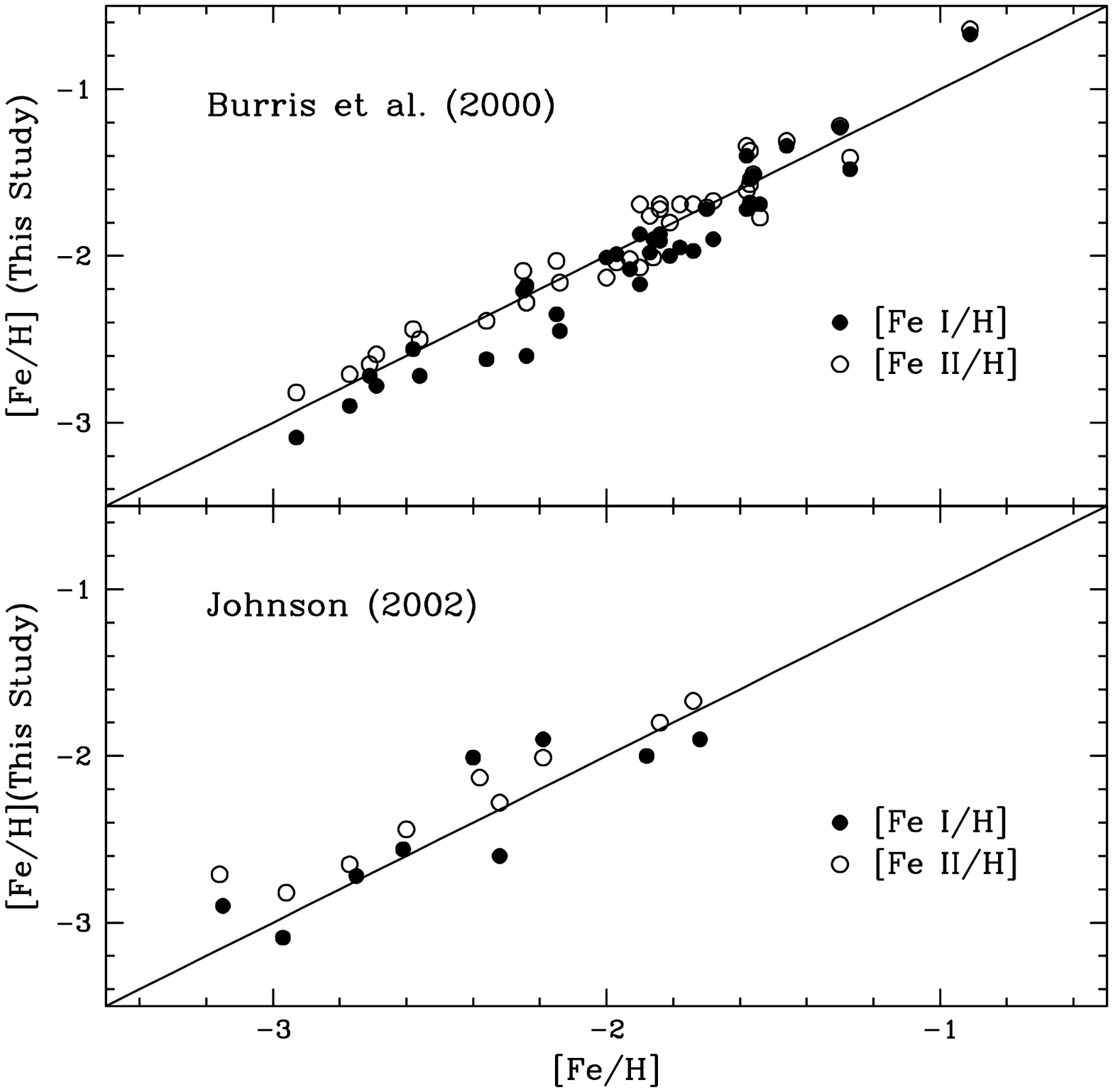}
\caption{ Comparison of derived [\ion{Fe}{1}/H] and [\ion{Fe}{2}/H] with \citet{Burris2000} (top) and \citet{Johnson2002} (bottom).\label{comp4}}
\end{figure}

\begin{figure}
\plotone{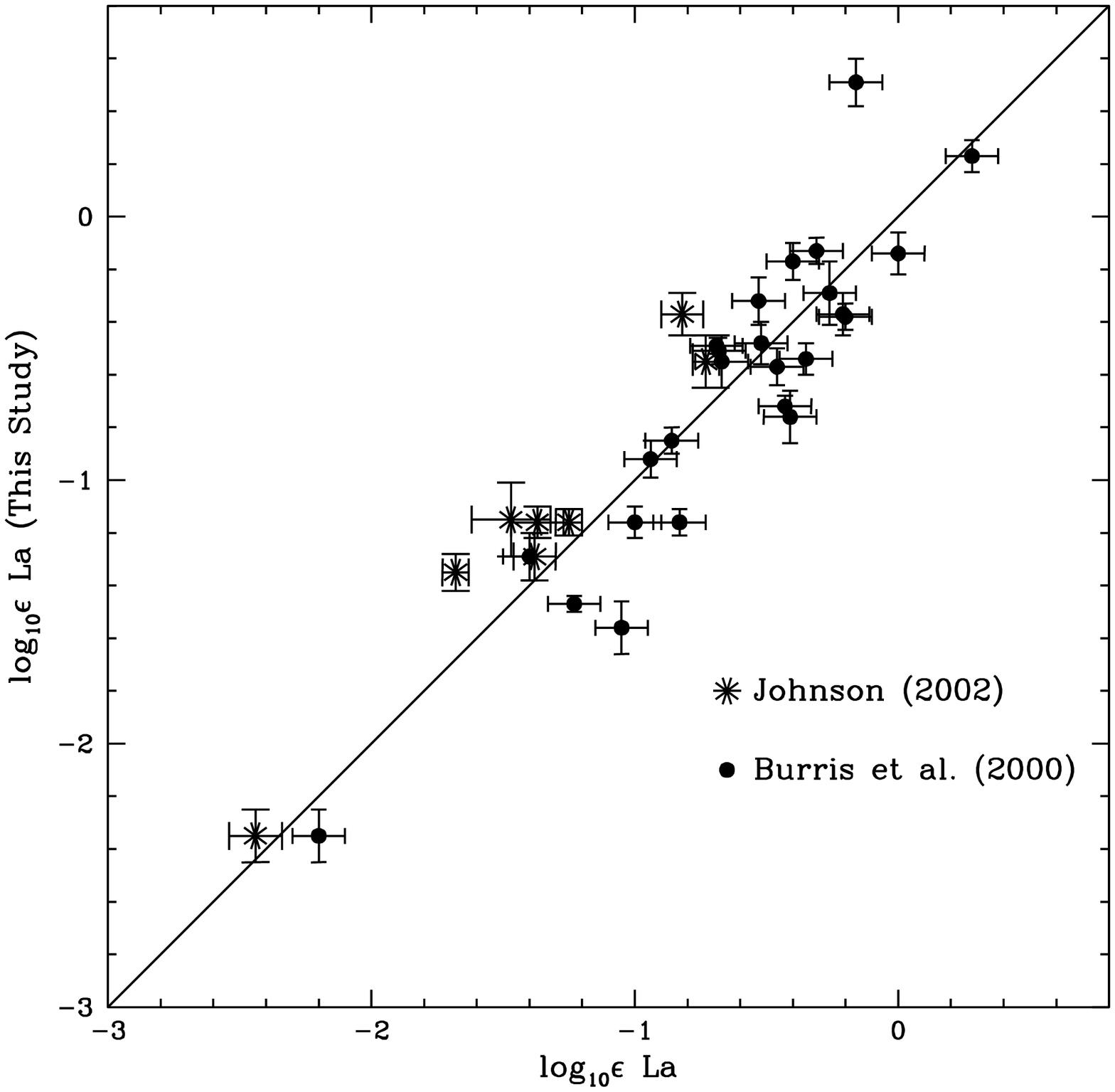}
\caption{log$\epsilon$ (\ion{La}{2})  as derived here compared with log$\epsilon$ (\ion{La}{2}) derived in \citet{Burris2000} and \citet{Johnson2002}.  
\label{comp1}}
\end{figure}

\begin{figure}
\plotone{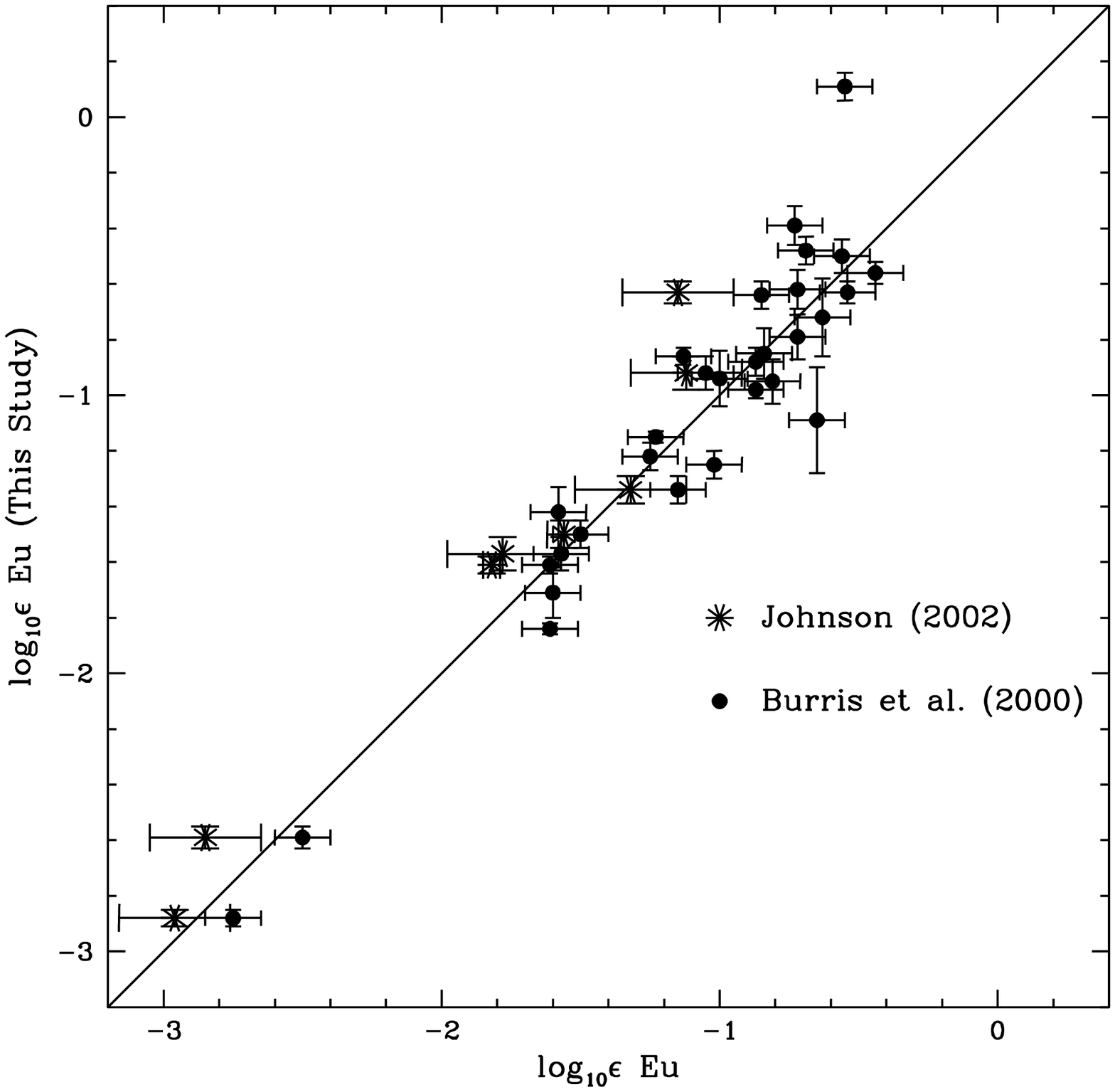}
\caption{log$\epsilon$ (\ion{Eu}{2})  as derived here compared with log$\epsilon$ (\ion{Eu}{2})  derived in \citet{Burris2000} and \citet{Johnson2002}.
\label{comp2}}
\end{figure}

\begin{figure}
\plotone{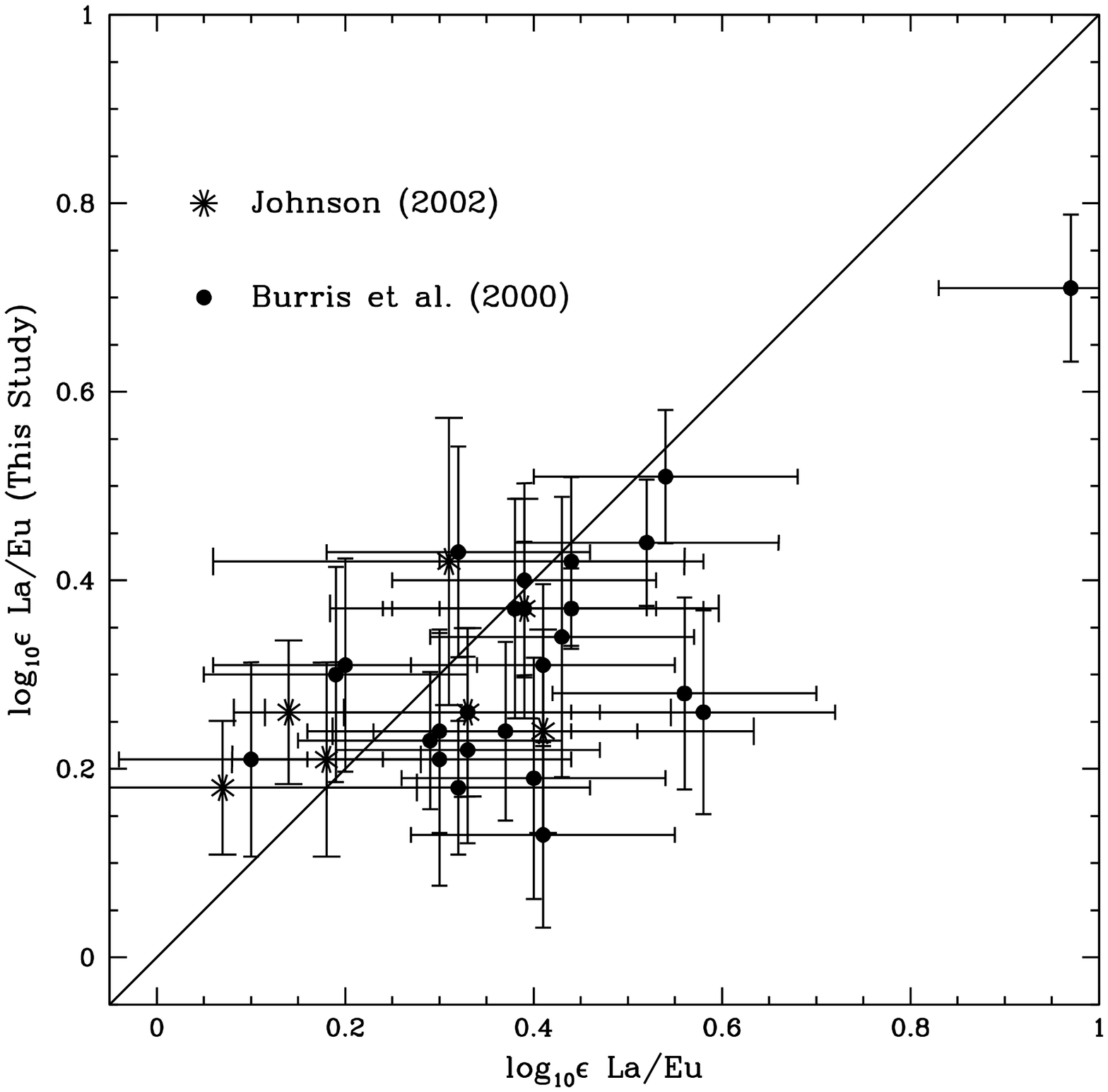}
\caption{log$\epsilon$(La/Eu) as derived here  compared with log$\epsilon$(La/Eu) derived in \citet{Burris2000} and \citet{Johnson2002}.
\label{comp3}}
\end{figure}

\begin{figure}
\plotone{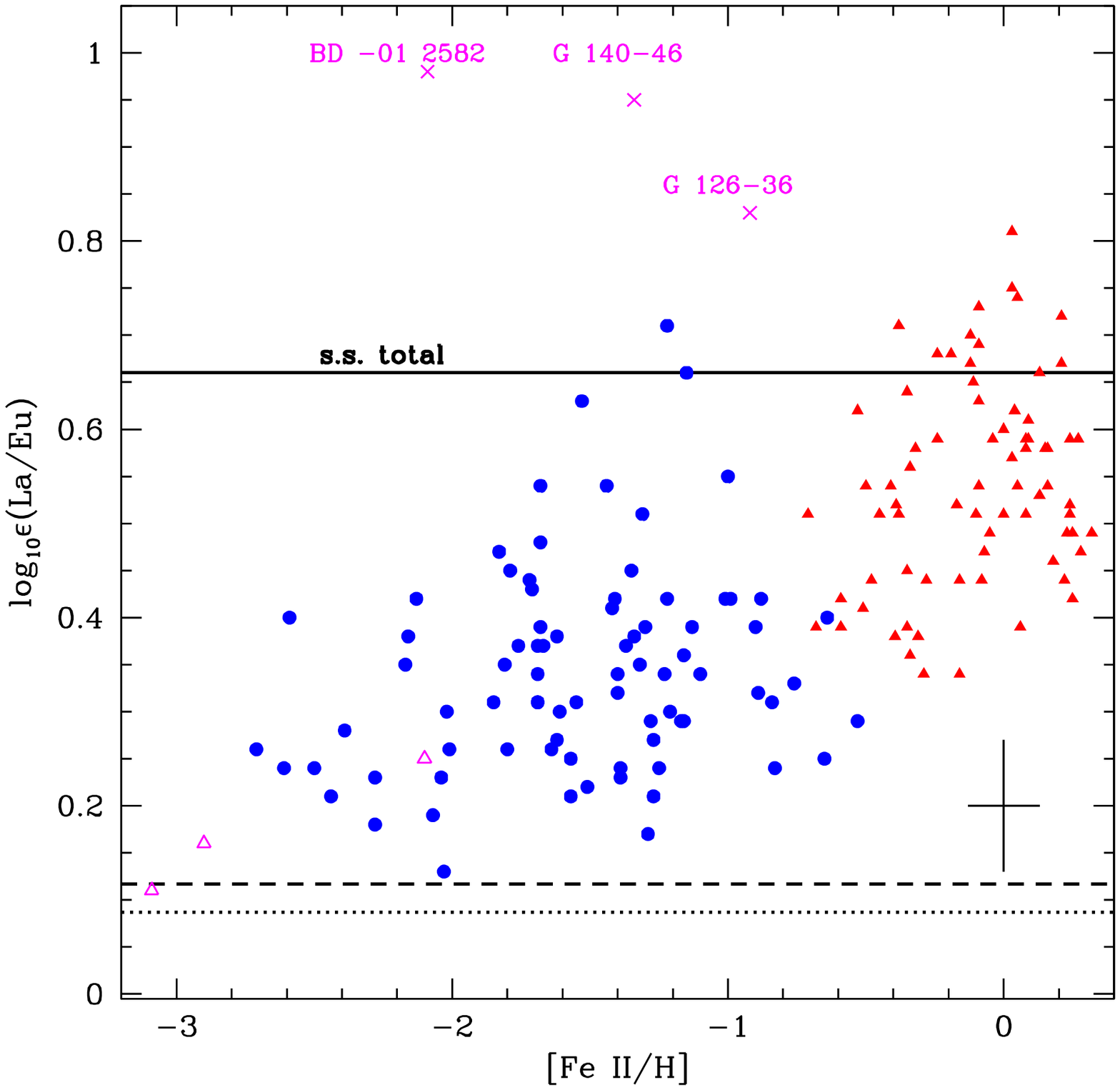}
\caption{log$\epsilon$(La/Eu) as a function of [Fe/H].   The solid line is the total solar system La/Eu ratio (from \citealt{Lodders2003}).  The broken lines indicate the solar system $s$- and $r$-process abundance breakdowns (see text for descriptions), where the dotted line is from \citet{Burris2000} and the dashed line is from \citet{Arlandini1999}.  Three metal-poor but La-rich stars are labeled.  
Symbols are as follows: circles, this study; filled triangles, \citet{Woolf1995}; open 
triangles, three $r$-process enhanced stars (BPS CS 31082-001 \citealt{Hill2002}, \bd17  \citealt{BD17}, and \cs22 \citealt{CS22892}); crosses, $s$-process enhanced stars.  A typical error is shown.
\label{feh}}
\end{figure}

\begin{figure}
\plotone{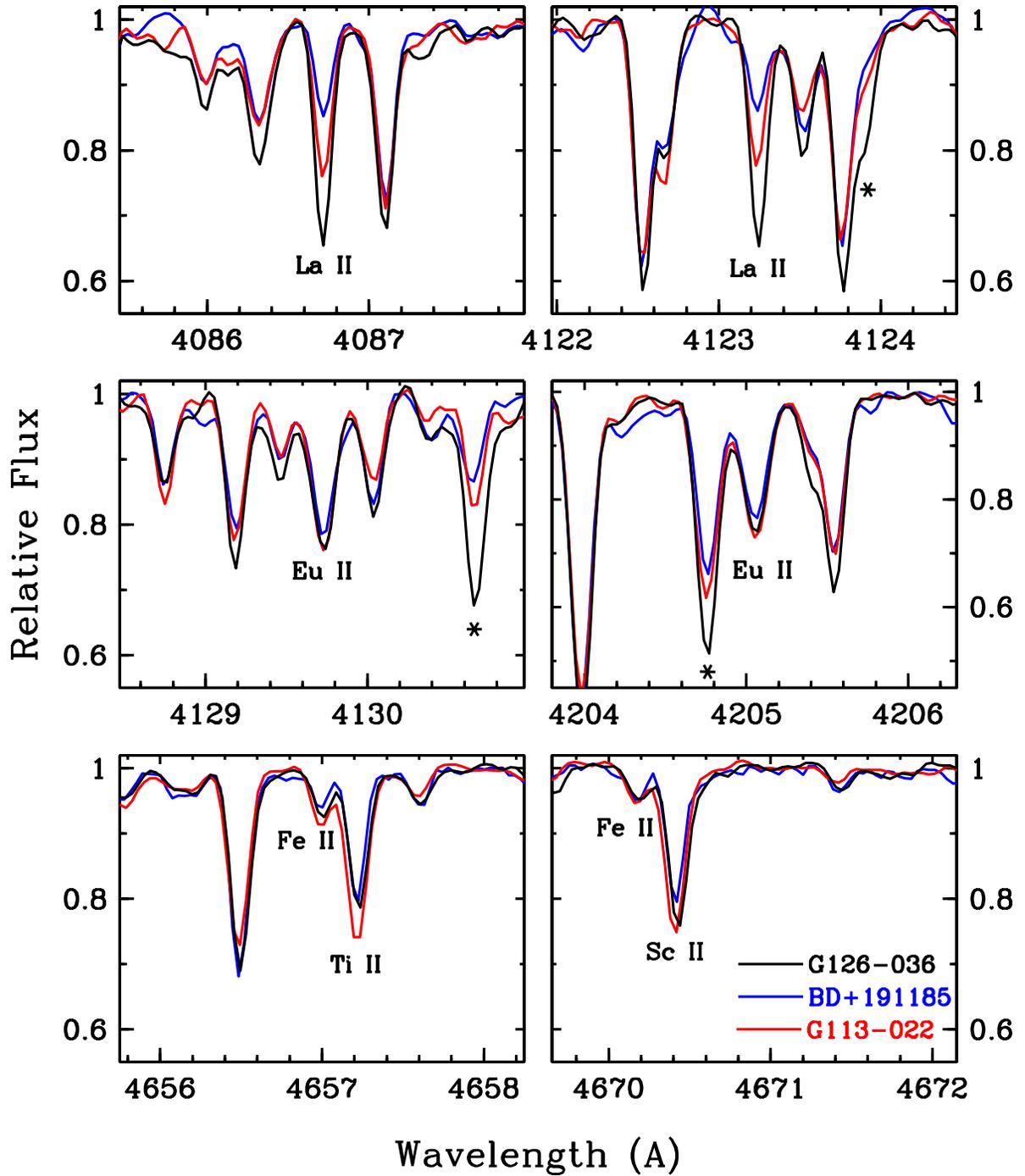}
\caption{Three stars with identical stellar parameters but very different La abundances.  Other $s$-process element features are marked with an asterisk (*).
\label{trio}}
\end{figure}

\begin{figure}
\plotone{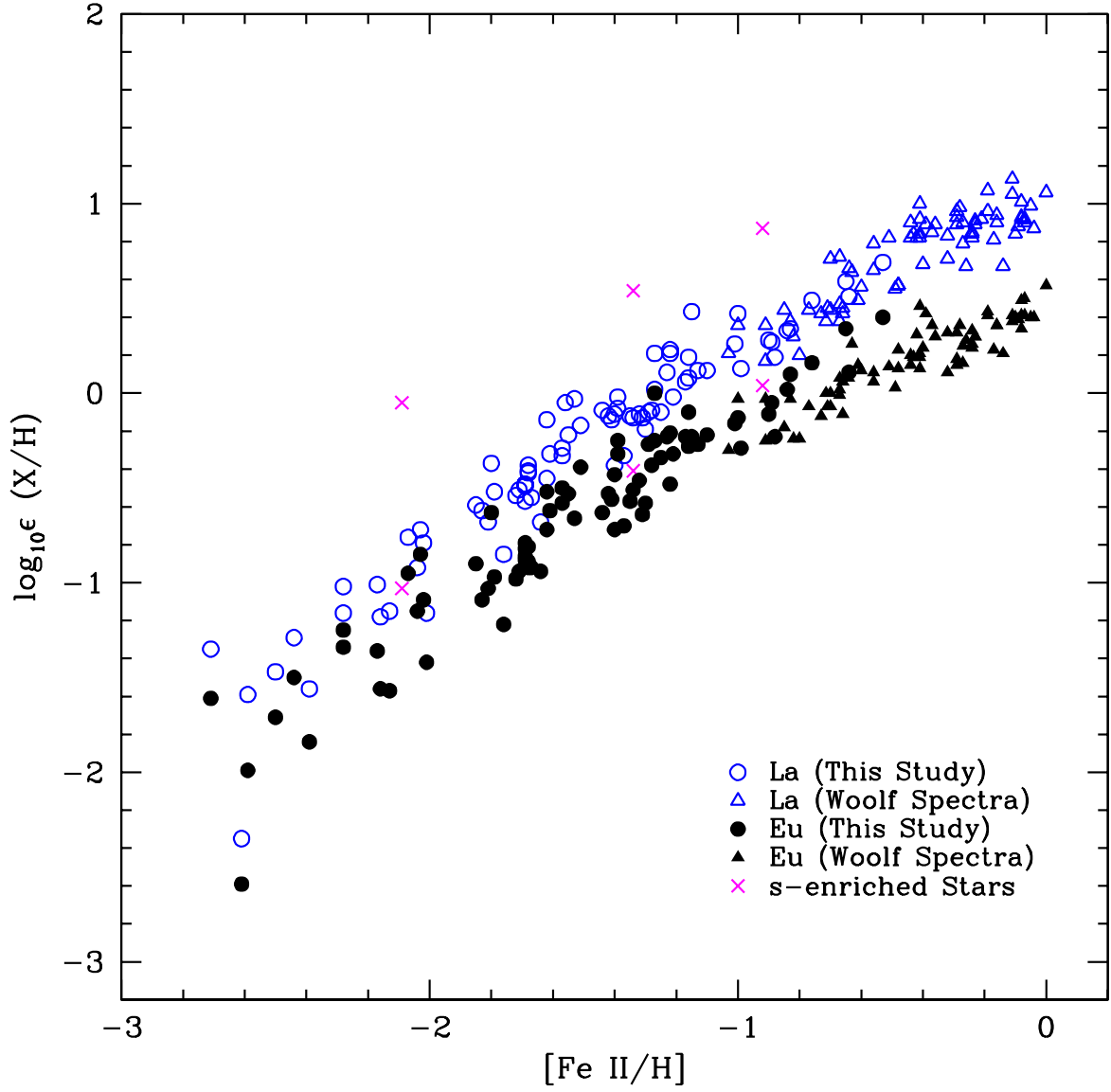}
\caption{La (open symbols)  and Eu (filled symbols) separately.  The ``$s$-enhanced'' stars are marked.
\label{logeps}}
\end{figure}

\begin{figure}
\plotone{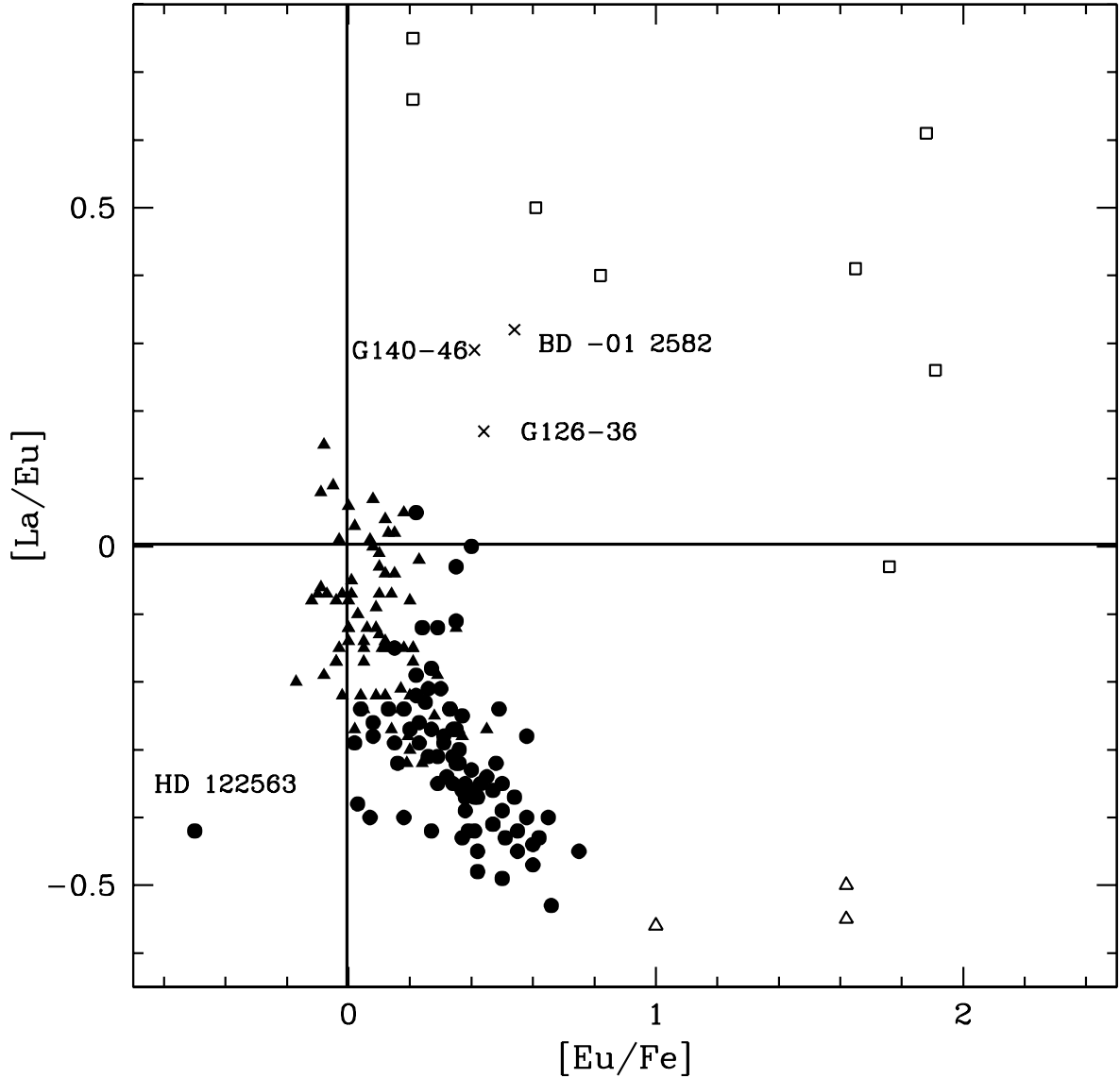}
\caption{[La/Eu] as a function of [Eu/Fe].  Symbols are as in Fig. \ref{feh}, where the open triangles again are the known $r$-process rich stars, BD$+17^{\circ} 3248$ \citep{BD17}, BPS CS 22892-052 \citep{CS22892}, and BPS CS 31082-001 \citep{Hill2002}.  In addition, the $s$-process rich but metal-poor stars for which Pb abundances have been measured by \citet{Aoki2002} are plotted with open squares.
\label{flin}}
\end{figure}

\begin{figure}
\plotone{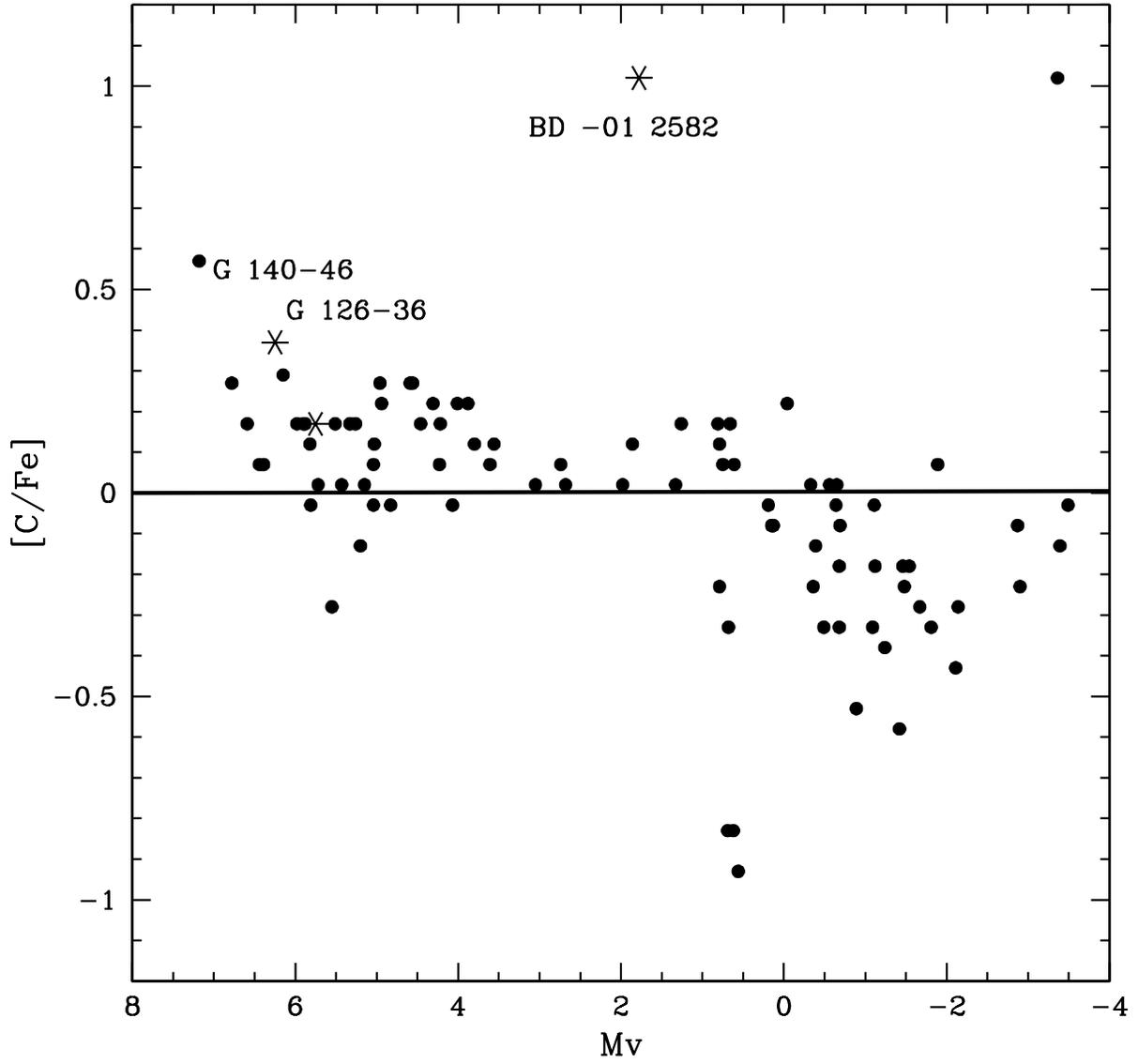}
\caption{Our derived [C/Fe] as a function of M$_V$, i.e., evolutionary state.
\label{c}}
\end{figure}

\begin{figure}
\plotone{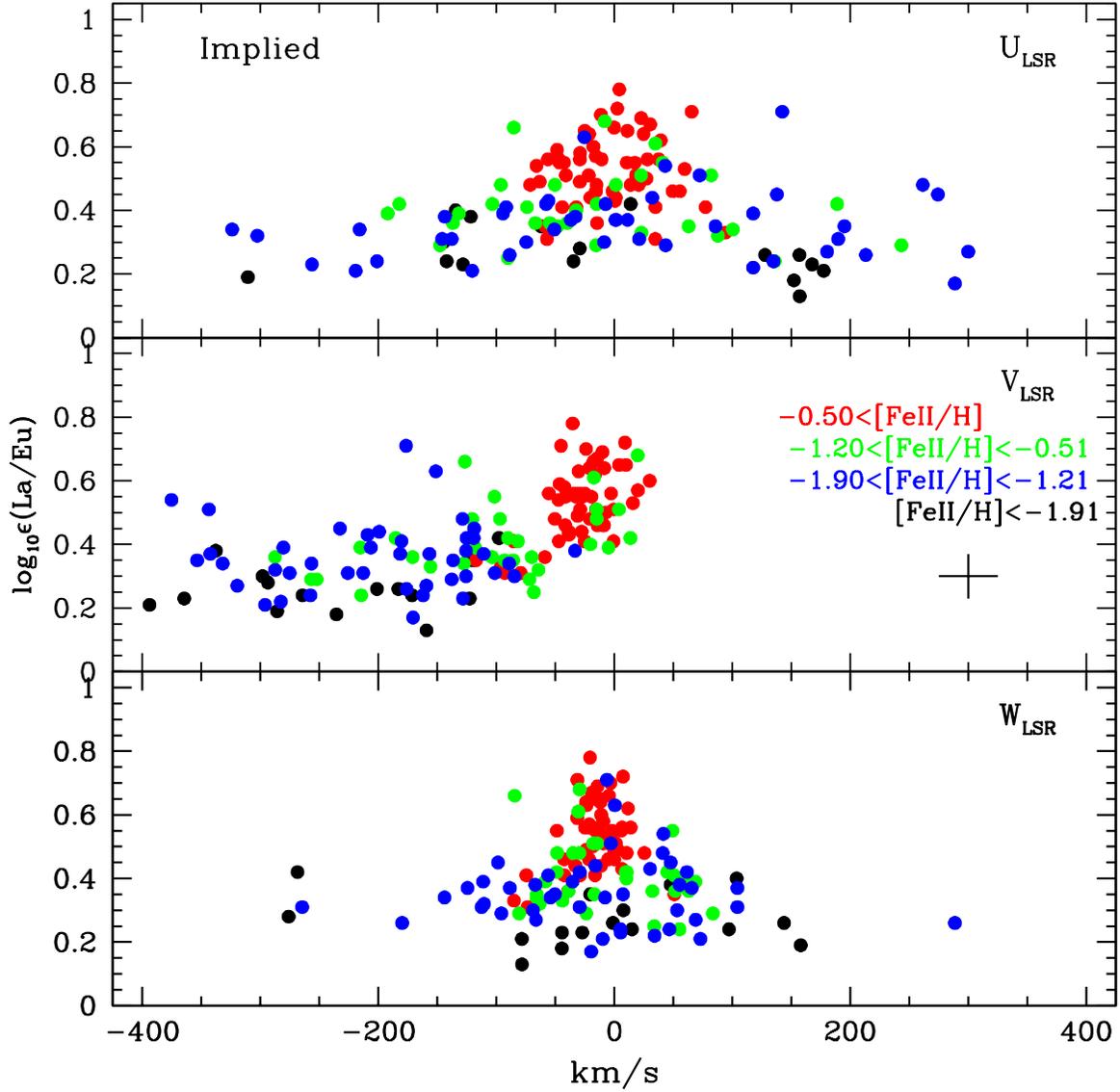}
\caption{La/Eu as a function of space velocity components.  Only stars with errors in U, V, and W less than 100 km/s are shown.  These velocities are computed from distances that are in accord with the spectroscopically derived stellar parameters.  The C-enhanced, $s$-process rich stars are not shown.  The cross indicates a typical errorbar.
\label{mot}}
\end{figure}

\begin{figure}
\plotone{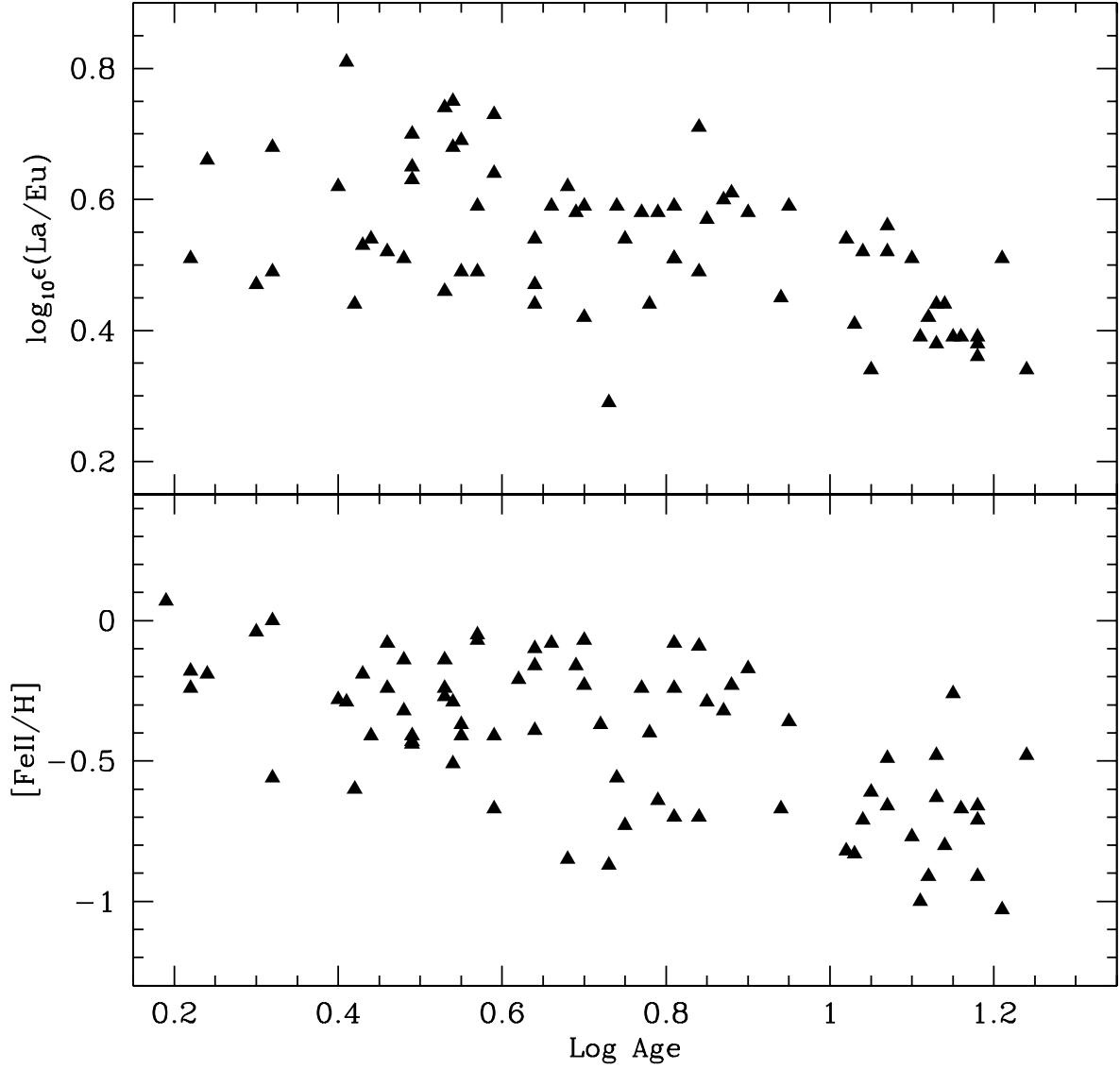}
\caption{[\ion{Fe}{2}/H] abundances and La/Eu ratios  as a function of stellar age.  Only stars from \citet{Woolf1995} are shown.
\label{bdpvin1}}
\end{figure}

\begin{figure}
\plotone{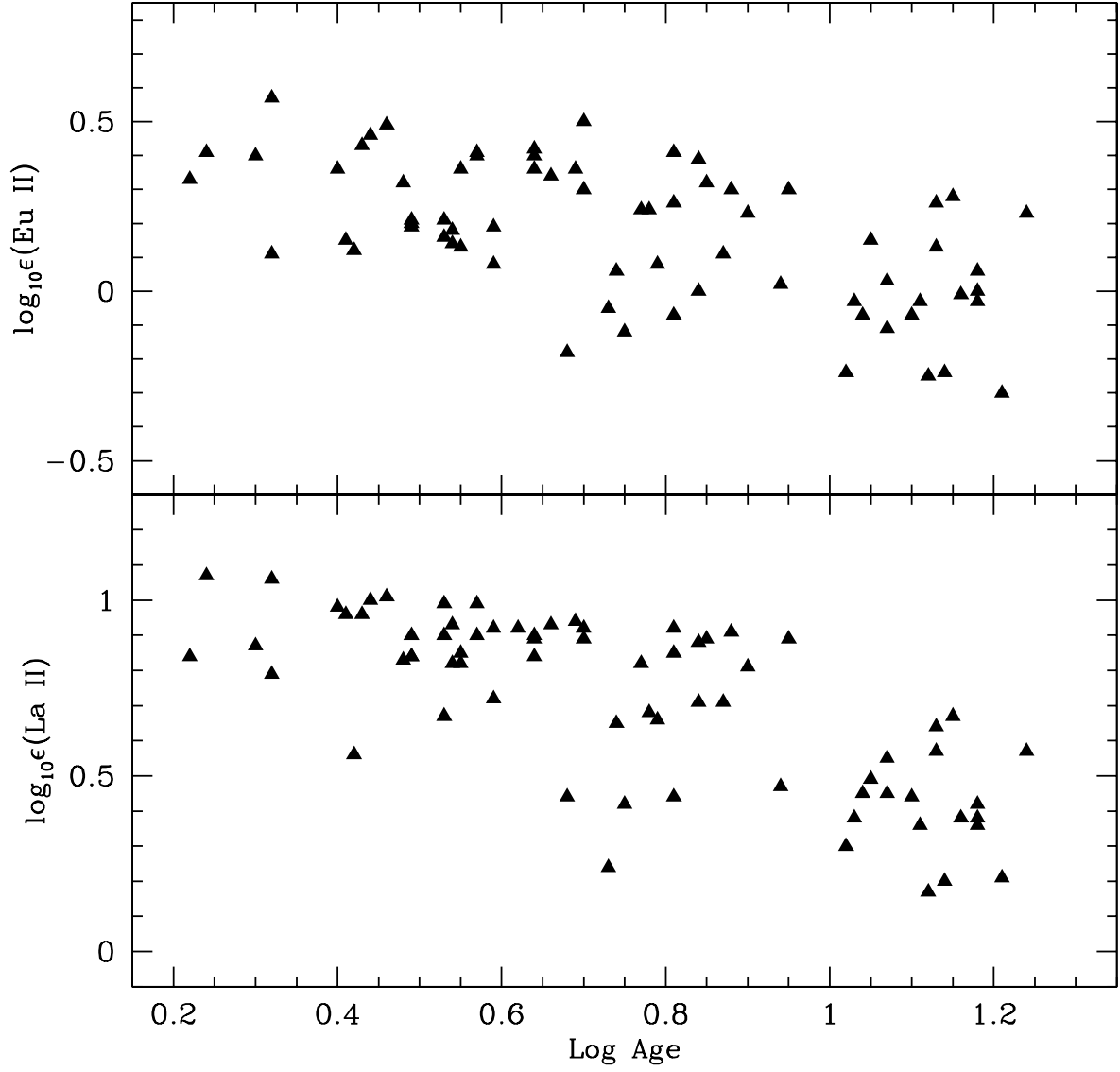}
\caption{La and Eu abundances as a function of stellar age.  Only stars from \citet{Woolf1995} are shown.
\label{bdpvin2}}
\end{figure}

\clearpage
\begin{figure}
\plotone{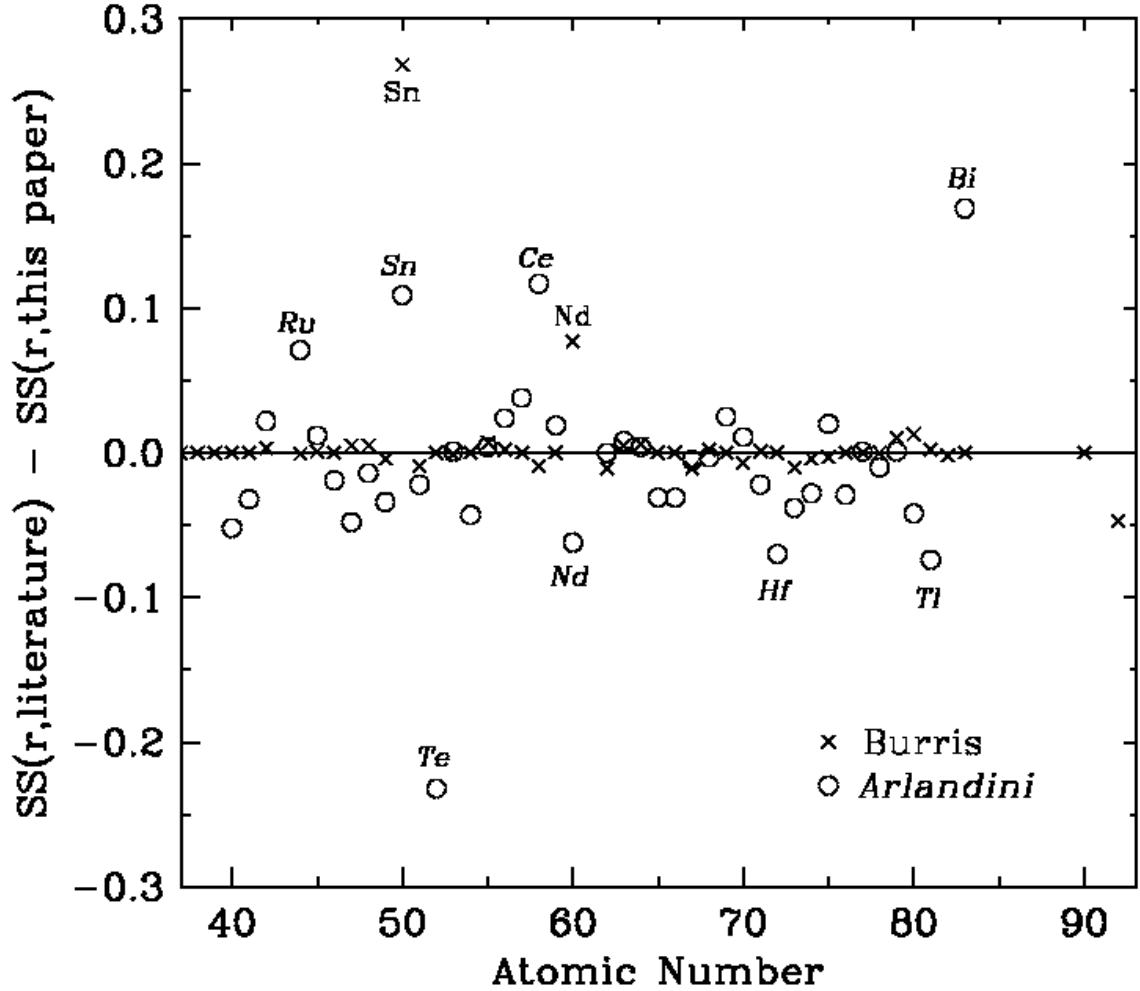}
\caption{Comparison of the solar system $r$-process abundances re-evaluated
in this paper (Table A) with previously-published values.  For
each atomic number, the ordinate value is the $r$-process abundance
from the literature (X-symbols, Burris et al. 2000; open circles,
Arlandini et al. 1999) minus the present value.
Abundance discrepancies greater than 0.05 dex are labeled in the figure,
and discussed in the Appendix.
\label{newfig}}
\end{figure}

\end{document}